\documentclass[11pt,a4paper]{article}
\usepackage{authblk}
\usepackage{lmodern}
\usepackage{jstyle}
\usepackage{amsmath}
\usepackage{mathtools}
\usepackage{tikz}
\usetikzlibrary{decorations.pathmorphing,arrows.meta,decorations.markings}
\tikzset{snake it/.style={decorate, decoration=snake}}

\usepackage{empheq}
\usepackage{enumitem}
\usepackage{graphics}
\usepackage{wrapfig}
\usepackage{caption}
\usepackage{natbib}
\usepackage{xcolor}
\usepackage{framed}
\usepackage{array}
\definecolor{shadecolor}{gray}{0.925}

\usepackage[cbgreek]{textgreek}

\def\sideremark#1{\ifvmode\leavevmode\fi\vadjust{\vbox to0pt{\vss
 \hbox to 0pt{\hskip\hsize\hskip1em
 \vbox{\hsize3cm\tiny\raggedright\pretolerance10000
 \noindent #1\hfill}\hss}\vbox to8pt{\vfil}\vss}}}%

\newcommand{\bi}{\begin{itemize}}
\newcommand{\ei}{\end{itemize}}
\newcommand{\bea}{\begin{align}}
\newcommand{\eea}{\end{align}}
\newcommand{\be}{\begin{equation}}
\newcommand{\ee}{\end{equation}}

\newcommand{\dd}{\text{d}}

\newcommand{\hs}{\hspace{0.35cm}}

\makeatletter
\renewcommand*\env@matrix[1][\arraystretch]{%
  \edef\arraystretch{#1}%
  \hskip -\arraycolsep
  \let\@ifnextchar\new@ifnextchar
  \array{*\c@MaxMatrixCols c}}
\makeatother

\author{\, Ernesto BIANCHI}
\author{\, Sacha PLOYET}
\author{\, Charlotte SLEIGHT}
\author{\, Massimo TARONNA}

\affiliation{Dipartimento di Fisica ``Ettore Pancini'', Universit\`a degli Studi di Napoli Federico II, \\Monte S. Angelo, Via Cintia, 80126 Napoli, Italy}

\affiliation{INFN, Sezione di Napoli, Monte S. Angelo, Via Cintia, 80126 Napoli, Italy}

\emailAdd{ernesto.bianchipalacios@unina.it,sacha.ployet@unina.it,\\ charlotte.sleight@na.infn.it, massimo.taronna@unina.it}

\title{\centering \fontsize{25pt}{25pt}\selectfont Late-time electric screening\\ in dS from EAdS loops}

\abstract{\noindent We compute the one-loop gauge-field two-point functions analytically in scalar QED
on dS$_{d+1}$ for general dimension and scalar mass, using the EAdS
reformulation of the in-in formalism
\cite{Sleight:2020obc,Sleight:2021plv}. The Dirichlet EAdS gauge field acquires a
Schwinger-type mass, with its longitudinal mode supplied by a composite
of charged scalars obeying opposite boundary conditions, while the
Neumann mode remains protected by gauge invariance. In dS, the
Dirichlet correction induces an anomalous dimension $\gamma_J$ for the
boundary current and modifies the late-time falloff of the electric
field. This bulk calculation reproduces the boundary
multiplet-recombination result of \cite{Sleight:2025dmt}. Summing over the in-in contour branches gives the retarded dS
bubble kernel, making direct contact with the Bros-Moschella framework. We establish a spectral
Cutkosky rule that expresses the discontinuity of this kernel
across the principal-series cut in the Casimir plane in terms
of a two-particle cut, with both internal scalar lines placed
on shell. Although the full
bubble coefficient involves a ${}_7F_6$ hypergeometric function, the
cut collapses to a closed product of Gamma functions and determines the complete non-local
part of the kernel through a dispersion relation. Analytic
continuation of the same cut reproduces the complete one-loop result
for $\gamma_J$. In the complex Casimir plane, the loop-corrected
Dirichlet photon appears as a pole on the second sheet and is therefore
naturally interpreted as a dS resonance, or quasinormal mode, whose
frequency is shifted at one loop.}

\begin{document}

\begin{flushright}    
\texttt{}
\end{flushright}

\maketitle

\section{Introduction}\label{sec::Intro}

Late-time correlation functions are central observables in quantum field theory (QFT) on de Sitter (dS) space. They encode the statistics of primordial fluctuations generated during inflation \cite{Mukhanov:1981xt,Maldacena:2002vr,Bartolo:2004if,Weinberg:2005vy} and furnish the boundary data for dS holography \cite{Witten:2001kn,Strominger:2001pn,Maldacena:2002vr,McFadden:2009fg}. At future infinity, the dS isometries act as conformal transformations, organising the field content and interactions of the theory into conformal data. Extracting these data from interacting theories is nevertheless difficult, particularly beyond tree level, where the Schwinger-Keldysh structure of the correlators and the appearance of infrared and secular effects complicate perturbation theory.

\vskip 4pt
These questions have been approached from several complementary directions. These include foundational studies of the analytic and spectral structure of dS QFT \cite{Bros:1994dn,Bros:1995js,Bros:1998ik,Bros:2010rku}; approaches based on the wavefunction of the universe \cite{Maldacena:2002vr,Harlow:2011ke,McFadden:2009fg,Anninos:2014lwa} and its consistency conditions \cite{Salcedo:2022aal,Goodhew:2024eup}; conformal symmetry \cite{Antoniadis:2011ib,Maldacena:2011nz,Bzowski:2012ih,Mata:2012bx,Ghosh:2014kba} and the cosmological bootstrap \cite{Arkani-Hamed:2015bza,Arkani-Hamed:2018kmz,Baumann:2022jpr}; Mellin-Barnes representation \cite{Sleight:2019mgd,Sleight:2021plv,Qin:2022lva}; geometric approaches \cite{Arkani-Hamed:2017fdk,Arkani-Hamed:2023kig,Baumann:2025qjx,CarrilloGonzalez:2025qjk,Arundine:2026fbr}; and Euclidean methods based on Wick rotation to the sphere \cite{Marolf:2010nz,Higuchi:2010xt,LopezNacir:2019ord,Anninos:2024fty} or to Euclidean anti-de Sitter (EAdS) space \cite{Sleight:2019mgd,Sleight:2019hfp,Sleight:2020obc}. In addition to providing new computational tools, these and other approaches offer complementary insights into the analytic and physical properties of dS observables.

\vskip 4pt
In this work, we use the EAdS reformulation of the in-in formalism developed in \cite{Sleight:2020obc,Sleight:2021plv,Abhishek:2025oki}, in which the two branches of the Schwinger-Keldysh contour are Wick rotated separately to EAdS. Each dS field $\varphi$ is then represented by two independent EAdS fields $\varphi_{\Delta_\pm}$ obeying Dirichlet and Neumann boundary conditions. Late-time dS correlators are then expressed in terms of EAdS diagrams, with interactions inherited from the original dS theory. This reformulation has allowed techniques developed in AdS/CFT to be imported into the study of dS late-time correlators, see e.g. \cite{Sleight:2020obc,Sleight:2021plv,Hogervorst:2021uvp,DiPietro:2021sjt,Heckelbacher:2022hbq,Bissi:2023bhv,Loparco:2023rug,Loparco:2023akg,Chowdhury:2023arc,DiPietro:2023inn,Bzowski:2023nef,Dey:2025kci,Chowdhury:2025ohm,Nowinski:2025cvw}. 

\vskip 4pt
This formulation has also helped uncover features of dS late-time correlators that depart from familiar AdS/CFT expectations. In gauge and gravity theories, for example, the shadow structure of the boundary spectrum provides the composite operators needed for multiplet recombination, allowing boundary currents to acquire anomalous dimensions \cite{Sleight:2025dmt}. In dS scalar QED, with gauge field $A$ and charged scalar $\varphi$, the relevant late-time falloffs take the form
\begin{align}
    A_i(\eta \to 0^-,\vec{x})
    &\sim
    \tilde a_i(\vec{x})
    +
    (-\eta)^{d-2+\gamma_J}J_i(\vec{x}),\\
    \varphi(\eta \to 0^-,\vec{x})
    &\sim (-\eta)^{\Delta^\varphi_+}{\cal O}_{\Delta^\varphi_+}(\vec{x})+(-\eta)^{\Delta^\varphi_-}{\cal O}_{\Delta^\varphi_-}(\vec{x}),
\end{align}
where $\tilde a_i$ is the boundary gauge boson and $J_i$ the boundary current with anomalous dimension $\gamma_J$, and
${\cal O}_{\Delta^\varphi_\pm}$ are the charged scalar operators associated with the two scalar falloffs. At leading order, the non-conservation of the current is sourced by the double-trace composite of shadow scalar operators 
\begin{equation}\label{eq::multrecomb}
    \partial \cdot J \sim
e\left(
{\cal O}_{\Delta^\varphi_+}^{\dagger}{\cal O}_{\Delta^\varphi_-}
-
{\cal O}_{\Delta^\varphi_+}{\cal O}_{\Delta^\varphi_-}^{\dagger}
\right)
+O(e^2), \qquad \Delta^\varphi_++\Delta^\varphi_- = d.
\end{equation}
The leading anomalous dimension $\gamma_J\sim O(e^2)$ was previously extracted in \cite{Sleight:2025dmt} using this multiplet recombination relation. In the present work, we recover the same result from an explicit one-loop computation of the gauge-field two-point function and from a spectral Cutkosky rule, which determines
the anomalous dimension through the analytic continuation
of the two-particle cut.

\begin{figure}[t]
    \centering
    \includegraphics[width=0.3\textwidth]{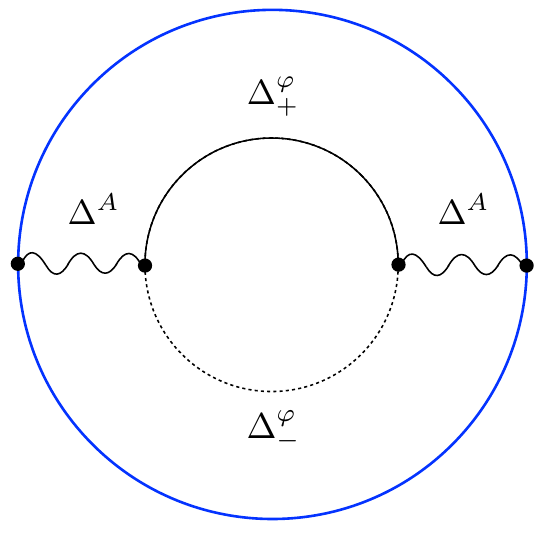}
    \caption{For the EAdS Dirichlet field, $\Delta^A=\Delta^A_+$, the mixed scalar bubble generates the mass associated with multiplet recombination. Its double cut places the shadow-related scalar lines $\Delta^\varphi_\pm$ on shell and determines the current anomalous dimension $\gamma_J$.}
    \label{fig::rebub}
\end{figure}

\vskip 4pt
The mechanism is particularly transparent in the EAdS formulation, where the Dirichlet gauge field $A_{\Delta^A_+}$ is associated with the boundary current $J_i$, while the Neumann field $A_{\Delta^A_-}$ is associated with the boundary gauge field ${\tilde a}_i$. An anomalous dimension for the current therefore appears as a mass for the Dirichlet field $A_{\Delta^A_+}$, which is generated through a Schwinger-type mechanism \cite{Schwinger:1962tn,Schwinger:1962tp}. A composite of charged scalars $\varphi_{\Delta^\varphi_+}$ and $\varphi_{\Delta^\varphi_-}$ obeying opposite boundary conditions supplies the longitudinal mode required for multiplet recombination, as illustrated in figure \ref{fig::rebub}. This EAdS mass, however, has a qualitatively different meaning in dS. Since it shifts only the current falloff while the boundary gauge field
remains protected, it does not describe a conventional massive photon. Instead, as we discuss further below, the corrected mode is naturally interpreted as a dS resonance.

\vskip 4pt
While analogous mechanisms can occur for suitable choices of mixed boundary conditions in AdS \cite{Manvelyan:2008ks,Rattazzi:2009ux,Porrati:2009dy,Giombi:2011ya,Karch:2023wui,Porrati:2024zvi}
and for the dS wavefunction of the universe \cite{Baumann:2025tkm}, in
the EAdS representation of dS late-time correlators both boundary conditions are simultaneously present and coupled, making the presence of a recombination partner intrinsic to the construction. The remaining question is whether this recombination survives the interference between the two in-in branches required by unitary time evolution. Whether it does is already encoded in the corresponding three-point function \cite{Sleight:2025dmt}.

\vskip 4pt
We compute analytically the one-loop two-point functions of the Dirichlet and Neumann EAdS gauge fields in scalar QED for general dimension and scalar mass.\footnote{For previous works on bubble and other loop diagrams in AdS, see e.g \cite{Giombi:2017hpr,Carmi:2018qzm,Carmi:2019ocp,Albayrak:2020bso,Carmi:2021dsn,Ankur:2023lum,Cacciatori:2024zbe,Cacciatori:2024zrv,Ciccone:2024guw,Ciccone:2025dqx,Ankur:2026ylr,Castellani:2026wwh}.} The derivative structure of the gauge-field bubble relates it to a
scalar bubble in boundary dimension $D=d+2$, with the two internal scalar dimensions shifted as $\Delta_i\to\Delta_i+1$ and the external spin-one dimension mapped as $\Delta^A\to\Delta^A+1$. This is the EAdS
counterpart of the dimension-shift identities for tensor loop integrals familiar from flat-space amplitudes
\cite{Bern:1993kr,Tarasov:1996br}. Applied to the EAdS representation of dS late-time correlators, the Dirichlet result reproduces the anomalous dimension of the boundary current derived in \cite{Sleight:2025dmt} using the multiplet recombination \eqref{eq::multrecomb}, while the Neumann result confirms that the boundary gauge field is protected, as required by boundary gauge invariance.

\vskip 4pt
From the dS perspective, the result has a direct late-time interpretation. The anomalous dimension $\gamma_J$ modifies the falloff of the contribution to the electric field carried by the boundary current. For $\gamma_J>0$, it therefore decays more rapidly than in the free theory, an effect we refer to as late-time electric screening.\footnote{Here ``late-time electric screening'' refers to the falloff of the boundary current contribution to the electric field, not to Debye screening of the long-wavelength electrostatic response; the corresponding one-loop Debye mass vanishes in the Bunch-Davies vacuum of dS scalar QED \cite{Popov:2017xut}.} For $\gamma_J<0$, it is instead enhanced, corresponding to electric anti-screening. The Neumann branch remains unshifted, so that the leading magnetic contribution retains its free-theory falloff, as required by boundary gauge invariance. For principal series scalars in dS$_4$ there is always electric screening $\gamma_J>0$.

\vskip 4pt
Since only one of the two gauge-field late-time falloffs is corrected, they no longer form a shadow pair, and the effect cannot be described by a local Proca mass in dS---which would shift both falloffs while preserving their shadow relation. Its dS interpretation is clarified by the causal and analytic structure of the scalar bubble. Summing over the two branches of the in-in contour yields the retarded bubble kernel of Bros-Moschella \cite{Bros:1990cu,Bros:1995js}. The corrected Dirichlet mode then appears as a pole of the analytically continued full gauge-field response, identifying it as a dS resonance rather than a stable particle.

\vskip 4pt
We also establish a spectral Cutkosky rule for dS: the discontinuity of the retarded kernel
across the principal-series cut in the Casimir plane is obtained
by replacing internal EAdS propagators by on-shell harmonic
functions. Although the full bubble coefficient involves a ${}_7F_6$ hypergeometric function,
its double cut reduces to a closed-form product of Gamma functions, from
which the complete non-local part of the retarded bubble can be reconstructed through a single dispersion integral. We denote by $C^{(D)}_{\Delta_1,\Delta_2;\Delta_3}$ the scalar-bubble coefficient in boundary dimension $D$, where $\Delta_1$ and $\Delta_2$ are the dimensions of the two scalar
fields running in the loop, and $\Delta_3$ is the external scaling dimension. The dimension-shifting relation between vector and scalar bubbles then gives the schematic relation
\begin{equation}
\gamma_J
\;\propto\;
e^2
\left.
\operatorname{Cut}_{\Delta_1\Delta_2}
C^{(d+2)}_{\Delta_1,\Delta_2;\Delta_3}
\right|_{\substack{
\Delta_1=\Delta^\varphi_++1,\,
\Delta_2=\Delta^\varphi_-+1,\,\Delta_3=d
}} .
\end{equation}
In scalar QED, boundary gauge invariance protects the Neumann
mode, so that the complete one-loop current anomalous dimension
is fixed by the analytic continuation of this cut. Local
ultraviolet terms cancel in the Dirichlet-Neumann difference.

\vskip 4pt
\paragraph{Outline.} The paper is organised as follows. Section~\ref{sec::eadssqed} reviews
the EAdS reformulation of the Schwinger-Keldysh formalism and sets up
dS scalar QED in this language. Section~\ref{sec::scalarloops} computes
the relevant EAdS scalar loop integrals in position space, the tadpole
and the bubble. Section~\ref{sec::loopssQEDEAdS} assembles them into
the one-loop gauge-field two-point functions of EAdS scalar QED, and
section~\ref{sec::dSloop} translates the result into the dS late-time
two-point functions, extracting the anomalous dimension $\gamma_J$.
Section~\ref{sec::interpretation} discusses the physical
interpretation: Schwinger mass generation in EAdS
(section~\ref{subsec::schweads}), late-time electric screening in dS
(section~\ref{subsec::electscreen}), the Schwinger-Keldysh
interpretation in terms of resonances (section~\ref{subsec::SKinterp}), and the spectral Cutkosky rule and dispersive reconstruction of the retarded dS bubble (section~\ref{sec::retarded}). Hypergeometric identities, the isometry constraints on EAdS bubble diagrams, the comparison with flat-space renormalisation and the dimension-shifting relations are collected in appendices~\ref{app:hyper}--\ref{app::FlatDimensionShift}.

\subsection{Notations and conventions}

We work in the expanding Poincar\'e patch of $\mathrm{dS}_{d+1}$ and the Poincar\'e patch of $\mathrm{EAdS}_{d+1}$, with unit curvature radii and coordinates $x^\mu=(\eta,x^i)$ and $x^\mu=(z,x^i)$, respectively:
\begin{align}\label{poincare}
{\rm d}s^2_{\mathrm{EAdS}}
=
\frac{{\rm d}z^2+{\rm d}\vec{x}^{2}}{z^2},
\qquad z>0, \qquad
\
{\rm d}s^2_{\mathrm{dS}}
=
\frac{-{\rm d}\eta^2+{\rm d}\vec{x}^{2}}{\eta^2},
\qquad \eta<0.
\end{align}
Their conformal boundaries are approached as $z\to0^+$ and $\eta\to0^-$, respectively. Greek indices $\mu,\nu,\ldots=0,\ldots,d$ denote bulk directions, with $0$ corresponding to $z$ in EAdS and to $\eta$ in dS. Latin indices $i,j,\ldots=1,\ldots,d$ denote boundary directions. Bulk indices are contracted with the corresponding bulk metric and boundary indices with $\delta_{ij}$.

\vskip 4pt
For sums of scaling dimensions, we use the shorthand 
\begin{equation}
    \Delta_{12 \ldots n}\equiv\Delta_1+\Delta_2+ \ldots + \Delta_n.
\end{equation}

We denote EAdS bulk-to-bulk and bulk-to-boundary propagators by
$G^{(d)}_\Delta$ and $K^{(d)}_\Delta$, respectively, with tensor
indices displayed when needed. The superscript $(d)$ specifies
the boundary dimension, so these propagators are defined on
$\mathrm{EAdS}_{d+1}$. We omit this superscript when the dimension is clear from the context.

\newpage

\section{EAdS reformulation of dS scalar QED}
\label{sec::eadssqed}

In this section we review the EAdS reformulation of dS late-time correlators developed in~\cite{Sleight:2020obc,Sleight:2021plv}, together with its extension to gauge and gravity theories, including scalar QED, in~\cite{Abhishek:2025oki}. We first recall the Schwinger-Keldysh generating functional and its rotation to EAdS, and then collect the resulting EAdS Feynman rules for scalar QED in the form needed for the one-loop calculation below.

\subsection{Schwinger-Keldysh generating functional}

Late-time correlators in dS are naturally computed using the Schwinger-Keldysh, or
in-in, formalism
\cite{Schwinger:1960qe,Feynman:1963fq,Keldysh:1964ud}.
For early path-integral applications in curved spacetime see
\cite{Calzetta:1986ey}, and for applications to cosmological correlators see \cite{Maldacena:2002vr,Weinberg:2005vy}. In the path-integral representation (for a review see e.g. \cite{Chen:2017ryl}) the field content is doubled,
\begin{equation}
    \Phi \longrightarrow \Phi_+,\Phi_- ,
\end{equation}
where $\Phi$ collectively denotes the field content of the theory, and the subscripts denote the
two branches of the Schwinger-Keldysh contour. For a theory with dS Lagrangian density
${\cal L}^{\rm dS}$, the Schwinger-Keldysh generating functional with bulk sources ${\cal J}_\pm$ and
final time $\eta_0$ in the Bunch-Davies vacuum \cite{Chernikov:1968zm,Schomblond:1976xc,Gibbons:1977mu,Bunch:1978yq} is 
\begin{multline}\label{SK generating scalar QED}
Z_{\rm SK}[{\cal J}_+,{\cal J}_-]
=
\int_{\Phi_+(\eta_0,\vec{x})=\Phi_-(\eta_0,\vec{x})}
{\cal D}\Phi_+{\cal D}\Phi_-
\\
\times
\exp\left[
i\int_{-\infty(1-i\epsilon)}^{\eta_0}
\frac{{\rm d}\eta\,{\rm d}^d\vec{x}}{(-\eta)^{d+1}}
\left(
{\cal L}^{\rm dS}[\Phi_+]+{\cal J}_+\cdot\Phi_+
\right)
\right]
\\
\times
\exp\left[
-i\int_{-\infty(1+i\epsilon)}^{\eta_0}
\frac{{\rm d}\eta\,{\rm d}^d\vec{x}}{(-\eta)^{d+1}}
\left(
{\cal L}^{\rm dS}[\Phi_-]+{\cal J}_-\cdot\Phi_-
\right)
\right],
\end{multline}
where $\eta_0\to0$ is the late-time boundary. Functional derivatives with respect to the bulk sources insert fields on the corresponding
Schwinger-Keldysh branch. For example,
\begin{equation}
    \left\langle
    \Phi_{a_1}(\eta_1,\vec{x}_1)
    \cdots
    \Phi_{a_n}(\eta_n,\vec{x}_n)
    \right\rangle_{\rm SK}
    =
    \left.
    \prod_{i=1}^{n}
    \left(
        \frac{1}{i a_i}
        \frac{\delta}{\delta{\cal J}_{a_i}(\eta_i,\vec{x}_i)}
    \right)
    Z_{\rm SK}[{\cal J}_+,{\cal J}_-]
    \right|_{{\cal J}_\pm=0},
\end{equation}
where $a_i = \pm$ label the two Schwinger-Keldysh branches. Late-time correlators are obtained by taking the insertion times to the final time slice,
$\eta_i\to\eta_0$, and then sending $\eta_0\to0$.\footnote{Since the two branches are identified at
$\eta_0$, the external late-time insertions may be represented on either branch.} In this framework, the Schwinger-Keldysh propagators connecting points on the contour are given by
\begin{equation}\label{SKprop}
    G_{ab}
    (\eta,\bar\eta;\vec{x},\vec{y})
    =
    \left.
    -\frac{1}{a b}
    \frac{\delta}{\delta{\cal J}_{a}(\eta,\vec{x})}
    \frac{\delta}{\delta{\cal J}_{b}(\bar\eta,\vec{y})}
    Z_{\rm SK}[{\cal J}_+,{\cal J}_-]
    \right|_{{\cal J}_\pm=0}, \qquad a,b=\pm.
\end{equation}

\vskip 4pt
\paragraph{Late-time boundary data.} 

At late times, the components of a free spin-$s$ field tangent to the
future boundary behave as
\begin{equation}
    \Phi_{i_1\ldots i_s}(\eta,\vec{x})
    \underset{\eta\to0^-}{\sim}
    (-\eta)^{\Delta_+-s}
    {\cal O}_{\Delta_+,i_1\ldots i_s}(\vec{x})
    +
    (-\eta)^{\Delta_--s}
    {\cal O}_{\Delta_-,i_1\ldots i_s}(\vec{x}).
\label{eq::lateTimeSpinJFalloffs}
\end{equation}
Here ${\cal O}_{\Delta_\pm,i_1\ldots i_s}$ are spin-$s$
conformal primaries of scaling dimensions $\Delta_\pm$, which follows from the fact that the dS isometry group $SO(d+1,1)$ acts on the future boundary as the Euclidean conformal group in $d$ dimensions. The two scaling dimensions are shadow related and satisfy
\begin{equation}
    \Delta_++\Delta_-=d,
    \qquad
    m_s^2=\Delta_+\Delta_-+s,
\label{eq::spinJMassDimensionRelation}
\end{equation}
where $m_s^2$ denotes the mass parameter appearing in the
Fierz-Pauli equations. Correlators of the operators ${\cal O}_{\Delta_\pm,i_1\ldots i_s}$ are extracted from bulk correlators by
isolating the corresponding late-time falloffs. For this purpose, we define the Schwinger-Keldysh bulk-to-boundary propagators $K^a_{\Delta_\pm}$ through the asymptotic expansion
\begin{multline}
   \left(G_{a\hat a}\right)_{\mu_1 \ldots \mu_s;i_1 \ldots i_s}(\eta,\bar{\eta};\vec{x},\vec{y})
    \underset{\bar{\eta}\to0^-}{\sim}
    (-\bar{\eta})^{\Delta_+-s}
    \left(K^a_{\Delta_+}\right)_{\mu_1 \ldots \mu_s;i_1 \ldots i_s}(\eta,\vec{x};\vec{y})
   \\ +
    (-\bar{\eta})^{\Delta_--s}
    \left(K^a_{\Delta_-}\right)_{\mu_1 \ldots \mu_s;i_1 \ldots i_s}(\eta,\vec{x};\vec{y}),
\label{SKbubo}
\end{multline}
where $a,\hat a=\pm$ label the two branches of the
Schwinger-Keldysh contour. Since the two Schwinger-Keldysh branches are identified at the final
time, the bulk-to-boundary propagators are independent of the branch
label $\hat a$ associated with the late-time endpoint.

\subsection{Rotation to EAdS}
\label{subsec::rottoeads}

The Poincar\'e patches \eqref{poincare} of dS and EAdS are related by a Wick rotation of the conformal-time coordinate. This has long been exploited in studies of the Bunch-Davies wavefunction \cite{Maldacena:2002vr,McFadden:2009fg,Harlow:2011ke,Anninos:2014lwa}, and was later extended to late-time correlators in the Schwinger-Keldysh formalism \cite{Sleight:2019mgd,Sleight:2019hfp,Sleight:2020obc,Sleight:2021plv,Abhishek:2025oki}. In this work, we apply the latter framework to the one-loop gauge field two-point function in dS scalar QED.

\begin{figure}[t]
    \centering
    \includegraphics[width=0.7\textwidth]{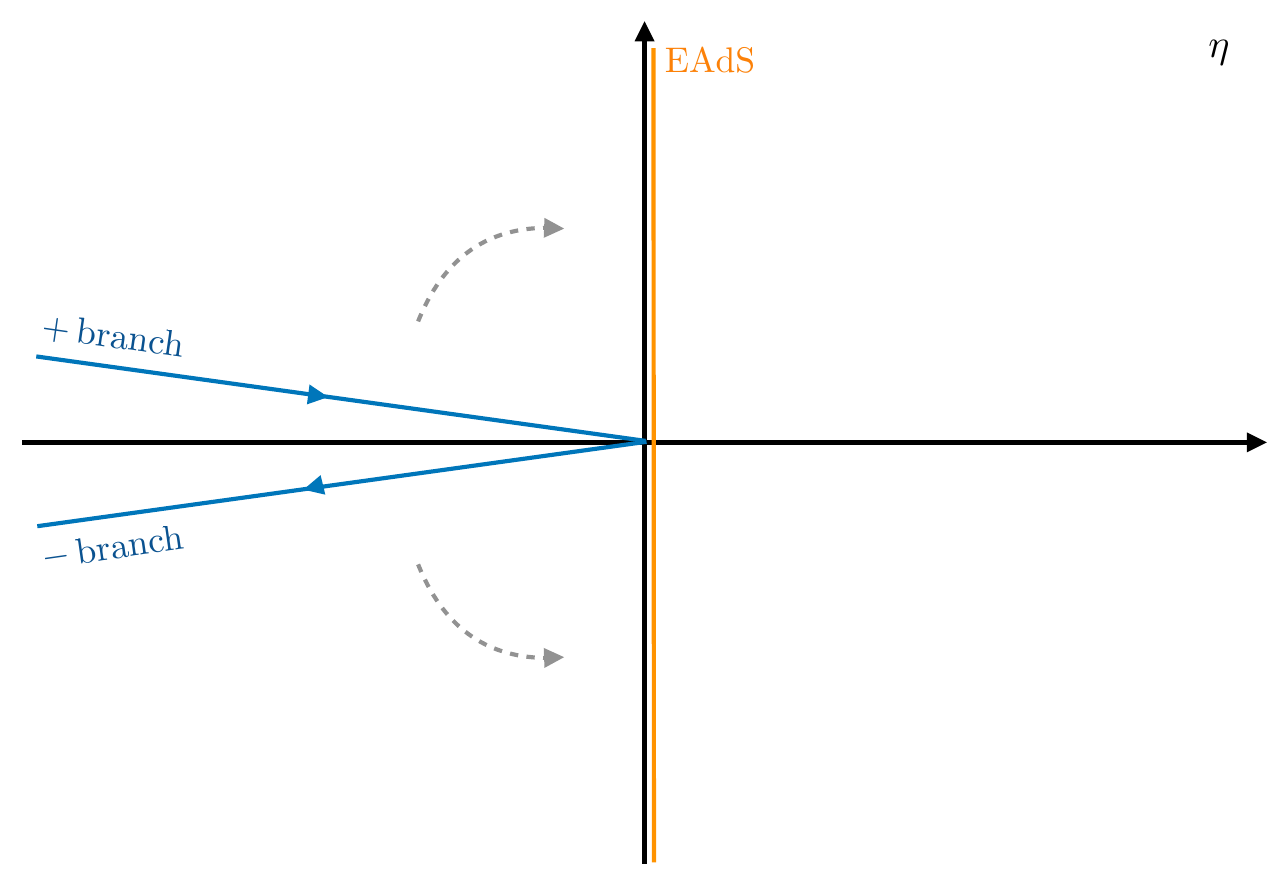}
    \caption{Wick rotation of the $+$ and $-$ Schwinger-Keldysh branches to the two sheets of the EAdS hyperboloid. Figure adapted from \cite{Sleight:2019mgd}.}
    \label{fig::Wick}
\end{figure}

\vskip 4pt
The two branches of the Schwinger-Keldysh contour may be rotated to EAdS according to \cite{Sleight:2019mgd,Sleight:2019hfp}
\begin{equation}
    \eta_\pm
    \to 
    e^{\pm \frac{\pi i}{2}}z,
    \qquad
    z\in(0,\infty).
\label{wickzeta}
\end{equation}
The two rotations map the Schwinger-Keldysh branches to the two sheets of the EAdS hyperboloid. For the Bunch-Davies vacuum, the absence of antipodal singularities allows these sheets to be identified and parametrised by a single EAdS Poincar\'e
patch with $z>0$.\footnote{See e.g. \cite{Chopping:2024oiu} for the extension of this to cases in which antipodal singularities are present.} 

\vskip 4pt
Under the rotation \eqref{wickzeta}, the Schwinger-Keldysh propagators \eqref{SKprop} are mapped to linear combinations
of EAdS propagators $G_{\Delta_\pm}$ obeying the two shadow-related boundary conditions $\Delta_\pm$ \cite{Sleight:2020obc}, 
\begin{multline}\label{general SK propagator map}
    (G_{ab})_{\mu_1\ldots\mu_s;\nu_1\ldots\nu_s}(\eta, \vec{x};\bar\eta,\vec{y})
    \longrightarrow
    c^{\text{dS-AdS}}_{\Delta_+}
    e^{-a(\Delta_+-s)\frac{\pi i}{2}}
    e^{-b(\Delta_+-s)\frac{\pi i}{2}}
    (G_{\Delta_+})_{\mu_1\ldots\mu_s;\nu_1\ldots\nu_s}
    (z,\vec{x};\bar z,\vec{y})
    \\
    +
    c^{\text{dS-AdS}}_{\Delta_-}
    e^{-a(\Delta_--s)\frac{\pi i}{2}}
    e^{-b(\Delta_--s)\frac{\pi i}{2}}
    (G_{\Delta_-})_{\mu_1\ldots\mu_s;\nu_1\ldots\nu_s}
    (z,\vec{x};\bar z,\vec{y}).
\end{multline}
Taking one endpoint to the late-time boundary gives the corresponding map for the Schwinger-Keldysh bulk-to-boundary propagators \cite{Sleight:2019hfp,Sleight:2020obc}: 
\begin{equation}\label{bubowick}
    \left(K^a_{\Delta}\right)_{\mu_1 \ldots \mu_s;i_1 \ldots i_s}(\eta,\vec{x};\vec{y}) \to e^{-a\left(\Delta-s\right)\frac{\pi i}{2}} c^{\text{dS-AdS}}_{\Delta}(K_{\Delta})_{\mu_1 \ldots \mu_s;i_1 \ldots i_s}(z,\vec{x};\vec{y}).
\end{equation} 
Here $K_\Delta$ denotes the EAdS bulk-to-boundary
propagator. The coefficients
$c^{\text{dS-AdS}}_\Delta$ account for the change in two-point-function normalisation
between dS and EAdS, and are
\begin{equation}\label{dS AdS c coefficient scalar loops}
    c^{\text{dS-AdS}}_\Delta
    =
    \frac12
    \csc\left[
        \left(
            \frac d2-\Delta
        \right)\pi
    \right], \qquad c^{\text{dS-AdS}}_{\Delta_+}=-c^{\text{dS-AdS}}_{\Delta_-}.
\end{equation}
The propagator map \eqref{general SK propagator map} is equivalent to the field redefinition
\begin{multline}\label{fieldredef}
    \left(\Phi_a\right)_{\mu_1 \ldots \mu_s}
    \longrightarrow
    e^{-a\left(\Delta_+-s\right)\frac{\pi i}{2}}
    \sqrt{\left|c^{\text{dS-AdS}}_{\Delta_+}\right|}
    \left(\Phi_{\Delta_+}\right)_{\mu_1 \ldots \mu_s}
    \\
    +
    e^{-a\left(\Delta_--s\right)\frac{\pi i}{2}}
    \sqrt{\left|c^{\text{dS-AdS}}_{\Delta_-}\right|}
    \left(\Phi_{\Delta_-}\right)_{\mu_1 \ldots \mu_s},
    \qquad
    a=\pm,
\end{multline}
where the doubled fields $\Phi_a$ of the Schwinger-Keldysh formalism are replaced by two independent EAdS fields $\Phi_{\Delta_+}$ and $\Phi_{\Delta_-}$ obeying $\Delta_+$ and $\Delta_-$ boundary conditions respectively.

\vskip 4pt
We now use this rotation to translate the dS
Schwinger-Keldysh path integral into a corresponding EAdS generating functional. Consider a theory in dS with Lagrangian
\begin{equation}
    {\cal L}_{\rm dS}
    =
    {\cal L}^{(2)}_{\rm dS}
    +
    {\cal L}^{\rm int}_{\rm dS},
\end{equation}
where ${\cal L}^{(2)}_{\rm dS}$ is the standard two-derivative quadratic
part, e.g. for a scalar field 
\begin{equation}\label{freeds}
    {\cal L}^{(2)}_{\rm dS}
    \left[
        \Phi
    \right]
    =
    -
    \frac12
    \nabla_\mu\Phi
    \nabla^\mu\Phi
    -
    \frac12
    m^2_{\rm dS}
    \Phi^2,
\end{equation}
and ${\cal L}^{\rm int}_{\rm dS}$ contains the remaining local
structures which we write as\footnote{Quadratic local terms not
included in the free operator \eqref{freeds} can also be
included in \eqref{eq:genlocint} and treated as insertion vertices.}
\begin{equation}\label{eq:genlocint}
    {\cal L}_{\rm dS}^{\rm int}
    =
    \sum_v
    \lambda_v
    V_v^{\rm dS}(\Phi),
\end{equation}
where $v$ labels the local structures treated as vertices, $\lambda_v$ denotes the
corresponding coupling, and $V_v^{\rm dS}$ contains $n_v$ fields. Applying the above map to the Schwinger-Keldysh generating functional \eqref{SK generating scalar QED} gives an EAdS generating functional that computes the same dS correlators\footnote{The sources ${\cal J}_\pm$ in the Schwinger-Keldysh path integral on the rotated contours can be
written in terms of the EAdS sources ${\cal J}_{\Delta_\pm}$. For a spin-$s$ field, under a rotation $\eta_\pm =e^{\pm \pi i/2}z$ we have 
    \begin{equation}\label{inverse source map scalar loops}
    {\cal J}_\pm(\eta_\pm,\vec{x})
    = e^{\mp \left(\Delta_++s\right)\frac{\pi i}{2}}
    \frac{
        c^{\text{dS-AdS}}_{\Delta_+}
    }{
        \sqrt{
            \left|
            c^{\text{dS-AdS}}_{\Delta_+}
            \right|
        }
    } 
        {\cal J}_{\Delta_+}(z,\vec{x}) + (\Delta_+ \to \Delta_-).   
\end{equation}}
\begin{equation}\label{SK EAdS generating functional scalar loops}
    Z_{\rm SK}[{\cal J}_+,{\cal J}_-]
    =
    Z_{\rm EAdS}[{\cal J}_{\Delta_+},{\cal J}_{\Delta_-}],
\end{equation}
where
\begin{multline}\label{EAdS generating functional scalar loops}
Z_{\rm EAdS}[{\cal J}_{\Delta_+},{\cal J}_{\Delta_-}]
=
\int
{\cal D}\Phi_{\Delta_+}
{\cal D}\Phi_{\Delta_-}
\,
\exp\left[
    -S_{\rm EAdS}[\Phi_{\Delta_+},\Phi_{\Delta_-}]
\right]
\\
\times
\exp\left[
-\int_0^\infty
\frac{{\rm d}z\,{\rm d}^d\vec{x}}{z^{d+1}}
\left(
    {\cal J}_{\Delta_+}\cdot\Phi_{\Delta_+}
    +
    {\cal J}_{\Delta_-}\cdot\Phi_{\Delta_-}
\right)
\right],
\end{multline}
with 
\begin{equation}
    S_{\rm EAdS}
    =
    \int_0^\infty
    \frac{{\rm d}z\,{\rm d}^d\vec{x}}{z^{d+1}}\,
    {\cal L}_{\rm EAdS}.
\end{equation}
The EAdS Lagrangian takes the form
\begin{equation}
    {\cal L}_{\rm EAdS}
    =
    {\cal L}^{(2)}_{\rm EAdS}
    +
    {\cal L}^{\rm int}_{\rm EAdS},
\end{equation} 
where the free quadratic part is given by
\begin{equation}\label{phase weighted quadratic}
{\cal L}^{(2)}_{\rm EAdS}
= \sum_{\pm}
\kappa^{-1}_{\Delta_\pm}
{\cal L}^{(2)}_{\rm EAdS}
\left[
    \Phi_{\Delta_\pm}
\right],
\end{equation}
where $\kappa_{\Delta}$ denotes the phase
\begin{equation}
    \kappa_{\Delta} = \frac{c^{\text{dS-AdS}}_{\Delta}}{|c^{\text{dS-AdS}}_{\Delta}|}.
\end{equation}
Here ${\cal L}^{(2)}_{\rm EAdS}\left[
    \Phi
\right]$ denotes the standard
two-derivative EAdS quadratic Lagrangian for the field $\Phi$. For a scalar field this is
\begin{equation}
    {\cal L}^{(2)}_{\rm EAdS}
    \left[
        \Phi
    \right]
    =
    \frac12
    \nabla_\mu\Phi
    \nabla^\mu\Phi
    +
    \frac12
    m^2_{\rm AdS}
    \Phi^2,
\end{equation}
with\footnote{This relative sign is the same Wick-rotation phase that, in the vertex rule below, is
encoded by $N_v$: the two-derivative quadratic term contains one more inverse-metric
contraction than the mass term.} 
\begin{equation}
    m^2_{{\rm AdS}}
    =
    -
    m^2_{{\rm dS}} .
\end{equation}
Since $c^{\text{dS-AdS}}_{\Delta_+}=-c^{\text{dS-AdS}}_{\Delta_-}$, the two members of
a shadow pair always carry opposite phases,
\begin{equation}
    \kappa_{\Delta_+}=-\kappa_{\Delta_-},
    \qquad
    \kappa_{\Delta_+}^{2}=\kappa_{\Delta_-}^{2}=-\kappa_{\Delta_+}\kappa_{\Delta_-},
\label{eq::ShadowPhaseRelation}
\end{equation}
so that their quadratic terms enter with opposite signs. For real scaling
dimensions these are opposite signs, while for principal-series fields they are
opposite imaginary phases. This reflects the non-sign-definite structure of the
Schwinger-Keldysh path integral \eqref{SK generating scalar QED}, in which the
actions on the two branches enter with opposite signs.

\vskip 4pt
The interaction terms are obtained from the corresponding local structures \eqref{eq:genlocint} in the original
dS Lagrangian. For each vertex $v$, the corresponding EAdS vertex is obtained by applying the field
replacement \eqref{fieldredef} and summing over the two Schwinger-Keldysh branches. This
gives \cite{Sleight:2019mgd,Sleight:2019hfp,Sleight:2020obc,Sleight:2021plv}
\begin{multline}\label{general EAdS Lagrangian scalar loops}
{\cal L}^{\rm int}_{\rm EAdS}
=
-2
\sum_v
\lambda_v
\sum_{\sigma_1,\ldots,\sigma_{n_v}=\pm}
\left(
    \prod_{i=1}^{n_v}
    \sqrt{
        \left|
        c^{\text{dS-AdS}}_{\Delta_{i,\sigma_i}}
        \right|
    }
\right)
\\
\times
\sin\left[
    \left(
        -d+N_v+
        \sum_{i=1}^{n_v}
        \left(
            \Delta_{i,\sigma_i}-s_i
        \right)
    \right)
    \frac{\pi}{2}
\right]
V_v^{\rm EAdS}
\left(
    \Phi_{1,\Delta_{1,\sigma_1}},
    \ldots,
    \Phi_{n_v,\Delta_{n_v,\sigma_{n_v}}}
\right).
\end{multline}
Here $s_i$ is the spin of the $i$-th field in the vertex, $N_v/2$ is the number of inverse metric contractions. The EAdS vertex $V_v^{\rm EAdS}$
denotes the same local tensor structure as $V_v^{\rm dS}$, but evaluated on the
corresponding EAdS fields after the Wick rotation.

\subsection{Scalar QED}
\label{subsec::sQEDEAdS}

\begin{figure}[t]
    \centering
    \includegraphics[width=\textwidth]{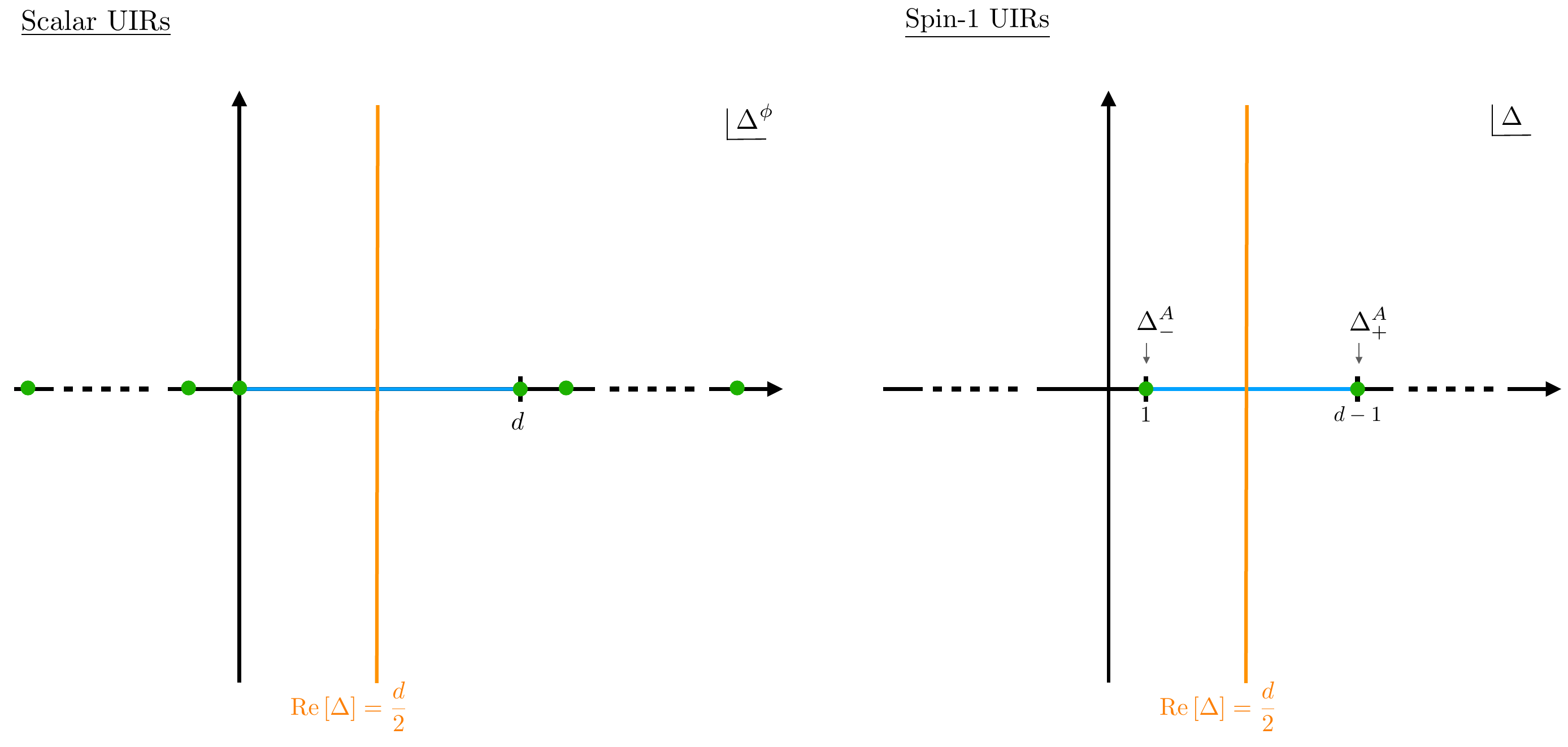}
    \caption{Scalar and spin-one unitary irreducible representations of $SO(d+1,1)$ in the complex $\Delta$ plane. See \eqref{unitaryscalarreps}. The orange lines denote the principal series and the blue intervals the complementary series. For scalars, the green points denote the exceptional series; for spin one, they mark the massless dimensions
$\Delta^A_-=1$ and $\Delta^A_+=d-1$. Shadow-related dimensions are
paired under $\Delta\leftrightarrow d-\Delta$.}
    \label{fig::UIRs}
\end{figure}

We now specialise the above EAdS reformulation of late-time correlators to scalar QED on dS$_{d+1}$, which were first given in \cite{Abhishek:2025oki}. The dS Lagrangian is
\begin{equation}\label{dS scalar QED Lagrangian}
    {\cal L}^{\rm dS}
    =
    -\frac14
    F_{\mu\nu}F^{\mu\nu}
    -
    g^{\mu\nu}
    \left(D_\mu\varphi\right)^*
    D_\nu\varphi
    -
    m^2_{\rm dS}
    \varphi^*\varphi,
    \qquad
    D_\mu=\nabla_\mu+ieA_\mu .
\end{equation}
Writing the complex scalar as
\begin{equation}
    \varphi
    =
    \frac{1}{\sqrt2}
    \left(
        \phi_1+i\phi_2
    \right),
\end{equation}
the interaction terms are
\begin{equation}\label{dS scalar QED interactions}
    {\cal L}_{\rm dS}^{\rm int}
    =
    e A_\mu
    \left(
        \phi_2\nabla^\mu\phi_1
        -
        \phi_1\nabla^\mu\phi_2
    \right)
    -
    \frac{e^2}{2}
    A_\mu A^\mu
    \left(
        \phi_1^2+\phi_2^2
    \right).
\end{equation}
For the gauge field, the two free theory falloffs are
\begin{equation}
    \Delta^A_+=d-1,
    \qquad
    \Delta^A_-=1.
\end{equation}
The scalar falloffs are parametrised as 
\begin{equation}
    \Delta^\varphi_\pm
    =
    \frac d2
    \pm i\mu,
    \qquad
    m^2_{\rm dS}
    =
    \Delta^\varphi_+\Delta^\varphi_-.
\end{equation}
Unitary values for scalar representations fall into the following categories \cite{Dobrev:1977qv,Basile:2016aen} (see figure \ref{fig::UIRs}):
\begin{align} \nonumber
  &\bullet \:\: \text{Principal Series:} \:\:\: \text{Massive Particles},\:\:\: \mu \in \mathbb{R}, \:\:\: m^2 \geq \left(\tfrac{d}{2}\right)^2.  \\ \nonumber
  &\bullet \:\: \text{Complementary Series:} \:\:\: \text{Light Particles}, \:\:\: \mu \in i\mathbb{R}, \:\:\: |\mu| \in \left(0,\tfrac{d}{2}\right), \:\:\: 0 < m^2 < \left(\tfrac{d}{2}\right)^2. \\ 
  &\bullet \:\: \text{Exceptional Series:} \:\:\: \mu = \pm i\left(\tfrac{d}{2}+n\right), \:\:\: n \in \mathbb{N}_0.  \label{unitaryscalarreps}
\end{align}
Massless scalar particles correspond to the first exceptional value $n=0$, with $ \Delta^\varphi_+ = d$ and $ \Delta^\varphi_- = 0$. 

\vskip 4pt Under the rotation to EAdS, each dS field is replaced by a pair of independent EAdS fields obeying the two shadow-related boundary conditions---see \eqref{fieldredef}. Scalar QED is therefore reformulated in terms of the pair of vector fields 
\begin{equation} 
A \quad\longrightarrow\quad A_{\Delta^A_+}, \, A_{\Delta^A_-}, 
\end{equation} 
together with two pairs of real scalar fields 
\begin{align} 
\phi_1 &\quad\longrightarrow\quad \phi_{1,\Delta^\varphi_+}, \, \phi_{1,\Delta^\varphi_-}, \\ \phi_2 &\quad\longrightarrow\quad \phi_{2,\Delta^\varphi_+}, \, \phi_{2,\Delta^\varphi_-}. 
\end{align} 
For the gauge field, we refer to the boundary conditions associated with $\Delta^A_+=d-1$ and $\Delta^A_-=1$ as Dirichlet and Neumann, respectively. The transverse spatial components of a free gauge boson in EAdS have the schematic near-boundary expansion
\begin{equation} A_i(z,\vec{x}) \sim \tilde a_i(\vec{x}) + z^{d-2}J_i(\vec{x}) + \ldots, \label{eq::EAdSMaxwellFalloffs} 
\end{equation} 
where $J_i$ is, up to normalisation, the radial canonical momentum conjugate to $\tilde a_i$. The two boundary conditions correspond to the dictionary \begin{align} 
A_{\Delta^A_+} \quad &\longleftrightarrow \quad \text{Dirichlet:} \qquad \tilde a_i\ \text{fixed}, \qquad \tilde a_i\ \text{sources}\ J_i, \qquad \Delta[J_i]=d-1, \label{eq::DirichletGaugeDictionary} \\ A_{\Delta^A_-} \quad &\longleftrightarrow \quad \text{Neumann:} \qquad J_i\ \text{fixed}, \qquad \tilde a_i\ \text{dynamical}, \qquad \Delta[\tilde a_i]=1. \label{eq::NeumannGaugeDictionary} 
\end{align} 
Thus $A_{\Delta^A_+}$ is associated with a conserved boundary current $J_i$, whereas $A_{\Delta^A_-}$ is associated with a dynamical boundary gauge field ${\tilde a}_i$.

\vskip 4pt
Applying \eqref{general EAdS Lagrangian scalar loops}, the EAdS Lagrangian entering
the generating functional \eqref{EAdS generating functional scalar loops} for dS
Schwinger-Keldysh correlators is \cite{Abhishek:2025oki}
\begin{equation}\label{scalarQEDL}
    {\cal L}_{\text{EAdS}}
    =
    {\cal L}^{(2)}_{\text{EAdS}}
    +
    \underbrace{{\cal L}^{(3)}_{\text{EAdS}}
    +
    {\cal L}^{(4)}_{\text{EAdS}}}_{{\cal L}^{\text{int}}_{\text{EAdS}}},
\end{equation}
with quadratic terms
\begin{multline}
{\cal L}^{(2)}_{\text{EAdS}}
=
\frac14
\operatorname{sgn}\!\left[c^{\text{dS-AdS}}_{\Delta^A_+}\right]
\left[
F^{\mu\nu}\!\left[A_{\Delta^A_+}\right]
F_{\mu\nu}\!\left[A_{\Delta^A_+}\right]
-
F^{\mu\nu}\!\left[A_{\Delta^A_-}\right]
F_{\mu\nu}\!\left[A_{\Delta^A_-}\right]
\right]
\\
+
\sum_{a=1}^{2}
\kappa^{-1}_{\Delta^\varphi_+}
\left(
    \frac{1}{2}
    \nabla_\mu\phi_{a,\Delta^\varphi_+}
    \nabla^\mu\phi_{a,\Delta^\varphi_+}
    -
    \frac{1}{2}m^2_{\rm dS}
    \phi_{a,\Delta^\varphi_+}^{\,2}
\right)
\\
+
\sum_{a=1}^{2}
\kappa^{-1}_{\Delta^\varphi_-}
\left(
    \frac{1}{2}
    \nabla_\mu\phi_{a,\Delta^\varphi_-}
    \nabla^\mu\phi_{a,\Delta^\varphi_-}
    -
    \frac{1}{2}m^2_{\rm dS}
    \phi_{a,\Delta^\varphi_-}^{\,2}
\right).
\end{multline}
The phases multiplying the quadratic terms depend on the representation \eqref{unitaryscalarreps}. For principal-series scalars, 
\begin{equation}
    \Delta^\varphi_\pm
    =
    \frac{d}{2}\pm i\mu,
    \qquad
    \mu>0
    \qquad
    \Longrightarrow
    \qquad
    \kappa_{\Delta^\varphi_\pm}
    = \frac{c_{\Delta^\varphi_\pm}^{\mathrm{dS-AdS}}}{|c_{\Delta^\varphi_\pm}^{\mathrm{dS-AdS}}|} =
    \pm i.
\label{eq::PrincipalSeriesScalarPhases}
\end{equation}
For real scaling dimensions, as in the complementary and exceptional
series, and for the gauge field, we have
\begin{equation}
    \kappa_{\Delta^\varphi_\pm}
    =
    \operatorname{sgn}\!\left[
        c_{\Delta^\varphi_\pm}^{\mathrm{dS-AdS}}
    \right]
    \in\{-1,+1\},
    \qquad
    \kappa_{\Delta^A_\pm}
    =
    \operatorname{sgn}\!\left[
        c_{\Delta^A_\pm}^{\mathrm{dS-AdS}}
    \right]
    \in\{-1,+1\}.
\label{eq::RealScalingDimensionPhases}
\end{equation}

\vskip 4pt
The cubic interaction becomes
\begin{multline}\label{scalarQEDL3}
{\cal L}^{(3)}_{\text{EAdS}}
=
-2e
\sqrt{\left|c^{\text{dS-AdS}}_{\Delta^A_+}\right|}
\left(A_{\Delta^A_+}\right)_\mu
\Bigg[
\left|c^{\text{dS-AdS}}_{\Delta^\varphi_+}\right|
\sin\left(\pi\Delta^\varphi_+\right)
J^\mu_{++}
\\
+
\sqrt{
\left|c^{\text{dS-AdS}}_{\Delta^\varphi_+}\right|
\left|c^{\text{dS-AdS}}_{\Delta^\varphi_-}\right|
}
\sin\left(\frac{d\pi}{2}\right)
\left(
    J^\mu_{+-}
    +
    J^\mu_{-+}
\right)
\\
+
\left|c^{\text{dS-AdS}}_{\Delta^\varphi_-}\right|
\sin\left(\pi\Delta^\varphi_-\right)
J^\mu_{--}
\Bigg]
\\
-
2e
\sqrt{\left|c^{\text{dS-AdS}}_{\Delta^A_-}\right|}
\left(A_{\Delta^A_-}\right)_\mu
\Bigg[
\left|c^{\text{dS-AdS}}_{\Delta^\varphi_+}\right|
\sin\left[
    \pi\left(
        \Delta^\varphi_+ - \frac{d}{2} + 1
    \right)
\right]
J^\mu_{++}
\\
+
\left|c^{\text{dS-AdS}}_{\Delta^\varphi_-}\right|
\sin\left[
    \pi\left(
        \Delta^\varphi_- - \frac{d}{2} + 1
    \right)
\right]
J^\mu_{--}
\Bigg],
\end{multline}
in terms of the current
\begin{equation}\label{scalarQEDcurrent}
    J^\mu_{\alpha\beta}
    =
    \phi_{2,\Delta^\varphi_\beta}
    \nabla^\mu
    \phi_{1,\Delta^\varphi_\alpha}
    -
    \phi_{1,\Delta^\varphi_\alpha}
    \nabla^\mu
    \phi_{2,\Delta^\varphi_\beta},
    \qquad
    \alpha,\beta=\pm.
\end{equation}
As noted in \cite{Abhishek:2025oki}, for the Neumann mode
$A_{\Delta^A_-}$ the contributions from the two branches of the in-in contour to the mixed-current couplings cancel. Consequently,
no $A_{\Delta^A_-}J^\mu_{+-}$ or $A_{\Delta^A_-}J^\mu_{-+}$ cubic vertex is generated in the EAdS action. This cancellation is encoded in the corresponding sine factor, which reduces to $\sin\pi=0$. The absence of these mixed couplings prevents multiplet recombination for the Neumann mode, which therefore acquires no anomalous dimension at one loop, as we verify explicitly in section \ref{sec::dSloop}.

\vskip 4pt
Similarly, the quartic interaction becomes
\begin{multline}\allowdisplaybreaks \label{scalarQEDL4}
{\cal L}^{(4)}_{\text{EAdS}}
=
-e^2
\left|c^{\text{dS-AdS}}_{\Delta^A_+}\right|
\left(A_{\Delta^A_+}\right)_\mu
\left(A_{\Delta^A_+}\right)^\mu
\Bigg[
\left|c^{\text{dS-AdS}}_{\Delta^\varphi_+}\right|
\sin\left[
    \pi
    \left(
        \Delta^\varphi_+ + \frac{d}{2}
    \right)
\right]
\left(
    \phi_{1,\Delta^\varphi_+}^{\,2}
    +
    \phi_{2,\Delta^\varphi_+}^{\,2}
\right)
\\ \allowdisplaybreaks
+
\left|c^{\text{dS-AdS}}_{\Delta^\varphi_-}\right|
\sin\left[
    \pi
    \left(
        \Delta^\varphi_- + \frac{d}{2}
    \right)
\right]
\left(
    \phi_{1,\Delta^\varphi_-}^{\,2}
    +
    \phi_{2,\Delta^\varphi_-}^{\,2}
\right)
\\  \allowdisplaybreaks
+
2
\sqrt{
\left|c^{\text{dS-AdS}}_{\Delta^\varphi_+}\right|
\left|c^{\text{dS-AdS}}_{\Delta^\varphi_-}\right|
}
\sin\left(
    d \pi
\right)
\left(
    \phi_{1,\Delta^\varphi_+}\phi_{1,\Delta^\varphi_-}
    +
    \phi_{2,\Delta^\varphi_+}\phi_{2,\Delta^\varphi_-}
\right)
\Bigg]
\\ \allowdisplaybreaks
-
e^2
\left|c^{\text{dS-AdS}}_{\Delta^A_-}\right|
\left(A_{\Delta^A_-}\right)_\mu
\left(A_{\Delta^A_-}\right)^\mu
\Bigg[
\left|c^{\text{dS-AdS}}_{\Delta^\varphi_+}\right|
\sin\left[
    \pi
    \left(
        \Delta^\varphi_+ - \frac{d}{2}
    \right)
\right]
\left(
    \phi_{1,\Delta^\varphi_+}^{\,2}
    +
    \phi_{2,\Delta^\varphi_+}^{\,2}
\right)
\\ \allowdisplaybreaks
+
\left|c^{\text{dS-AdS}}_{\Delta^\varphi_-}\right|
\sin\left[
    \pi
    \left(
        \Delta^\varphi_- - \frac{d}{2}
    \right)
\right]
\left(
    \phi_{1,\Delta^\varphi_-}^{\,2}
    +
    \phi_{2,\Delta^\varphi_-}^{\,2}
\right)
\Bigg]
\\ \allowdisplaybreaks
+
2e^2
\sqrt{
\left|c^{\text{dS-AdS}}_{\Delta^A_+}\right|
\left|c^{\text{dS-AdS}}_{\Delta^A_-}\right|
}
\left(A_{\Delta^A_+}\right)_\mu
\left(A_{\Delta^A_-}\right)^\mu
\Bigg[
\left|c^{\text{dS-AdS}}_{\Delta^\varphi_+}\right|
\sin\left(
    \pi\Delta^\varphi_+
\right)
\left(
    \phi_{1,\Delta^\varphi_+}^{\,2}
    +
    \phi_{2,\Delta^\varphi_+}^{\,2}
\right)
\\ \allowdisplaybreaks
+
\left|c^{\text{dS-AdS}}_{\Delta^\varphi_-}\right|
\sin\left(
    \pi\Delta^\varphi_-
\right)
\left(
    \phi_{1,\Delta^\varphi_-}^{\,2}
    +
    \phi_{2,\Delta^\varphi_-}^{\,2}
\right)
\\ \allowdisplaybreaks
+
2
\sqrt{
\left|c^{\text{dS-AdS}}_{\Delta^\varphi_+}\right|
\left|c^{\text{dS-AdS}}_{\Delta^\varphi_-}\right|
}
\sin\left(
    \frac{d\pi}{2}
\right)
\left(
    \phi_{1,\Delta^\varphi_+}\phi_{1,\Delta^\varphi_-}
    +
    \phi_{2,\Delta^\varphi_+}\phi_{2,\Delta^\varphi_-}
\right)
\Bigg].
\end{multline}
An analogous cancellation between the two branches of the in-in
contour occurs at quartic order. After summing over the two branches,
the quartic coupling between two Neumann gauge fields $A_{\Delta^A_-}$ and a mixed
scalar-shadow bilinear vanishes. This cancellation is encoded in
the corresponding sine factor, which reduces to $\sin\pi=0$.

\vskip 4pt
The Lagrangian \eqref{scalarQEDL} gives a set of EAdS Feynman rules for perturbative
Schwinger-Keldysh correlators in scalar QED \cite{Abhishek:2025oki}. Each dS in-in diagram is expressed as a linear combination of EAdS Witten diagrams built from the fields with falloffs $\Delta^\varphi_\pm$ and $\Delta^A_\pm$. In the next sections we apply this prescription to the one-loop two-point diagrams of scalar QED.

\section{EAdS scalar loop integrals in position space}
\label{sec::scalarloops}

In this section we compute scalar loop diagrams with scalar external legs in EAdS$_{d+1}$ with arbitrary scaling dimensions and arbitrary boundary dimension $d$. The results of this section for specific values of the scaling dimensions will then be used as building blocks in sections \ref{sec::vectorTadpole} and \ref{subsec::bubblediagramEAdSsQED} to compute scalar QED loop diagrams, whose coefficients take the same functional form but with shifted parameters. 

\subsection{Scalar propagators in position space}

We begin by collecting the position-space representations of the scalar propagators that will be used throughout this section. 

\vskip 4pt
For a scalar field of mass $m^2=\Delta(\Delta-d)$, the bulk-to-bulk propagator for scaling dimension $\Delta$ is \cite{Burgess:1984ti} 
\begin{equation}
    G_\Delta(x,x')
    =
    \frac{c_\Delta}{2^{\Delta+1}\beta}
    \xi^\Delta\,
    {}_2F_1\left(
        \begin{matrix}
            \frac{\Delta}{2},
            \frac{\Delta+1}{2}
            \\
            \beta+1
        \end{matrix}
        ;
        \xi^2
    \right),
\label{eq:Gprop}
\end{equation}
with 
\begin{equation}\label{eq::sc2ptnorm}
    c_\Delta
    =
    \frac{\Gamma(\Delta)}
    {\pi^{\frac d2}\Gamma(\beta)},
    \qquad
    \beta
    =
    \Delta-\frac d2,
\end{equation}
and where
\begin{equation}
    \xi(x,x')
    =
    \frac{
        2zz'
    }{
        z^2+z'^2+
        (\vec{x}-\vec{x}^{\,\prime})^2
    },
\label{eq::AdSInvariantXi}
\end{equation}
is the EAdS-invariant chordal variable, where the coincident limit corresponds to
$\xi\to1$.

\vskip 4pt
The corresponding scalar bulk-to-boundary propagator reads\footnote{When $\beta=0$, the normalization factor $c_\Delta$ is replaced by $\Gamma(\frac{d}{2})/(2\pi^\frac{d}{2})$.} \cite{Witten:1998qj,Freedman:1998tz}
\begin{equation}
    K_\Delta(x,\vec{y})
    =
    c_\Delta
    \left[
        \frac{z}{z^2+(\vec{x}-\vec{y})^2}
    \right]^\Delta.
\end{equation}

\subsection{Tadpole diagram}
\begin{figure}[h]
    \centering
    \begin{tikzpicture}
        \draw[blue,thick] (0,0) circle (2cm);
        \draw (0,0.7) circle (0.7cm);
        \draw (-2,0) node[left]{$\vec{y}_1$} -- (2,0) node[right]{$\vec{y}_2$};
        \node [above] at (-1.2,0) {$\Delta_3$};
        \node [above] at (1.2,0) {$\Delta_3$};
        \node [below] at (0,0) {$x$};
        \node [above] at (0,1.4) {$\Delta$};
        \filldraw[black] (-2,0) circle (1.5pt);
        \filldraw[black] (0,0) circle (1.5pt);
        \filldraw[black] (2,0) circle (1.5pt);
    \end{tikzpicture}
    \caption{Scalar Tadpole diagram in EAdS\label{fig:tadpole}}
\end{figure}
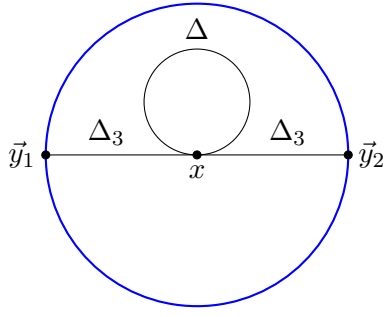

The first relevant diagram is the non-derivative scalar tadpole with scaling dimension $\Delta$ running in the loop and $\Delta_3$ in the external legs. See Figure \ref{fig:tadpole}. In position space this simply corresponds to 
\begin{equation}
    I^\text{T}_{\Delta;\Delta_3} = \int\dd^{d+1}x\sqrt{g}\ K_{\Delta_3}(x,\vec{y}_1)G_{\Delta}(x,x)K_{\Delta_3}(x,\vec{y}_2)\,.
\end{equation}

At coincident points $x=x'$, the variable $\xi$ becomes $1$ and $G_\Delta(x,x)$ is just a number, independent of the bulk point $x$
\begin{equation}
    G_\Delta(x,x) = \frac{c_\Delta}{2^{\Delta+1}\beta}{}_2F_1\left(\begin{matrix}
        \frac{\Delta}{2},\ \frac{\Delta+1}{2}\\
        \beta+1
    \end{matrix}\,;\,1\right)\,.
\end{equation}
The hypergeometric function converges for $d<1$ and formally diverges outside this region. Computing it for small $d$ with the formula \eqref{eq:2f1unit} and then adopting dimensional regularization, the resulting expression for the tadpole can be simplified with the duplication formula for gamma functions
\begin{equation}
    \Gamma(z)\Gamma\left(z+\frac{1}{2}\right) = 2^{1-2z}\sqrt{\pi}\,\Gamma(2z)\,,
\end{equation}
obtaining for any $d$
\begin{equation}\label{eq:tadpole}
    G_\Delta(x,x) = \frac{1}{(4\pi)^\frac{d+1}{2}}\frac{\Gamma(\Delta)\Gamma\big(\frac{1-d}{2}\big)}{\Gamma(\Delta-d+1)}\,.
\end{equation}
In this scheme, the tadpole is finite for even $d$ and it has simple poles for odd $d$.

\subsection{Bubble diagram}
\label{sec:scalar-bubble}

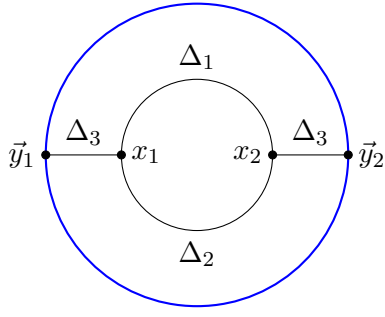
\begin{figure}[h]
    \centering
    \begin{tikzpicture}
        \draw[blue,thick] (0,0) circle (2cm);
        \draw (0,0) circle (1cm);
        \draw (-2,0) node[left]{$\vec{y}_1$} -- (-1,0) node[right]{$x_1$};
        \node [above] at (-1.5,0) {$\Delta_3$};
        \node [above] at (1.5,0) {$\Delta_3$};
        \node [above] at (0,1) {$\Delta_1$};
        \node [below] at (0,-1) {$\Delta_2$};
        \draw (1,0) node[left]{$x_2$} -- (2,0) node[right]{$\vec{y}_2$};
        \filldraw[black] (-2,0) circle (1.5pt);
        \filldraw[black] (-1,0) circle (1.5pt);
        \filldraw[black] (1,0) circle (1.5pt);
        \filldraw[black] (2,0) circle (1.5pt);
    \end{tikzpicture}
    \caption{Scalar Bubble diagram in EAdS\label{fig:scalarloops}}
\end{figure}

The second scalar loop building block is the non-derivative bubble diagram, with scalar fields of dimensions $\Delta_1$ and $\Delta_2$ running in the loop and external legs of dimension $\Delta_3$. See figure \ref{fig:scalarloops}. In position space this corresponds to
\begin{multline}\label{eq:bubbleint}
    I^\text{B}_{\Delta_1,\Delta_2;\Delta_3}(\vec{y}_1,\vec{y}_2) = \int\dd^{d+1}x_1\sqrt{g_1}\ K_{\Delta_3}(x_1,\vec{y}_1)\\ \times \int\dd^{d+1}x_2\sqrt{g_2}\ G_{\Delta_1}(x_1,x_2)G_{\Delta_2}(x_1,x_2)K_{\Delta_3}(x_2,\vec{y}_2)\,.
\end{multline}

In the following we first determine the convergence conditions of~\eqref{eq:bubbleint}. We then use EAdS isometries to show that the integral over either interaction point is proportional to a scalar bulk-to-boundary propagator\footnote{In \cite{Banados:2022nhj}, similar results were found for scalar loop integrals relevant for quartic interactions.},
\begin{equation}
    \int \dd^{d+1}x_2\,\sqrt{g_2}\,
    G_{\Delta_1}(x_1,x_2)
    G_{\Delta_2}(x_1,x_2)
    K_{\Delta_3}(x_2,\vec{y}_2)
    =
    C^{(d)}_{\Delta_1,\Delta_2;\Delta_3}\,
    K_{\Delta_3}(x_1,\vec{y}_2),
\label{eq::scalarBubbleReducedForm}
\end{equation}
so that the diagram is fixed up to the scalar coefficient
$C^{(d)}_{\Delta_1,\Delta_2;\Delta_3}$. We compute this coefficient exactly, analyse its short-distance expansion and associated UV poles, and finally specialise the result to the values of the dimensions required for the scalar QED calculation.

\subsubsection{Convergence}

Focusing on the $x_2$ vertex of the bubble integral \eqref{eq:bubbleint}
\begin{equation}\label{eq:bubblevert}
    \tilde{I}^\text{B}_{\Delta_1,\Delta_2;\Delta_3}(x_1,\vec{y}_2) = \int\dd^{d+1}x_2\sqrt{g_2}\ G_{\Delta_1}(x_1,x_2)G_{\Delta_2}(x_1,x_2)K_{\Delta_3}(x_2,\vec{y}_2)\,,
\end{equation}
one encounters two types of divergences: infrared (IR) and ultraviolet (UV). IR divergences are related to long-distance singularities due to the infinite volume of AdS and come from the lower and upper limits of the radial integral $z_2$. At the conformal boundary $z_2=0$, the AdS measure and the propagators scale as
\begin{equation}
    \sqrt{g_2}\sim z_2^{-d-1}, \quad G_{\Delta_i}\sim z_2^{\Delta_i}, \qquad K_{\Delta_3}\sim z_2^{d-\Delta_3},
\end{equation}
and the vertex $\tilde{I}^\text{B}_{\Delta_1,\Delta_2;\Delta_3}$ goes as
\begin{equation}
    \tilde{I}^\text{B}_{\Delta_1,\Delta_2;\Delta_3} \sim \int_{z_2\rightarrow0}\dd z_2\ z_2^{-1+\Delta_1+\Delta_2-\Delta_3}\,.
\end{equation}
Whereas for the point at infinity $z_2=\infty$, it can be brought to $z_2=0$ after a translation of $\vec{y}_2$ followed by an inversion, with $K_{\Delta_3}$ now scaling instead as $z_2^{\Delta_3}$ and implying for the vertex
\begin{equation}
    \tilde{I}^\text{B}_{\Delta_1,\Delta_2;\Delta_3} \sim \int_{z_2\rightarrow0}\dd z_2\ z_2^{-1+\Delta_1+\Delta_2+\Delta_3-d}\,.
\end{equation}
Convergence of the bubble vertex in the IR requires then the conditions between scaling dimensions
\begin{equation}\label{eq:IRconv}
    \Delta_1+\Delta_2>\Delta_3\,,\hs \Delta_1+\Delta_2>d-\Delta_3\,.
\end{equation}

In turn, UV divergences are related to short-distance singularities in loop
diagrams and come from the coincident bulk region $x_2\to x_1$, or equivalently
$\xi\to1$. It is convenient to set
\begin{equation}
    s=1-\xi^2 .
\end{equation}
Near $s=0$, the measure and propagators scale as
\begin{equation}
    \dd^{d+1}x_2\sqrt{g_2} \sim \dd s\,s^{\frac{d-1}{2}},
    \qquad
    G_{\Delta_i}\sim s^{\frac{1-d}{2}},
    \qquad
    K_{\Delta_3}\sim 1 .
\end{equation}
Therefore the leading short-distance behaviour of the bubble is 
\begin{equation}
     \tilde{I}^\text{B}_{\Delta_1,\Delta_2;\Delta_3}
    \sim
    \int_0 \dd s\,s^{\frac{1-d}{2}} .
\end{equation}
Local convergence of the unregularised integral requires
\begin{equation}\label{eq:uvconv}
    d<3 .
\end{equation}
For $d\geq3$ the integral must therefore be defined by a UV regulator. In
dimensional regularisation, the full short-distance expansion resolves this
leading convergence criterion into a sequence of local poles at
$d=3,5,7,\ldots$, as discussed in section \ref{subsub:UVandSDexp}.

\subsubsection{Exact computation}

Assuming one is in the convergent window \eqref{eq:IRconv}, so there is no IR regulator that explicitly breaks the AdS symmetry, using AdS isometries as change of variables in the integrated point $x_2$ one can completely factorise the dependence on the external points $x_1$ and $\vec{y}_2$ from the integral \eqref{eq:bubblevert}:
\begin{equation}\label{eq:bubbleform}
    \tilde{I}^\text{B}_{\Delta_1,\Delta_2;\Delta_3} = C^{(d)}_{\Delta_1,\Delta_2;\Delta_3}K_{\Delta_3}(x_1,\vec{y}_2)\,,
\end{equation}
with numerical factor $C^{(d)}_{\Delta_1,\Delta_2;\Delta_3}$. For details see appendix \ref{subsec::expectisomsc}.

\vskip 4pt
To compute the scalar bubble integral $ \tilde{I}^\text{B}_{\Delta_1,\Delta_2;\Delta_3}$ and determine the value of $C^{(d)}_{\Delta_1,\Delta_2;\Delta_3}$, we start from the known decomposition for the bulk-to-bulk propagator \cite{Bros:2011vh} 
\begin{equation}\label{321}
    G_{\Delta_1}(\xi)G_{\Delta_2}(\xi) = \sum_{k=0}^\infty a^{(d)}_{\Delta_1,\Delta_2}(k)G_{\Delta_{12}+2k}(\xi)\,,
\end{equation}
where $\Delta_{12}=\Delta_1+\Delta_2$, and where the coefficient $a^{(d)}_{\Delta_1,\Delta_2}(k)$ is given by
\begin{equation}\label{eq:akcoef}
    a^{(d)}_{\Delta_1,\Delta_2}(k) = \frac{\big(\frac{d}{2}\big)_k}{2\pi^\frac{d}{2}k!}\frac{(\Delta_{12}+2k)_{1-\frac{d}{2}}(\Delta_{12}-d+1+k)_k}{(\Delta_1+k)_{1-\frac{d}{2}}(\Delta_2+k)_{1-\frac{d}{2}}\big(\Delta_{12}-\frac{d}{2}+k\big)_k}\,.
\end{equation}
By expressing the propagator on the RHS in its series representation, one can write the product of propagators as a double series in $\xi$
\begin{equation}\label{eq:GGdouble}
    G_{\Delta_1}(\xi)G_{\Delta_2}(\xi) = \sum_{k=0}^\infty\sum_{l=0}^\infty g^{(d)}_{\Delta_1,\Delta_2}(k,l)\xi^{\Delta_{12}+2k+2l}\,,
\end{equation}
with
\begin{equation}\label{eq:gklcoef}
    g^{(d)}_{\Delta_1,\Delta_2}(k,l) = \frac{a^{(d)}_{\Delta_1,\Delta_2}(k)}{2^{\Delta_{12}+1+2k}\pi^\frac{d}{2}(\Delta_{12}+2k)_{1-\frac{d}{2}}}\frac{\big(\frac{\Delta_{12}}{2}+k\big)_l\big(\frac{\Delta_{12}+1}{2}+k\big)_l}{\big(\Delta_{12}-\frac{d}{2}+1+2k\big)_l\,l!}\,.
\end{equation}
Then replacing \eqref{eq:GGdouble} in \eqref{eq:bubblevert}, the resulting integral can be performed with the formula (E.13) in \cite{Bianchi:2025hvy}
\begin{equation}\label{eq:xikformula}
    \int\dd^{d+1}x_2\sqrt{g_2}\ \xi^{\Delta_i}(x_1,x_2)K_{\Delta_j}(x_2,\vec{y}_2) = \pi^\frac{d+1}{2}\frac{\Gamma\Big(\frac{\Delta_i+\Delta_j-d}{2}\Big)\Gamma\Big(\frac{\Delta_i-\Delta_j}{2}\Big)}{\Gamma\big(\frac{\Delta_i}{2}\big)\Gamma\big(\frac{\Delta_i+1}{2}\big)}K_{\Delta_j}(x_1,\vec{y}_2)\,.
\end{equation}
This confirms the expected form \eqref{eq:bubbleform}, and computes $C^{(d)}_{\Delta_1,\Delta_2;\Delta_3}$ to
\begin{equation}\label{eq:Ctaucoeff1}
    C^{(d)}_{\Delta_1,\Delta_2;\Delta_3} = \pi^\frac{d+1}{2}\sum_{k=0}^\infty\sum_{l=0}^\infty g^{(d)}_{\Delta_1,\Delta_2}(k,l)\frac{\Gamma\big(\frac{\Delta_{123}-d}{2}+k+l\big)\Gamma\big(\frac{\Delta_{12}-\Delta_3}{2}+k+l\big)}{\Gamma\big(\frac{\Delta_{12}}{2}+k+l\big)\Gamma\big(\frac{\Delta_{12}+1}{2}+k+l\big)}\,,
\end{equation}
where we called $\Delta_{123}=\Delta_1+\Delta_2+\Delta_3$. The $l$-series sums into a $_2F_1$ which is reduced using \eqref{eq:2f1unit}, and the remaining $k$-series sums into a $_9F_8$, finally obtaining
\begin{align}\label{eq:bubbleC}
    &C^{(d)}_{\Delta_1,\Delta_2;\Delta_3} = \frac{1}{2\pi^\frac{d}{2}(\Delta_{123}-d)(\Delta_{12}-\Delta_3)}\frac{(\Delta_{12})_{1-\frac{d}{2}}}{(\Delta_1)_{1-\frac{d}{2}}(\Delta_2)_{1-\frac{d}{2}}}\\
    &\times{}_9F_8\left(\begin{matrix}
        \frac{d}{2},\ \frac{\Delta_{123}-d}{2},\ \frac{\Delta_{12}-\Delta_3}{2},\ \frac{\Delta_{12}}{2}-\frac{d}{4}+1,\ \Delta_1,\ \Delta_2,\ \Delta_{12}-\frac{d}{2},\ \frac{\Delta_{12}-d+1}{2},\ \frac{\Delta_{12}-d}{2}+1\\
        \frac{\Delta_{12}}{2},\ \frac{\Delta_{12}+1}{2},\ \frac{\Delta_{123}-d}{2}+1,\ \frac{\Delta_{12}-\Delta_3}{2}+1,\ \Delta_{12}-d+1,\ \beta_1+1,\ \beta_2+1,\ \frac{\Delta_{12}}{2}-\frac{d}{4}
    \end{matrix}\,;\,1\right).\nonumber
\end{align}
It has poles at $\Delta_{12}=\Delta_3$ and $\Delta_{12}=d-\Delta_3$ as expected from the IR analysis in \eqref{eq:IRconv}, and the hypergeometric diverges for $d\geq3$ as expected from the UV analysis in \eqref{eq:uvconv}. The expression is symmetric under $\Delta_3\rightarrow d-\Delta_3$, implying the same coefficient is obtained for the bubble with the external propagators satisfying the opposite boundary condition.

\vskip 4pt
The $_9F_8$ may be further reduced as a sum of two $_7F_6$ upon use of the formula \eqref{eq:abc+1rel} for $a=\frac{\Delta_{123}-d}{2}$, $b=\frac{\Delta_{12}-\Delta_3}{2}$, $c=\frac{\Delta_{12}}{2}-\frac{d}{4}$
\begin{align}\label{eq::Bubble9F8To7F6}
    _9F_8 =&\, \frac{\Delta_{12}-\Delta_3}{2\Delta_{12}-d}{}_7F_6\left(\begin{matrix}
        \frac{d}{2},\ \frac{\Delta_{123}-d}{2},\ \Delta_1,\ \Delta_2,\ \Delta_{12}-\frac{d}{2},\ \frac{\Delta_{12}-d+1}{2},\ \frac{\Delta_{12}-d}{2}+1\\
        \frac{\Delta_{12}}{2},\ \frac{\Delta_{12}+1}{2},\ \frac{\Delta_{123}-d}{2}+1,\ \Delta_{12}-d+1,\ \beta_1+1,\ \beta_2+1
    \end{matrix}\,;\,1\right)\nonumber\\
    &+ (\Delta_3\rightarrow d-\Delta_3)\,,
\end{align}
recovering the expression obtained in \cite{Sleight:2021plv,DiPietro:2021sjt}. Further
simplifications\footnote{Other cases of simplification include $C^{(d=2)}_{\Delta_1,\Delta_2;\Delta_3}=\frac{\psi\left(\frac{\Delta_{123}}{2}-1\right)-\psi\left(\frac{\Delta_{12}-\Delta_3}{2}\right)}{8\pi(\Delta_3-1)}$, recovering the $d=2$ bubble of \cite{Giombi:2017hpr}, and $C^{(d)}_{\Delta_1,\Delta_2;\Delta_3=0}=-\frac{G_{\Delta_1}(x,x)-G_{\Delta_2}(x,x)}{m_1^2-m_2^2}$, also recovering the 1-loop banana computation of \cite{Cacciatori:2024zbe}.} occur for the scalar-QED values $\Delta_3=d-2$ considered in section~\ref{sec::ScalarQEDSummedForms}. When $\Delta_1=\Delta_2$, the series reduces to a very-well-poised ${}_5F_4(1)$ summable by the Rogers-Dougall formula \eqref{eq:RogersDougall5F4}. When $\Delta_1+\Delta_2=d$, the original ${}_9F_8(1)$ representation reduces to the two-parameter very-well-poised
${}_7F_6(1)$ summed using formula \eqref{eq:VeryWellPoised7F6} in appendix~\ref{app::z=1}. In both cases, the bubble coefficient reduces to a closed expression in Gamma functions, as we show in section~\ref{sec::ScalarQEDSummedForms}.

\subsubsection{Short distance expansion and UV poles}
\label{subsub:UVandSDexp}

The closed-form bubble coefficient \eqref{eq:bubbleC} is defined by analytic continuation in $d$ and contains the UV divergences of the loop as poles at odd $d$. Renormalisation requires isolating these poles and identifying them with local counterterms. In appendix~\ref{app::MaxwellRenormalisation} we discuss how the pole at $d=5$ fixes the Maxwell counterterm and the scalar QED beta function. In this section we therefore extract the UV poles directly from the short-distance expansion of the integrand, which makes their locality manifest and gives their residues in closed form.

\vskip 4pt
UV divergences are local and arise from the coincident bulk region
\begin{equation}
    x_2\to x_1,
    \qquad
    \xi\to 1.
\end{equation}
To extract them it is useful to return to the integral representation \eqref{eq:bubblevert} and expand the scalar propagators around $\xi=1$, where only a finite number of terms contribute to the divergence and are integrated into simpler functions. The representation \eqref{eq:Gprop} for the propagator $G_\Delta(\xi)$ expresses it as a series around $\xi=0$, so in order to expand the integral around $\xi=1$ one needs the transformation that relates the hypergeometric of argument $z$ to $1-z$. For generic $d$, the transformation~\eqref{eq:hyptr1} gives the local decomposition\footnote{At odd integer dimensions $d=3,5,\ldots$, the two local solutions entering this decomposition become degenerate. The corresponding hypergeometric transformation \eqref{eq:hyptr2} contains logarithmic terms. When inserted into the bubble integral, terms involving these
logarithms are locally integrable and therefore contribute only to its finite part. For integrals involving higher powers of the propagator, however, terms containing these logarithms need not be locally integrable and can contribute to UV divergences.}
\begin{equation}
    G_\Delta(\xi)
    =
    G_{\Delta,\mathrm{sing}}(\xi)
    +
    G_{\Delta,\mathrm{reg}}(\xi),
\end{equation}
where, introducing $s=1-\xi^2$, the singular part is
\begin{equation}
    G_{\Delta,\mathrm{sing}}(\xi)
    =
    \frac{\Gamma\left(\frac{d-1}{2}\right)}
    {4\pi^{\frac{d+1}{2}}}
    s^{\frac{1-d}{2}}
    (1-s)^{\frac{\Delta}{2}}
    {}_2F_1\left(
    \begin{matrix}
    \frac{\Delta-d+2}{2},\ \frac{\Delta-d+1}{2}\\
    \frac{3-d}{2}
    \end{matrix}
    ;s
    \right),
\label{eq:GsingExact}
\end{equation}
and the regular part is
\begin{equation}
    G_{\Delta,\mathrm{reg}}(\xi)
    =
    \frac{
    \Gamma(\Delta)\Gamma\left(\frac{1-d}{2}\right)
    }
    {
    2^{\Delta+1}\pi^\frac{d}{2}
    \Gamma\left(\frac{\Delta-d+2}{2}\right)
    \Gamma\left(\frac{\Delta-d+1}{2}\right)
    }
    (1-s)^{\frac{\Delta}{2}}
    {}_2F_1\left(
    \begin{matrix}
    \frac{\Delta}{2},\ \frac{\Delta+1}{2}\\
    \frac{d+1}{2}
    \end{matrix}
    ;s
    \right).
\label{eq:GregExact}
\end{equation}

\vskip 4pt
For extracting UV pole residues, only the singular part is needed. The regular part is locally integrable in the bubble integral. Near coincidence the measure behaves as
\begin{equation}
    \int_{x_2}
    \sim
    \int \dd s\,s^{\frac{d-1}{2}},
\end{equation}
while
\begin{equation}
    G_{\mathrm{sing}}\sim s^{\frac{1-d}{2}},
    \qquad
    G_{\mathrm{reg}}\sim s^0.
\end{equation}
Therefore
\begin{align}
    G_{\mathrm{sing}}G_{\mathrm{sing}}
    &\sim
    s^{1-d},
    \\
    G_{\mathrm{sing}}G_{\mathrm{reg}}
    &\sim
    s^{\frac{1-d}{2}},
    \\
    G_{\mathrm{reg}}G_{\mathrm{reg}}
    &\sim
    s^0 .
\end{align}
After including the measure, only $G_{\mathrm{sing}}G_{\mathrm{sing}}$ can generate UV poles. The mixed and regular-regular terms are locally integrable at $s=0$ and affect only the finite part of the bubble coefficient. We therefore expand the singular branch as
\begin{equation}
    G_{\Delta,\mathrm{sing}}(\xi)
    =
    \frac{\Gamma\left(\frac{d-1}{2}\right)}
    {4\pi^{\frac{d+1}{2}}}
    s^{\frac{1-d}{2}}
    \sum_{n=0}^{\infty}
    a_n(\Delta,d)s^n .
\label{eq:GsingSeries}
\end{equation}
The coefficients $a_n$ are defined by
\begin{equation}
    \sum_{n=0}^{\infty}a_n(\Delta,d)s^n
    =
    (1-s)^{\frac{\Delta}{2}}
    {}_2F_1\left(
    \begin{matrix}
    \frac{\Delta-d+2}{2},\ \frac{\Delta-d+1}{2}\\
    \frac{3-d}{2}
    \end{matrix}
    ;s
    \right),
\label{eq:anGenerating}
\end{equation}
or, equivalently,
\begin{equation}
    a_n(\Delta,d)
    =
    \sum_{r=0}^{n}
    \frac{
    \left(-\frac{\Delta}{2}\right)_{n-r}
    }
    {(n-r)!}
    \frac{
    \left(\frac{\Delta-d+2}{2}\right)_r
    \left(\frac{\Delta-d+1}{2}\right)_r
    }
    {
    \left(\frac{3-d}{2}\right)_r r!
    } ,
\label{eq:anGeneral}
\end{equation}
which can be expressed as a terminating $_3F_2$
\begin{equation}
    a_n(\Delta,d)
    = \frac{\left(-\frac{\Delta}{2}\right)_n}{n!}{}_3F_2\left(
        \begin{matrix}
            -n,\, \frac{\Delta-d+2}{2},\, \frac{\Delta-d+1}{2}
            \\
            \frac{3-d}{2},\, \frac{\Delta-2n+2}{2}
        \end{matrix}
        \,;\,1
    \right).
\end{equation}
The coefficients $a_n$ can have poles at odd dimensions $d\geq3$, with the pole at $d=3+2n$ first appearing at order $s^{n+1}$, reflecting the degeneration of the generic short-distance expansion at that odd dimension. The first coefficients are
\begin{equation}
    a_0(\Delta,d)=1,
\end{equation}
\begin{equation}
    a_1(\Delta,d)
    =
    -\frac{
    m^2+(d-1)(d-2)
    }
    {2(d-3)},
    \qquad
    m^2=\Delta(\Delta-d),
\label{eq:a1Coefficient}
\end{equation}
and
\begin{equation}
    a_2(\Delta,d)
    =
    \frac{
    (m^2)^2
    +2(d^2-6d+10)m^2
    +(d-1)(d-2)(d-3)(d-4)
    }
    {8(d-3)(d-5)} .
\label{eq:a2Coefficient}
\end{equation}
Thus the singular part begins as
\begin{equation}
    G_{\Delta,\mathrm{sing}}(\xi)
    =
    \frac{\Gamma\left(\frac{d-1}{2}\right)}
    {4\pi^{\frac{d+1}{2}}}
    s^{\frac{1-d}{2}}
    \left[
        1
        -
        \frac{
        m^2+(d-1)(d-2)
        }
        {2(d-3)}
        s
        +\cdots
    \right].
\label{eq:GshortFirstTerms}
\end{equation}

\vskip 4pt
The singular part of the product of two propagators can therefore be written as
\begin{equation}
    G_{\Delta_1}(\xi)G_{\Delta_2}(\xi)
    =
    \frac{\Gamma\left(\frac{d-1}{2}\right)^2}
    {16\pi^{d+1}}
    s^{1-d}
    \sum_{n=0}^{\infty}
    b_n(\Delta_1,\Delta_2;d)s^n
    +
    \text{UV-finite},
\label{eq:GGshort}
\end{equation}
with
\begin{equation}
    b_n(\Delta_1,\Delta_2;d)
    =
    \sum_{r=0}^{n}
    a_r(\Delta_1,d)a_{n-r}(\Delta_2,d).
\label{eq:bnGeneral}
\end{equation}
Substituting into the bubble integral \eqref{eq:bubbleint} gives
\begin{equation}
    \tilde{I}^{\text{B},\,\mathrm{div}}_{\Delta_1,\Delta_2;\Delta_3}
    =
    \frac{\Gamma\left(\frac{d-1}{2}\right)^2}
    {16\pi^{d+1}}
    \sum_{n=0}^{\infty}
    b_n(\Delta_1,\Delta_2;d)
    \int_{x_2}
    (1-\xi^2)^{1-d+n}
    K_{\Delta_3}(x_2,\vec y_2),
\label{eq:IBshortGeneral}
\end{equation}
up to finite contributions. This expression organises the UV divergences. The term proportional to $b_n$ has the local behaviour
\begin{equation}
    \int \dd s\,s^{\frac{d-1}{2}}s^{1-d+n}
    =
    \int \dd s\,s^{n+\frac{1-d}{2}}.
\end{equation}
It is logarithmically divergent when $d=3+2n$. Thus the pole at $d=3+2N$ is determined by the finite set
\begin{equation}
    b_0,b_1,\ldots,b_N.
\end{equation}
To evaluate the residues explicitly, we expand each term using \begin{equation}
    \left(1-\xi^2\right)^{-a}
    =
    \sum_{k=0}^{\infty}
    \frac{(a)_k}{k!}\,
    \xi^{2k},
\end{equation}
and integrate each power of $\xi$ using
\eqref{eq:xikformula}. Retaining the terms with
$n=0,\ldots,N$ gives
\begin{align}
    C_{\Delta_1,\Delta_2;\Delta_3}^{(d)}
    ={}&
    \frac{
        \Gamma\left(\frac{d-1}{2}\right)^2
        \Gamma\left(1+\frac{\Delta_3-d}{2}\right)
        \Gamma\left(1-\frac{\Delta_3}{2}\right)
    }{
        8\pi^{\frac{d}{2}+1}
    }
    \sum_{n=0}^{N}
    b_n(\Delta_1,\Delta_2;d)
    (d-1-n)
    \nonumber\\
    &\times
    {}_3F_2\left(
        \begin{matrix}
            d-n,\,
            1+\frac{\Delta_3-d}{2},\,
            1-\frac{\Delta_3}{2}
            \\
            2,\,
            \frac{3}{2}
        \end{matrix}
        \,;\,1
    \right)
    +
    \text{terms finite as }d\to3+2N.
\label{eq::BubblePoleHypergeometric}
\end{align}
The parametric excess of the $n$th hypergeometric function is $(3+2n-d)/2$. Consequently, at $d=3+2N$, precisely the terms with $n\leq N$ can contribute to the UV pole. Their residues can be extracted using \eqref{eq:3f2div}, and the remaining hypergeometrics evaluated with \eqref{eq:3f2PositiveShift}.

\vskip 4pt
Near $d=3$, only the $n=0$ term contributes, giving
\begin{equation}
    C_{\Delta_1,\Delta_2;\Delta_3}^{(d)}
    =
    -\frac{1}{8\pi^2(d-3)}
    +
    \text{finite},
    \qquad
    d\to3.
\label{eq:Cpole3}
\end{equation}
The residue is independent of the masses, as expected for the leading local UV singularity giving a scalar mass renormalisation. Setting $d=3-2\epsilon$ it reads $1/\!\left[(4\pi)^{2}\epsilon\right]$, which is
exactly the ultraviolet divergence of the scalar bubble in four flat dimensions,
itself independent of the masses and of the external momentum; cf.\
appendix~\ref{app::FlatDimensionShift}, where the analogous identification is
made for the pole at $d=5$.

\vskip 4pt
Near $d=5$, the terms with $n=0$ and $n=1$ contribute. Defining
\begin{equation}
    m_i^2
    =
    \Delta_i(\Delta_i-d),
\end{equation}
the corresponding coefficients are
\begin{equation}
    b_0=1,
    \qquad
    b_1
    =
    -\frac{
        m_1^2+m_2^2+2(d-1)(d-2)
    }{
        2(d-3)
    }.
\end{equation}
The two pole contributions are
\begin{align}
    \left.
        C_{\Delta_1,\Delta_2;\Delta_3}^{(d)}
    \right|_{n=0}
    &=
    -\frac{
        m_3^2+42
    }{
        192\pi^3(d-5)
    }
    +
    \text{finite},\qquad
    d\to5
    \nonumber\\
    \left.
        C_{\Delta_1,\Delta_2;\Delta_3}^{(d)}
    \right|_{n=1}
    &=
    \frac{
        m_1^2+m_2^2+24
    }{
        64\pi^3(d-5)
    }
    +
    \text{finite},\qquad
    d\to5.
\label{eq::Cpole5Contributions}
\end{align}
Adding them gives
\begin{equation}
    C_{\Delta_1,\Delta_2;\Delta_3}^{(d)}
    =
    \frac{
        m_1^2+m_2^2-\frac{1}{3}m_3^2+10
    }{
        64\pi^3(d-5)
    }
    +
    \text{finite},
    \qquad
    d\to5.
\label{eq:Cpole5}
\end{equation}
The residue is invariant under $\Delta_i\to d-\Delta_i$, as expected
for a local UV divergence.\footnote{The pole at $d=5$ will be relevant for the renormalisation of the Maxwell kinetic term in scalar QED. Indeed, the dimension-shifting relation \eqref{eq::vectorBubbleScalarBubbleRelation} for the vector bubble expresses the gauge field form factor in boundary dimension $d$ in terms of a
scalar bubble coefficient in boundary dimension $D=d+2$. The
logarithmic UV divergence of scalar QED in four bulk dimensions,
i.e. $d=3$, is therefore mapped to the pole of the scalar bubble for $d=5$. In particular, the dependence of the residue
\eqref{eq:Cpole5} on the external mass $m_3^2$ will determine the
coefficient of the Maxwell kinetic counterterm. We return to this
point in appendix \ref{app::MaxwellRenormalisation}.} 

\vskip 4pt
The same procedure can be continued to higher odd $d$. The pole at $d=3+2N$ is fixed by finitely many coefficients in the local expansion, namely $b_0,\ldots,b_N$. This makes manifest that the poles of the scalar bubble coefficient are local UV divergences. For the scalar QED specialisations considered below, the pole structure can
also be read off directly from the summed expressions
\eqref{eq:samebcresult} and \eqref{eq:diffbcresult}, in perfect agreement with the results above. The short-distance expansion, however, identifies these poles as local UV divergences generated by the coincident bulk region $x_2\to x_1$.

\subsubsection{Simplifications for scalar QED diagrams}
\label{sec::ScalarQEDSummedForms}

For the scalar QED bubble computation in section \ref{subsec::bubblediagramEAdSsQED}, we only need the 
bubble coefficient \eqref{eq:bubbleC} with external dimension 
\begin{equation}
    \Delta_3=d-2 .
\end{equation}
By the symmetry of the bubble coefficient under $\Delta_3\to d-\Delta_3$, all the results below also apply to the shadow value $\Delta_3=2$. There are two relevant choices for the boundary conditions of the internal
scalar lines: same boundary conditions, for which
\begin{equation}
    \Delta_1=\Delta_2=\Delta,
\end{equation}
and opposite boundary conditions, for which
\begin{equation}
    \Delta_1=\Delta_+,
    \qquad
    \Delta_2=\Delta_-=d-\Delta_+.
\end{equation}
It is useful first to impose the relation between $\Delta_1$ and $\Delta_2$ while keeping $\Delta_3$ arbitrary. In both cases, two upper parameters cancel against matching lower parameters in the original ${}_9F_8(1)$ representation, leaving a single very-well-poised ${}_7F_6(1)$. Setting $\Delta_3=d-2$ then gives specialisations that can be summed in terms of Gamma functions. The required summation identities are collected in appendix~\ref{app:hyper}.\footnote{The two generic ${}_7F_6(1)$ functions obtained below are
very-well-poised and have parametric excess $3-d$. They therefore converge for $\operatorname{Re}d<3$, in agreement with the UV criterion obtained directly from the bulk integral. For $d\geq3$, the resulting expressions are understood by analytic continuation in $d$, which defines the corresponding dimensionally regulated bubble coefficients.}

\paragraph{Same boundary conditions.}

We first take
\begin{equation}
    \Delta_1=\Delta_2=\Delta
\end{equation}
while leaving $\Delta_3$ arbitrary. In the original ${}_9F_8(1)$
representation, the matching upper and lower parameters $\Delta$
and $\Delta-\frac{d}{2}+1$ cancel. The bubble coefficient becomes
\begin{multline}
    C^{(d)}_{\Delta,\Delta;\Delta_3}
    ={}
    \frac{1}{
        2\pi^{\frac d2}
        \left(
            2\Delta+\Delta_3-d
        \right)
        \left(
            2\Delta-\Delta_3
        \right)
    }
    \frac{
        (2\Delta)_{1-\frac d2}
    }{
        (\Delta)_{1-\frac d2}^{\,2}
    }\\
    \times
    {}_7F_6\left(
    \begin{matrix}
        A,\,
        1+\frac{A}{2},\,
        B,\,
        C,\,
        D,\,
        E,\,
        F
        \\
        \frac{A}{2},\,
        1+A-B,\,
        1+A-C,\,
        1+A-D,\,
        1+A-E,\,
        1+A-F
    \end{matrix}
    ;1
    \right),
\label{eq:bubbleCequal}
\end{multline}
where 
\begin{equation}
\begin{aligned}
    A&=2\Delta-\frac d2,
    &\qquad
    B&=\frac d2,
    &\qquad
    C&=\Delta,
    \\
    D&=\Delta-\frac d2+\frac12,
    &
    E&=\Delta-\frac{d-\Delta_3}{2},
    &
    F&=\Delta-\frac{\Delta_3}{2}.
\end{aligned}
\label{eq::SameBoundary7F6Parameters}
\end{equation}

\vskip 4pt
Specialising now to the scalar QED value 
\begin{equation}
    \Delta_3=d-2,
\end{equation}
two upper parameters of the
${}_7F_6(1)$ cancel against identical lower parameters. The
remaining series can be written in the canonical Rogers-Dougall
form,
\begin{multline}
    C^{(d)}_{\Delta,\Delta;d-2}
    =
    \frac{1}
    {
        8\pi^{\frac d2}
        (\Delta-1)
        \left(
            \Delta-\frac d2+1
        \right)
    }
    \frac{
        (2\Delta)_{1-\frac d2}
    }{
        (\Delta)_{1-\frac d2}^{\,2}
    }
    \\
    \times
    {}_5F_4\left(
    \begin{matrix}
        a,\,
        1+\frac a2,\,
        b,\,
        c,\,
        e
        \\
        \frac a2,\,
        1+a-b,\,
        1+a-c,\,
        1+a-e
    \end{matrix}
    ;1
    \right),
\label{eq:CsameBC}
\end{multline}
where the parameters appearing in the hypergeometric function are
\begin{equation}
\begin{aligned}
    a&=2\Delta-\frac d2,
    &\qquad
    b&=\frac d2,
    \\
    c&=\Delta-1,
    &
    e&=\Delta-\frac d2+\frac12.
\end{aligned}
\label{eq::SameBoundary5F4Parameters}
\end{equation}
The series is manifestly very-well-poised and can therefore be
summed using the Rogers-Dougall formula
\eqref{eq:RogersDougall5F4}. Substituting the parameters
\eqref{eq::SameBoundary5F4Parameters} into that identity and
simplifying gives
\begin{equation}
    C^{(d)}_{\Delta,\Delta;d-2}
    =
    \frac{1}{(4\pi)^{\frac{d+1}{2}}}
    \frac{
        \Gamma(\Delta-1)
        \Gamma\left(\frac{3-d}{2}\right)
    }{
        \Gamma(\Delta-d+2)
    }.
\label{eq:samebcresult}
\end{equation}
Using the tadpole result \eqref{eq:tadpole}, this can be written as
\begin{equation}
    C^{(d)}_{\Delta,\Delta;d-2}
    =
    \frac{1}{4\pi}
    G^{(d-2)}_{\Delta-1}(x,x).
\label{eq:CsameBCTadpole}
\end{equation}

\paragraph{Opposite boundary conditions.}

We next impose
\begin{equation}
    \Delta_1=\Delta_+,
    \qquad
    \Delta_2=\Delta_-=d-\Delta_+,
\end{equation}
while leaving $\Delta_3$ arbitrary. In the original ${}_9F_8(1)$
representation, the matching upper and lower parameters
$\frac{d}{2}$ and $1$ cancel, giving
\begin{multline}
    C^{(d)}_{\Delta_+,\Delta_-;\Delta_3}
    =
    \frac{1}{
        2\pi^{\frac d2}
        \Delta_3
        \left(
            d-\Delta_3
        \right)
    }
    \frac{
        (d)_{1-\frac d2}
    }{
        (\Delta_+)_{1-\frac d2}
        (\Delta_-)_{1-\frac d2}
    }
    \\
    \times
    {}_7F_6\left(
    \begin{matrix}
        A,\,
        1+\frac{A}{2},\,
        B,\,
        C,\,
        D,\,
        E,\,
        F
        \\
        \frac{A}{2},\,
        1+A-B,\,
        1+A-C,\,
        1+A-D,\,
        1+A-E,\,
        1+A-F
    \end{matrix}
    ;1
    \right),
\label{eq:bubbleCopposite}
\end{multline}
where
\begin{equation}
\begin{aligned}
    A&=\frac d2,
    &\qquad
    B&=\frac12,
    &\qquad
    C&=\Delta_+,
    \\
    D&=\Delta_-,
    &
    E&=\frac{\Delta_3}{2},
    &
    F&=\frac{d-\Delta_3}{2}.
\end{aligned}
\label{eq::OppositeBoundary7F6Parameters}
\end{equation}

\vskip 4pt
For the scalar-QED application we set
\begin{equation}
    \Delta_3=d-2.
\end{equation}
The series in
\eqref{eq:bubbleCopposite} takes the two-parameter
very-well-poised form evaluated in
\eqref{eq:VeryWellPoised7F6}, with the parameters $A$ and $B$ in
that identity set to
\begin{equation}
    A=\frac d2,
    \qquad
    B=\frac{\Delta_+-\Delta_-}{2}.
\end{equation}
Applying \eqref{eq:VeryWellPoised7F6} directly to
\eqref{eq:bubbleCopposite}, we obtain
\begin{equation}
    C^{(d)}_{\Delta_+,\Delta_-;d-2}
    =
    \frac{
        \Gamma\left(\frac{3-d}{2}\right)
    }{
        (4\pi)^{\frac{d+1}{2}}(d-2)
    }
    \left[
        \frac{\Gamma(\Delta_+)}
        {\Gamma(\Delta_+-d+2)}
        +
        \frac{\Gamma(\Delta_-)}
        {\Gamma(\Delta_--d+2)}
    \right].
\label{eq:diffbcresult}
\end{equation}
This expression is manifestly invariant under
$\Delta_+\leftrightarrow\Delta_-$. Using the
same-boundary-condition result \eqref{eq:samebcresult}, it can also
be written as
\begin{equation}
    C^{(d)}_{\Delta_+,\Delta_-;d-2}
    =
    \frac{\Delta_+-1}{d-2}
    C^{(d)}_{\Delta_+,\Delta_+;d-2}
    +
    \frac{\Delta_--1}{d-2}
    C^{(d)}_{\Delta_-,\Delta_-;d-2}.
\label{eq:diffbcSameBCRelation}
\end{equation}
Equivalently, in terms of tadpoles,
\begin{equation}
    C^{(d)}_{\Delta_+,\Delta_-;d-2}
    =
    \frac{\Delta_+-1}{4\pi(d-2)}
    G^{(d-2)}_{\Delta_+-1}(x,x)
    +
    \frac{\Delta_--1}{4\pi(d-2)}
    G^{(d-2)}_{\Delta_--1}(x,x).
\label{eq:CdiffBCTadpole}
\end{equation}

\section{One-loop two-point functions in EAdS scalar QED}
\label{sec::loopssQEDEAdS}

In this section we compute the EAdS diagrams contributing to the
one-loop two-point function of the gauge field in scalar QED. The loop
integrals are evaluated for external spin-one bulk-to-boundary propagators
of generic dimension $\Delta_3$; the Dirichlet and Neumann gauge fields are
recovered by setting
\begin{equation}
    \Delta_3=\Delta^A_+=d-1,
    \qquad
    \Delta_3=\Delta^A_-=1,
\end{equation}
respectively, where the Neumann external leg is defined through the gauge
invariant projection discussed in section \ref{subsec::NeumannProjection}.
As we shall see, the loop coefficients are shadow symmetric in $\Delta_3$,
so that both cases follow from a single computation.

\subsection{Gauge boson propagators}

In this section we collect the gauge boson propagators needed for the loop calculations below. We consider Dirichlet and Neumann boundary conditions in turn.

\subsubsection{Dirichlet boundary condition}
For the Dirichlet boundary condition, the gauge boson bulk-to-bulk propagator takes a particularly simple form in Fried-Yennie gauge \cite{Ciccone:2024guw,Moga:2025gdy}: 
\begin{equation}
    G_{\mu\nu'}(x,x') = \frac{\Gamma\left(\frac{d+1}{2}\right)}{2\pi^\frac{d+1}{2}(d-2)}\frac{1}{\left[u(u+2)\right]^\frac{d-1}{2}}\left[\frac{u+1}{u(u+2)}\nabla_\mu u\nabla_{\nu'}u-\nabla_\mu\nabla_{\nu'}u\right]\,,
\end{equation}
where $x^\mu=(z,\vec{x})$, $x'^{\nu'}=(z',\vec{x}^{\,\prime})$, and where
\begin{equation}
    u(x,x')
    =
    \frac{
        (x-x')^2
    }{
        2zz'
    }
\label{eq::AdSChordalU}
\end{equation}
is the EAdS-invariant chordal-distance variable. Here $\nabla_\mu$ and $\nabla_{\nu'}$ act on the points $x^\mu$ and $x'^{\nu'}$, respectively. The derivatives of $u$ are
\begin{align}
    \nabla_\mu u &= \frac{1}{z}\left[\frac{1}{z'}(x-x')_\mu-u\delta_{\mu0}\right]\,,\\
    \nabla_{\nu'}u &= \frac{1}{z'}\left[\frac{1}{z}(x'-x)_{\nu'}-u\delta_{\nu'0}\right]\,,\\
    \nabla_\mu\nabla_{\nu'}u &= -\frac{1}{zz'}\left[\delta_{\mu\nu'}+\frac{1}{z'}(x-x')_\mu\delta_{\nu'0}+\frac{1}{z}(x'-x)_{\nu'}\delta_{\mu0}-u\delta_{\mu0}\delta_{\nu'0}\right]\,.
\end{align}
The corresponding gauge boson bulk-to-boundary propagator $K_{\mu i}(x,\vec{x}^{\,\prime})$ may be read from $G_{\mu\nu'}(x,x')$ by taking the bulk point $x'$ to the boundary: $z'\to0$, where to leading order in $z'$
\begin{align}
    u &\rightarrow \frac{(x-\vec{x}^{\,\prime})^2}{2zz'}\,,\\
    \nabla_\mu u &\rightarrow \frac{1}{zz'}\left[(x-\vec{x}^{\,\prime})_\mu-\frac{1}{2z}(x-\vec{x}^{\,\prime})^2\delta_{\mu0}\right]\,,\\
    \nabla_{\nu'}u &\rightarrow -\frac{1}{zz'}(x-\vec{x}^{\,\prime})_i\,,\\
    \nabla_\mu\nabla_{\nu'}u &\rightarrow -\frac{1}{zz'}\left[\delta_{\mu i}-\frac{1}{z}(x-\vec{x}^{\,\prime})_i\delta_{\mu0}\right]\,,
\end{align}
where $(x-\vec{x}^{\,\prime})^2=z^2+(\vec{x}-\vec{x}^{\,\prime})^2$. In this limit the propagator is identified by $G_{\mu\nu'}\to\frac{(z')^{d-2}}{d-2}K_{\mu i}$, obtaining
\begin{equation}\label{eq::spin1buboD}
    K_{\mu i}(x,\vec{x}^{\,\prime}) = \frac{C^{(d)}}{z}\left[\frac{z}{(x-\vec{x}^{\,\prime})^2}\right]^{d-1}I_{\mu i}(x-\vec{x}^{\,\prime})\,,\hs C^{(d)} = \frac{\Gamma(d)}{2\pi^\frac{d}{2}\Gamma\left(\frac{d}{2}\right)}\,,
\end{equation}
where $I_{\mu i}$ is the inversion tensor:
\begin{equation}\label{eq::InversionTensor} 
    I_{\mu i}(x-\vec{x}^{\,\prime}) = \delta_{\mu i}-2\frac{(x-\vec{x}^{\,\prime})_\mu(x-\vec{x}^{\,\prime})_i}{(x-\vec{x}^{\,\prime})^2}\,.
\end{equation}
This propagator coincides with eq. (47) of \cite{Freedman:1998tz}. It may also be expressed as
\begin{equation}
    K_{\mu i}(x,\vec{x}^{\,\prime}) = \left(\frac{d-1}{d-2}\right)\frac{1}{z}K_{d-1}(x,\vec{x}^{\,\prime})I_{\mu i}(x-\vec{x}^{\,\prime})\,,
\end{equation}
where $K_{d-1}(x,\vec{y})$ is the scalar bulk-to-boundary propagator with dimension $\Delta=d-1$.

\subsubsection{Neumann boundary condition}
\label{subsec::NeumannProjection}

For generic scaling dimension $\Delta$, the spin-1 bulk-to-boundary propagator reads \cite{Mueck:1998iz} 
\begin{equation}
    K_{\Delta,\mu i}(x,\vec y)
    =
    \frac{C_{\Delta,1}}{z}
    \left[
        \frac{z}{(x-\vec y)^2}
    \right]^\Delta
    I_{\mu i}(x-\vec y),
    \qquad
    C_{\Delta,1}
    =
    \frac{\Gamma(\Delta+1)}
    {\pi^{d/2}(\Delta-1)
    \Gamma\left(\Delta-\frac d2\right)} .
\label{eq::IntrinsicSpinOneBulkToBoundary}
\end{equation}
In terms of the scalar bulk-to-boundary normalisation \eqref{eq::sc2ptnorm} we have
\begin{equation}
    C_{\Delta,1}
    =
    \frac{\Delta}{\Delta-1}\,c_\Delta .
\label{eq::SpinOneScalarNormalisationRelation}
\end{equation}
This is the continuation to generic $\Delta$ of the conventions used for the Dirichlet gauge-boson bulk-to-boundary
propagator in~\eqref{eq::spin1buboD} of the previous section.
In particular, $C_{d-1,1}=C^{(d)}$ and $K_{d-1,\mu i}(x,\vec y)=K_{\mu i}(x,\vec y)$.

\vskip 4pt
The two boundary conditions for a Maxwell field correspond to
\begin{equation}
    \Delta^A_+=d-1,
    \qquad
    \Delta^A_-=1.
\end{equation}
At the Dirichlet point $\Delta=d-1$, the expression
\eqref{eq::IntrinsicSpinOneBulkToBoundary} reduces to the expression \eqref{eq::spin1buboD}. At the Neumann point $\Delta=1$, however, the normalisation $C_{\Delta,1}$ has a pole for generic $d$. Since
the one-loop mass matrix \eqref{eq::oneLoopGaugeMassMatrix} involves external legs with both boundary conditions, it remains to specify how this limit is taken.

\vskip 4pt
To identify the origin of the pole, strip off the normalisation and
define
\begin{equation}
    \hat K_{\Delta,\mu i}(x,\vec y)
    =
    \frac{1}{z}
    \left[
        \frac{z}{(x-\vec y)^2}
    \right]^\Delta
    I_{\mu i}(x-\vec y).
\end{equation}
At $\Delta=1$, this tensor degenerates into a double gradient,
\begin{equation}
    \hat K_{1,\mu i}(x,\vec y)
    =
    \frac{I_{\mu i}(x-\vec y)}{(x-\vec y)^2}
    =
    -\frac{1}{2}\,
    \partial^x_\mu\partial^{\vec y}_i
    \log\left[(x-\vec y)^2\right].
\label{eq::NeumannPotentialDoubleGradient}
\end{equation}
Introducing
\begin{equation}
    \widetilde C_{\Delta,1}
    \equiv
    (\Delta-1)C_{\Delta,1}
    =
    \frac{\Gamma(\Delta+1)}
    {\pi^{d/2}\Gamma\left(\Delta-\frac d2\right)},
\label{eq::RegularSpinOneNormalisation}
\end{equation}
the singular part of the bulk-to-boundary propagator is therefore
\begin{equation}
    K_{\Delta,\mu i}(x,\vec y)
    =
    -\frac{\widetilde C_{1,1}}
    {2(\Delta-1)}
    \partial^x_\mu\partial^{\vec y}_i
    \log\left[(x-\vec y)^2\right]
    +
    O\left((\Delta-1)^0\right).
\label{eq::NeumannPureGaugeDivergence}
\end{equation}
The residue is thus pure gauge with respect to both the bulk and
boundary indices. It is annihilated either by taking the bulk curl at
$x$ or by taking the boundary curl at $\vec y$. This is the EAdS
counterpart of the longitudinal $p^\mu p^\nu/(p^2)^2$ contribution to a
flat space covariant propagator, which drops out of field-strength
correlators.\footnote{In a covariant
gauge with gauge parameter $\zeta$, the massless vector propagator in flat space reads 
\begin{equation}
    \left\langle A^\mu(x)A^\nu(y)\right\rangle
    =
    \int \frac{{\rm d}^dp}{(2\pi)^d}\, e^{ip\cdot(x-y)}
    \left[
        \frac{1}{p^2}\left(\delta^{\mu\nu}-\frac{p^\mu p^\nu}{p^2}\right)
        +
        \frac{\zeta}{p^2}\frac{p^\mu p^\nu}{p^2}
    \right].
\label{eq::FlatCovariantGaugePropagator}
\end{equation}}

\vskip 4pt
This separation is manifest at the level of the helicity decomposition. Fourier transforming in the boundary directions, the spin-one propagator decomposes as 
\begin{equation}
    K_{\Delta,\mu i}(z,\vec p)
    =
    K^{{\rm T}}_{\Delta,\mu i}(z,\vec p)
    +
    K^{{\rm L}}_{\Delta,\mu i}(z,\vec p),
\end{equation}
where
\begin{equation}
    K^{{\rm T}}_{\Delta,\mu i}
    \equiv
    K_{\Delta,\mu j}\Pi^{\rm T}_{ji},
    \qquad
    K^{{\rm L}}_{\Delta,\mu i}
    \equiv
    K_{\Delta,\mu j}\Pi^{\rm L}_{ji}.
\label{eq::SpinOneHelicityDecomposition}
\end{equation}
The transverse $({\rm T})$ and longitudinal $({\rm L})$ projectors are
\begin{equation}
    \Pi^{\rm T}_{ij}(\vec p)
    =
    \delta_{ij}-\frac{p_i p_j}{\vec p^{\,2}},
    \qquad
    \Pi^{\rm L}_{ij}(\vec p)
    =
    \frac{p_i p_j}{\vec p^{\,2}} .
\label{eq::BoundaryTransverseLongitudinalProjectors}
\end{equation}
The component $K^{{\rm T}}_{\Delta,\mu i}$ carries the highest helicity,
whereas $K^{{\rm L}}_{\Delta,\mu i}$ is the longitudinal, helicity-zero
component. In particular,
\begin{equation}
    p^iK^{{\rm T}}_{\Delta,\mu i}(z,\vec p)=0.
\end{equation}

\vskip 4pt
The pure-gauge form of \eqref{eq::NeumannPureGaugeDivergence} implies
that in momentum space
\begin{equation}
    \underset{\Delta=1}{\operatorname{Res}}\,
    K_{\Delta,\mu i}(z,\vec p)
    =
    i p_i\,\partial^x_\mu\Lambda(z,\vec p)
\end{equation}
for some scalar function $\Lambda$. The residue is entirely
longitudinal in the boundary index and therefore
\begin{equation}
    \underset{\Delta=1}{\operatorname{Res}}\,
    K^{{\rm T}}_{\Delta,\mu i}(z,\vec p)
    =
    0.
\end{equation}
Although the full gauge field propagator is singular at $\Delta=1$,
its highest-helicity component consequently has a smooth Neumann limit. The Neumann external leg can therefore be obtained directly from the standard spin-one bulk-to-boundary propagator \eqref{eq::IntrinsicSpinOneBulkToBoundary}: one keeps $\Delta$ generic and retains its highest-helicity component, only then taking $\Delta\to1$.

\vskip 4pt
According to the dictionary \eqref{eq::NeumannGaugeDictionary}, the
boundary operator associated with the Neumann branch is the dynamical
gauge field $\tilde a_i$.  Its gauge invariant content is carried by the
boundary field strength
\begin{equation}
    f_{ij}
    =
    \partial_i\tilde a_j-\partial_j\tilde a_i .
\label{eq::BoundaryFieldStrength}
\end{equation}
For the Dirichlet branch no such projection is required, since the
associated operator $J_i$ is a gauge-invariant conserved current.

\vskip 4pt
To obtain the field strength bulk-to-boundary propagator, we first
take the boundary curl of
\eqref{eq::IntrinsicSpinOneBulkToBoundary} at generic $\Delta$:
\begin{equation}
    \partial^{\vec y}_kK_{\Delta,\mu i}(x,\vec y)
    -
    \partial^{\vec y}_iK_{\Delta,\mu k}(x,\vec y)
    =
    \frac{2\widetilde C_{\Delta,1}}{z}
    \left[
        \frac{z}{(x-\vec y)^2}
    \right]^\Delta
    \frac{
        (\vec x-\vec y)_k\delta_{\mu i}
        -
        (\vec x-\vec y)_i\delta_{\mu k}
    }{(x-\vec y)^2}.
\label{eq::BoundaryCurlKernel}
\end{equation}
The factor of $\Delta-1$ generated by the boundary curl cancels
the pole in $C_{\Delta,1}$, giving the smooth Neumann limit
\begin{equation}
\begin{aligned}
    K^{(f)}_{\mu|ki}(x,\vec y)
    & \equiv
    \lim_{\Delta\to1}
    \left(
        \partial^{\vec y}_kK_{\Delta,\mu i}(x,\vec y)
        -
        \partial^{\vec y}_iK_{\Delta,\mu k}(x,\vec y)
    \right)\\
    & =
    \frac{2}
    {\pi^{d/2}\Gamma\left(1-\frac d2\right)}
    \frac{
        (\vec x-\vec y)_k\delta_{\mu i}
        -
        (\vec x-\vec y)_i\delta_{\mu k}
    }{
        \left[(x-\vec y)^2\right]^2
    }.
\end{aligned}
\label{eq::BoundaryCurlKernelNeumann}
\end{equation}

\vskip 4pt
The boundary curl depends only on the highest-helicity component
identified above, since the longitudinal component drops out upon
antisymmetrisation. Thus, in momentum space,
\begin{equation*}
    K^{(f)}_{\mu|ki}(z,\vec p)
    =
    i\lim_{\Delta\to1}
    \left(
        p_kK^{{\rm T}}_{\Delta,\mu i}(z,\vec p)
        -
        p_iK^{{\rm T}}_{\Delta,\mu k}(z,\vec p)
    \right).
\end{equation*}
Conversely, for $\vec p\neq0$,
\begin{equation}
    \lim_{\Delta\to1}K^{{\rm T}}_{\Delta,\mu i}(z,\vec p)
    =
    \frac{p^k}{i\vec p^{\,2}}\,
    K^{(f)}_{\mu|ki}(z,\vec p).
\end{equation}
At non-zero momentum, the field strength bulk-to-boundary propagator
and the highest-helicity component of the Neumann gauge field
propagator therefore contain the same information. Working with
the boundary field strength is convenient because the boundary
curl is local in position space and commutes with the bulk integrations.

\vskip 4pt
Gauge invariant late-time correlators with Neumann external legs
can then be obtained by evaluating the diagram at generic external
dimensions, applying a boundary curl to each Neumann leg, and only
then taking the corresponding dimensions to one. For a two-point diagram with two external spin-one legs of
dimension $\Delta$, denote its contribution by
$\left(I_{\Delta\Delta}\right)_{ij}(\vec y_1,\vec y_2)$. The corresponding contribution
with two boundary field-strength insertions is
\begin{equation}
    \left(I_{ff}\right)_{ki|lj}(\vec y_1,\vec y_2)
    =
    \lim_{\Delta\to1}
    4\,
    \partial^{\vec y_1}_{[k}
    \partial^{\vec y_2}_{[l}
    \left(I_{\Delta\Delta}\right)_{i]\,j]}
    (\vec y_1,\vec y_2).
    \label{eq::NeumannBoundaryProjectedTwoPoint}
\end{equation}
For a mixed Dirichlet-Neumann amplitude, the projection and limit
are applied only to the Neumann leg.

\subsection{Tadpole diagram}
\label{sec::vectorTadpole}

We first evaluate the tadpole building block associated with a single real scalar field $\phi_{\Delta}$ with scaling dimension $\Delta$, generated by the local
quartic structure
\begin{equation}
    {\cal V}^{(4)}
    \left[
        A,\phi_{\Delta}
    \right]
    =
    \frac{1}{2}
    A_\mu A^\mu
    \phi^2_{\Delta}.
\label{eq::quarticTadpoleBuildingBlock}
\end{equation}

The contribution to the two-point function with external spin-one legs of dimension $\Delta_3$ is
\begin{equation}
    \left(I^{T}_{\Delta;\Delta_3}\right)_{ij}(\vec{y}_1,\vec{y}_2)
    =
    \int\dd^{d+1}x\sqrt{g}\
    g^{\mu\nu}
    K_{\Delta_3,\mu i}(x,\vec{y}_1)
    K_{\Delta_3,\nu j}(x,\vec{y}_2)
    G_\Delta(x,x).
\label{eq::vectorTadpoleIntegral}
\end{equation}
By EAdS invariance, the coincident scalar propagator
$G_\Delta(x,x)$ is independent of the bulk point and may therefore
be taken outside the integral. Defining
\begin{equation}
    \left(T_{\Delta\Delta}\right)_{ij}(\vec{y}_1,\vec{y}_2)
    =
     \int\dd^{d+1}x\sqrt{g}\
    g^{\mu\nu}
    K_{\Delta,\mu i}(x,\vec{y}_1)
    K_{\Delta,\nu j}(x,\vec{y}_2),
\label{eq::Tij}
\end{equation}
we obtain
\begin{equation}
   \left(I^{T}_{\Delta;\Delta_3}\right)_{ij}(\vec{y}_1,\vec{y}_2)
    =
    G_\Delta(x,x)\,
    \left(T_{\Delta_3\Delta_3}\right)_{ij}(\vec{y}_1,\vec{y}_2).
\label{eq::vectorTadpoleResult}
\end{equation}
The coincident propagator $G_\Delta(x,x)$ was evaluated in dimensional regularisation in~\eqref{eq:tadpole}. The tensor structure
$\left(T_{\Delta_3\Delta_3}\right)_{ij}(\vec{y}_1,\vec{y}_2)$ is that generated by a local mass
insertion for the spin-one field. As we show below, the bubble diagram is proportional to the same structure.

\subsection{Bubble diagram}
\label{subsec::bubblediagramEAdSsQED}

We next evaluate the bubble building block generated by two cubic
current vertices. For two real scalar fields $\phi_{1,\Delta_1}$ and
$\phi_{2,\Delta_2}$, the relevant local cubic structure is
\begin{equation}
    {\cal V}^{(3)}
    \left[
        A,\phi_{1,\Delta_1},\phi_{2,\Delta_2}
    \right]
    =
    A_\mu\,
    \phi_{2,\Delta_2}
    \overset{\leftrightarrow}{\nabla}{}^\mu
    \phi_{1,\Delta_1},
\label{eq::cubicBubbleBuildingBlock}
\end{equation}
where
\begin{equation}
    f\overset{\leftrightarrow}{\nabla}{}_\mu g
    \equiv
    f\nabla_\mu g
    -
    \left(\nabla_\mu f\right)g.
\label{eq::bidirectionalDerivativeDefinition}
\end{equation}

The resulting contribution to the two-point function with external spin-one legs of dimension $\Delta_3$ is
\begin{equation}
    \label{eq:SQED2p1loop}
    \left(I^\text{B}_{\Delta_1,\Delta_2;\Delta_3}\right)_{ij}(\vec{y}_1,\vec{y}_2) =  \int_{x_1}\int_{x_2}K_{\Delta_3,\mu i}(x_1,\vec{y}_1)G_{\Delta_1}(x_1,x_2)\overset{\leftrightarrow}{\nabla}{}_1^\mu\overset{\leftrightarrow}{\nabla}{}_2^\nu G_{\Delta_2}(x_1,x_2)K_{\Delta_3,\nu j}(x_2,\vec{y}_2)\,,
\end{equation}
where $\int_x\equiv\int\dd^{d+1}x\sqrt{g}$.

\begin{figure}[h]
    \centering
    \includegraphics[width=0.4\textwidth]{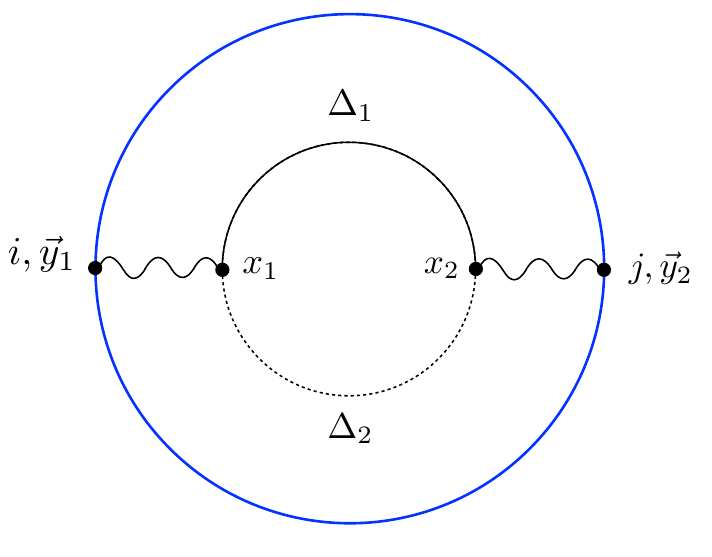}
    \caption{Scalar QED bubble diagram in EAdS\label{fig:SQED2p1loop}}
\end{figure}

\vskip 4pt
We evaluate the bubble integral \eqref{eq:SQED2p1loop} in
the following. We first note that EAdS isometries imply the contribution takes the form (see appendix \ref{subsec::expectisomGB})
\begin{equation}
    \left(I^\text{B}_{\Delta_1,\Delta_2;\Delta_3}\right)_{ij}(\vec{y}_1,\vec{y}_2)
    =
    4A^{(d)}_{\Delta_1,\Delta_2;\Delta_3}\,
    \left(T_{\Delta_3\Delta_3}\right)_{ij}(\vec{y}_1,\vec{y}_2),
\label{eq::bubbleA12}
\end{equation}
where $(T_{\Delta\Delta})_{ij}$ was defined in~\eqref{eq::Tij}. The
coefficient $A^{(d)}_{\Delta_1,\Delta_2;\Delta_3}$ is then related to the scalar bubble
coefficient introduced in section~\ref{sec:scalar-bubble} by\footnote{Here the superscript $d+2$ means that the scalar bubble coefficient is evaluated
in boundary dimension $d+2$, while the three dimensions entering the coefficient
are
\begin{equation}
    \Delta_1\to\Delta_1+1,
    \qquad
    \Delta_2\to\Delta_2+1,
    \qquad
    \Delta_3\to\Delta_3+1 .
\end{equation}}${}^{,}$\footnote{Relation \eqref{eq::vectorBubbleScalarBubbleRelation} is the EAdS
counterpart of a familiar identity in flat space: the trace part of a rank-two
tensor bubble in $D$ dimensions is $2\pi$ times the corresponding non-derivative scalar bubble in $D+2$ dimensions,
\begin{equation}
    \int
    \frac{{\rm d}^D\ell}{(2\pi)^D}\,
    \ell^\mu\ell^\nu
    f(\ell^2)
    =
    2\pi\,
    \delta^{\mu\nu}
    \int
    \frac{{\rm d}^{D+2}\ell}{(2\pi)^{D+2}}\,
    f(\ell^2).
\nonumber
\end{equation}
Note the same universal factor $2\pi$ that appears in
\eqref{eq:derivativeGshift}. Dimension-shift identities of this type are
standard in the amplitudes literature
\cite{Bern:1993kr,Tarasov:1996br}. We spell out these parallels in
appendix \ref{app::FlatDimensionShift}.}
\begin{equation}
    A^{(d)}_{\Delta_1,\Delta_2;\Delta_3}
    =
    2\pi\,
    C^{(d+2)}_{\Delta_1+1,\Delta_2+1;\Delta_3+1}.
\label{eq::vectorBubbleScalarBubbleRelation}
\end{equation}
To see this, it is useful to consider the trace of \eqref{eq:Aconstant}, which gives
\begin{equation}
    A^{(d)}_{\Delta_1,\Delta_2;\Delta_3}
    =
    \frac{1}{d}
    \int\dd^{d+1}x_2\sqrt{g_2}\,
    z_2^{\Delta_3-1}\vec{x}_2^{\,2}\xi^4
    \partial_\xi G_{\Delta_1}(\xi)
    \partial_\xi G_{\Delta_2}(\xi),
\label{eq:Atrace}
\end{equation}
where here and below $ \xi=\xi((1,\vec 0),x_2)$. The key observation is that a derivative of the scalar bulk-to-bulk propagator can
be written as a scalar propagator in two higher boundary dimensions:
\begin{align}
    \partial_\xi G^{(d)}_\Delta(\xi)
    &=
    \frac{\Delta c_\Delta}{2^{\Delta+1}\left(\Delta-\frac d2\right)}
    \xi^{\Delta-1}
    {}_2F_1\left(
    \begin{matrix}
        \frac{\Delta+1}{2},\ \frac{\Delta}{2}+1\\
        \Delta-\frac d2+1
    \end{matrix}
    ;\xi^2
    \right)
    \nonumber\\
    &=
    \frac{2\pi}{\xi^2}
    G^{(d+2)}_{\Delta+1}(\xi).
\label{eq:derivativeGshift}
\end{align}
Therefore
\begin{equation}
    \partial_\xi G^{(d)}_{\Delta_1}(\xi)
    \partial_\xi G^{(d)}_{\Delta_2}(\xi)
    =
    \frac{4\pi^2}{\xi^4}
    G^{(d+2)}_{\Delta_1+1}(\xi)
    G^{(d+2)}_{\Delta_2+1}(\xi).
\label{eq:derivativeProduct}
\end{equation}
Using the scalar product decomposition in boundary dimension $d+2$, this becomes
\begin{equation}
    \partial_\xi G^{(d)}_{\Delta_1}(\xi)
    \partial_\xi G^{(d)}_{\Delta_2}(\xi)
    =
    \frac{4\pi^2}{\xi^4}
    \sum_{k=0}^{\infty}
    \sum_{l=0}^{\infty}
    g^{(d+2)}_{\Delta_1+1,\Delta_2+1}(k,l)
    \xi^{\Delta_{12}+2+2k+2l},
\label{eq:derivativeProductSeries}
\end{equation}
with the coefficients $g^{(d)}_{\Delta_1,\Delta_2}(k,l)$ defined in
\eqref{eq:gklcoef}. Substituting \eqref{eq:derivativeProductSeries} into \eqref{eq:Atrace} gives
\begin{equation}
    A^{(d)}_{\Delta_1,\Delta_2;\Delta_3}
    =
    \frac{4\pi^2}{d}
    \sum_{k=0}^{\infty}
    \sum_{l=0}^{\infty}
    g^{(d+2)}_{\Delta_1+1,\Delta_2+1}(k,l)
    \int\dd^{d+1}x_2\sqrt{g_2}\,
    z_2^{\Delta_3-1}\vec{x}_2^{\,2}
    \xi^{\Delta_{12}+2+2k+2l}.
\label{eq:AseriesBeforeIntegral}
\end{equation}
The remaining integral is elementary. For generic parameters,
\begin{multline}
\int {\rm d}^{d+1}x_2\sqrt{g_2}\,
    \xi^{\Delta_i}((1,\vec 0),x_2)
    z_2^{\Delta_j}
    (\vec{x}_2^{\,2})^{\Delta_k} \\ =
    \pi^{\frac{d+1}{2}}
    \frac{\Gamma\left(\Delta_k+\frac d2\right)}
    {\Gamma\left(\frac d2\right)}
    \frac{
    \Gamma\left(\frac{\Delta_i+\Delta_j-d}{2}\right)
    \Gamma\left(\frac{\Delta_i-\Delta_j}{2}-\Delta_k\right)
    }
    {
    \Gamma\left(\frac{\Delta_i}{2}\right)
    \Gamma\left(\frac{\Delta_i+1}{2}\right)
    } .
\label{eq:xiIntegralWithPower}
\end{multline}
Applying \eqref{eq:xiIntegralWithPower} with
\begin{equation}
    \Delta_i=\Delta_{12}+2+2k+2l,
    \qquad
    \Delta_j=\Delta_3-1,
    \qquad
    \Delta_k=1,
\end{equation}
one obtains
\begin{align}
    A^{(d)}_{\Delta_1,\Delta_2;\Delta_3}
    =
    2\pi^{\frac{d+5}{2}}
    \sum_{k=0}^{\infty}
    \sum_{l=0}^{\infty}
    g^{(d+2)}_{\Delta_1+1,\Delta_2+1}(k,l)
    \frac{
    \Gamma\left(\frac{\Delta_{12}-\Delta_3+1}{2}+k+l\right)
    \Gamma\left(\frac{\Delta_{123}-d+1}{2}+k+l\right)
    }
    {
    \Gamma\left(\frac{\Delta_{12}}{2}+1+k+l\right)
    \Gamma\left(\frac{\Delta_{12}+3}{2}+k+l\right)
    }.
\label{eq:AseriesFinal}
\end{align}
Comparing \eqref{eq:AseriesFinal} with the scalar bubble coefficient
\eqref{eq:Ctaucoeff1}, we find the compact relation
\begin{equation}
    A^{(d)}_{\Delta_1,\Delta_2;\Delta_3}
    =
    2\pi\,
    C^{(d+2)}_{\Delta_1+1,\Delta_2+1;\Delta_3+1}.
\label{eq:AasScalarBubble}
\end{equation}

\vskip 4pt
The coefficient \eqref{eq:AseriesFinal} is manifestly invariant under
$\Delta_3\to d-\Delta_3$, as it must be: $\Delta_3+1$ and $d+1-\Delta_3$ are
shadow related in boundary dimension $d+2$, and the scalar bubble coefficient
is invariant under the external shadow transformation. In particular, the
Dirichlet and Neumann gauge fields, $\Delta_3=d-1$ and $\Delta_3=1$, give the
same coefficient,
\begin{equation}
    A^{(d)}_{\Delta_1,\Delta_2}
    \equiv
    A^{(d)}_{\Delta_1,\Delta_2;d-1}
    =
    A^{(d)}_{\Delta_1,\Delta_2;1}
    =
    2\pi\,C^{(d+2)}_{\Delta_1+1,\Delta_2+1;d}
    =
    2\pi\,C^{(d+2)}_{\Delta_1+1,\Delta_2+1;2},
\label{eq::Ashorthand}
\end{equation}
which we denote by $A^{(d)}_{\Delta_1,\Delta_2}$ in the following. Relation
\eqref{eq:AasScalarBubble} thus determines the scalar QED loop coefficient for
both external boundary conditions in terms of the scalar bubble coefficient
studied in section \ref{sec:scalar-bubble}. 

\vskip 4pt
This has a transparent boundary interpretation. Under the dimension shift a spin-one external leg of dimension $\Delta$ is traded for a scalar leg of dimension $\Delta+1$, which in either case is the dimension of one boundary
derivative acting on the associated boundary operator: $\Delta_3=d$ for the divergence of the current $J_i$ of dimension $d-1$, and $\Delta_3=2$ for the field strength $f_{ki}=\partial_k\tilde a_i-\partial_i\tilde a_k$ of the
dimension one boundary gauge field $\tilde a_i$. The Neumann value is thus precisely the dimension of the gauge-invariant boundary observable singled out by the projection of section~\ref{subsec::NeumannProjection}. Note also that
$d=D-2$ in boundary dimension $D=d+2$, so $(d,2)$ is exactly the
shadow-related pair $(D-2,2)$ for which the summed expressions of
section~\ref{sec::ScalarQEDSummedForms} apply.

\subsection{One loop two-point function in EAdS scalar QED}
\label{subsec::EAdSScalarQEDVacuumPolarisation}

The loop integrals evaluated above also determine the one-loop vacuum polarisation in scalar QED defined directly on
$\text{AdS}_{d+1}$. This has Euclidean action
\begin{equation}
    S_{\text{EAdS}}
    =
    \int_{\mathrm{EAdS}_{d+1}}
    {\rm d}^{d+1}x\,\sqrt{g}\,
    \left[
        \frac{1}{4}F_{\mu\nu}F^{\mu\nu}
        +
        (D_\mu\varphi)^*D^\mu\varphi
        +
        m^2\varphi^*\varphi
    \right],
\label{eq::OrdinaryEAdSScalarQEDAction}
\end{equation}
where
\begin{equation}
    m^2=\Delta(\Delta-d).
\end{equation}
We impose boundary conditions on the scalar that preserve the bulk
$U(1)$ symmetry. Writing
\begin{equation}
    \varphi
    =
    \frac{1}{\sqrt{2}}
    \left(
        \phi_1+i\phi_2
    \right),
\end{equation}
the two real components $\phi_1$ and $\phi_2$ therefore obey the
same scalar boundary condition $\Delta$. The interaction Lagrangian is
\begin{equation}
    {\cal L}^{\mathrm{int}}_{\mathrm{EAdS}}
    =
    -eA_\mu j^\mu
    +
    \frac{e^2}{2}A_\mu A^\mu
    \left(
        \phi_1^2+\phi_2^2
    \right),
    \qquad
    j^\mu
    =
    \phi_2\nabla^\mu\phi_1
    -
    \phi_1\nabla^\mu\phi_2 .
\label{eq::OrdinaryEAdSScalarQEDInteractions}
\end{equation}

\vskip 4pt
At one loop, the gauge field two-point function receives a bubble contribution
from two insertions of the cubic vertex and a tadpole contribution
from the quartic vertex. These are
\begin{align}
   e^2 \left(
        I^{\mathrm{B}}_{\Delta \Delta; \Delta^A}
    \right)_{ij}
    &=
    4 e^2 A^{(d)}_{\Delta,\Delta}\,
    \left(T_{\Delta^A \Delta^A}\right)_{ij},\\
   -2e^2 \left(
        I^{\mathrm{T}}_{\Delta;\Delta^A}
    \right)_{ij}
    &=
    -2e^2G_\Delta(x,x)\,
    \left(T_{\Delta^A \Delta^A}\right)_{ij},
\label{eq::OrdinaryEAdSBubbleTadpole}
\end{align}
where $\left(T_{\Delta^A \Delta^A}\right)_{ij}$ is the mass-insertion tensor \eqref{eq::Tij} introduced above. From \eqref{eq:CsameBCTadpole} we have
\begin{equation}
    A^{(d)}_{\Delta,\Delta}
    =
    2\pi\,
    C^{(d+2)}_{\Delta+1,\Delta+1;d}
    =
    \frac{1}{2}G^{(d)}_\Delta(x,x).
\label{eq::SameBoundaryBubbleTadpoleIdentity}
\end{equation}
It follows that
\begin{equation}
   e^2 \left(
        I^{\mathrm{B}}_{\Delta \Delta; \Delta^A}
        -2
        I^{\mathrm{T}}_{\Delta;\Delta^A}
    \right)_{ij}
    =
    0.
\label{eq::OrdinaryEAdSMassCancellation}
\end{equation}
Thus the scalar loop does not generate a local mass for the gauge
field. This is the AdS generalisation of the analogous cancellation for the flat space vacuum polarisation which we review in Appendix~\ref{subsec::sqedflatren}. 

\vskip 4pt
The cancellation \eqref{eq::OrdinaryEAdSMassCancellation} is in fact independent of the boundary condition imposed on the external gauge field. For the Neumann boundary condition
$\Delta^A_-=1$, the dimension-shifting relation involves a scalar
external leg of dimension $2$. Since $2$ and $d$ are shadow related in boundary dimension $d+2$, external shadow symmetry gives the same loop coefficient as for the Dirichlet boundary condition, as discussed after
\eqref{eq:AasScalarBubble}. External legs subject to the Neumann boundary condition are understood
in the gauge-invariant sense of
section~\ref{subsec::NeumannProjection}: one projects onto the boundary
field strength, or equivalently onto the highest-helicity component,
before taking $\Delta^A\to1$. Since this projection acts only on the
external points and commutes with the bulk integrations, it preserves
the bubble-tadpole cancellation. The combination \eqref{eq::OrdinaryEAdSMassCancellation} therefore vanishes for Neumann boundary conditions as well.

\vskip 4pt
The distinction from the EAdS theory \eqref{scalarQEDL} for dS Schwinger-Keldysh correlators is important. In the ordinary Euclidean formulation \eqref{eq::OrdinaryEAdSScalarQEDAction} of AdS scalar QED, Dirichlet and Neumann are alternative boundary conditions for a single gauge field, while a
$U(1)$-preserving scalar boundary condition assigns the same
dimension $\Delta$ to the two real components of the charged scalar. Every scalar bubble therefore contains two propagators with the same boundary condition and cancels against the corresponding tadpole. By contrast, the EAdS reformulation of the de Sitter theory contains independent scalar fields with shadow-related dimensions $\Delta^\varphi_+$ and $\Delta^\varphi_-$, together with interactions that couple the two. It therefore admits a bubble containing one scalar of each dimension, which has no counterpart in \eqref{eq::OrdinaryEAdSScalarQEDAction}. This mixed sector is not subject to the cancellation above and gives rise to the non-vanishing mass insertion found in the next section.

\section{One-loop two-point functions in dS scalar QED from EAdS}
\label{sec::dSloop}

\begin{figure}[t]
    \centering
    \includegraphics[width=0.9\textwidth]{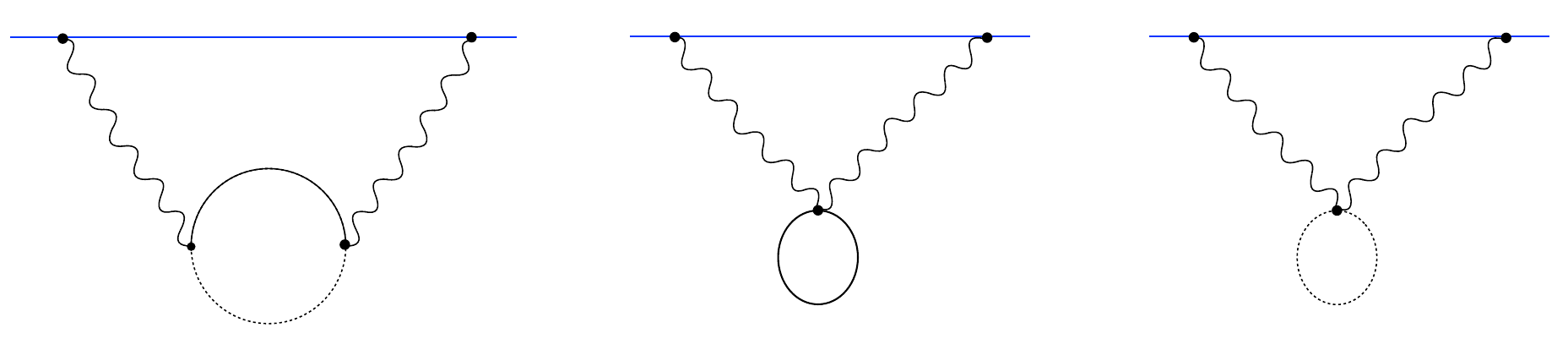}
    \caption{One-loop scalar-QED diagrams contributing to the late-time dS gauge field two-point function: the scalar bubble and the two tadpole diagrams. The blue line denotes the future boundary, wavy lines the gauge field, and solid and dotted lines the two real scalar fields.}
    \label{fig::dSloops}
\end{figure}

In this section we combine the EAdS Feynman rules for dS late-time correlators in scalar QED, derived in
section~\ref{sec::eadssqed}, with the EAdS one-loop integrals evaluated in sections~\ref{sec::scalarloops}
and~\ref{sec::loopssQEDEAdS}, to evaluate the late-time two-point functions of gauge fields in dS at one loop. See figure~\ref{fig::dSloops}.

\vskip 4pt
Before evaluating the diagrams, it is useful to reorganise the
interaction Lagrangian \eqref{scalarQEDL} in a form adapted to the one-loop gauge field
two-point function. In particular, 
\begin{align}
   - \mathcal{L}^{\mathrm{int}}_{\mathrm{EAdS}}
    ={}&
    \sum_{\alpha=\pm}
    \bigg[
        \left(
            g_{+\alpha}A_{\Delta^A_+}
            +
            g_{-\alpha}A_{\Delta^A_-}
        \right) \cdot 
        J_{\alpha\alpha}
        \nonumber\\
    &\hspace{1.3cm}
        -
        \frac{\kappa_{\Delta^\varphi_\alpha}}{2}
        \left(
            g_{+\alpha}A_{\Delta^A_+}
            +
            g_{-\alpha}A_{\Delta^A_-}
        \right) \cdot 
        \left(
            g_{+\alpha}A_{\Delta^A_+}
            +
            g_{-\alpha}A_{\Delta^A_-}
        \right)
        \left(\phi_{1,\Delta^\varphi_\alpha}^{\,2}+\phi_{2,\Delta^\varphi_\alpha}^{\,2}\right)
    \bigg]
    \nonumber\\
    &+
    g_{\mathrm{mix}}
    A_{\Delta^A_+} \cdot 
    \left(
        J_{+-}+J_{-+}
    \right)+
    \frac{g_{\mathrm{mix}}^2}{2}
    A_{\Delta^A_+} \cdot 
    A_{\Delta^A_+}
    \sum_{\alpha=\pm}
    \kappa_{\Delta^\varphi_\alpha}\left(\phi_{1,\Delta^\varphi_\alpha}^{\,2}+\phi_{2,\Delta^\varphi_\alpha}^{\,2}\right)
    \nonumber\\
    &+
    \mathcal{L}_{\mathrm{mixed}}^{(4)},
\label{eq::LoopAdaptedLagrangian}
\end{align}
with coefficients
\begin{align}
    g_{+\alpha}
    &\equiv
    2e
    \sqrt{
        \left|
            c_{\Delta^A_+}^{\mathrm{dS-AdS}}
        \right|
    }
    \left|
        c_{\Delta^\varphi_\alpha}^{\mathrm{dS-AdS}}
    \right|
    \sin\left(\pi\Delta^\varphi_\alpha\right),
    \nonumber\\
    g_{-\alpha}
    &\equiv
    2e
    \sqrt{
        \left|
            c_{\Delta^A_-}^{\mathrm{dS-AdS}}
        \right|
    }
    \left|
        c_{\Delta^\varphi_\alpha}^{\mathrm{dS-AdS}}
    \right|
    \sin\left[
        \pi\left(
            \Delta^\varphi_\alpha-\frac{d}{2}+1
        \right)
    \right],
    \qquad \alpha=\pm,
    \nonumber\\
    g_{\mathrm{mix}}
    &\equiv
    2e
    \sqrt{
        \left|
            c_{\Delta^A_+}^{\mathrm{dS-AdS}}
        \right|
    }
    \sqrt{
        \left|
            c_{\Delta^\varphi_+}^{\mathrm{dS-AdS}}
        \right|
        \left|
            c_{\Delta^\varphi_-}^{\mathrm{dS-AdS}}
        \right|
    }
    \sin\left(\frac{d\pi}{2}\right).
\label{eq::LoopAdaptedCouplings}
\end{align}
Here $\mathcal{L}_{\mathrm{mixed}}^{(4)}$ collects the quartic
interactions containing a bilinear of scalar fields with opposite boundary conditions, which therefore do not contribute to gauge field two-point functions at one-loop.

\vskip 4pt
The decomposition \eqref{eq::LoopAdaptedLagrangian} separates the
interactions whose one-loop contributions cancel from those that can
produce a non-vanishing contribution. Let $I,J=\pm$ label the boundary
conditions of the two external gauge fields, and define\footnote{For $I=-$ or $J=-$ this should be understood as projected in the sense of \eqref{eq::NeumannBoundaryProjectedTwoPoint}. Note that the vanishing of the corresponding mass insertions established in \eqref{eq::DiagonalScalarCancellation} does not, however, depend on this prescription.}
\begin{equation}
    \left(T_{\Delta^A_I \Delta^A_J}\right)_{ij}(\vec{y}_1,\vec{y}_2)
    \equiv
    \int\dd^{d+1}x\sqrt{g}
    g^{\mu\nu}
    K_{\Delta^A_I,\mu i}(x,\vec{y}_1)
    K_{\Delta^A_J,\nu j}(x,\vec{y}_2).
\label{eq::GeneralMassInsertionTensor}
\end{equation}
This is precisely the two-point tensor structure generated by a local
quadratic deformation in the gauge fields $A_{\Delta^A_\pm}$, of the form
\begin{equation}
    \delta\mathcal{L}_{\mathrm{EAdS}}^{(2)}
    =
    \frac{1}{2}
    \sum_{I,J=\pm} \kappa_{\Delta^A_I}
    \delta M_{\Delta^A_I \Delta^A_J}^{2}\,
    A_{\Delta^A_I}\cdot A_{\Delta^A_J}.
\label{eq::GeneralMassInsertion}
\end{equation}
Expanding the EAdS generating functional to
first order in \eqref{eq::GeneralMassInsertion}, the corresponding
two-point insertion is
\begin{equation}
    \left(I_{\Delta^A_I \Delta^A_J}\right)_{ij}
    =
    -
    \kappa_{\Delta^A_I}\,
    \delta M_{\Delta^A_I \Delta^A_J}^{2}\,
    \left(T_{\Delta^A_I \Delta^A_J}\right)_{ij}.
\label{eq::GeneralMassInsertionContribution}
\end{equation}

\begin{figure}[t]
    \centering
    \includegraphics[width=0.8\textwidth]{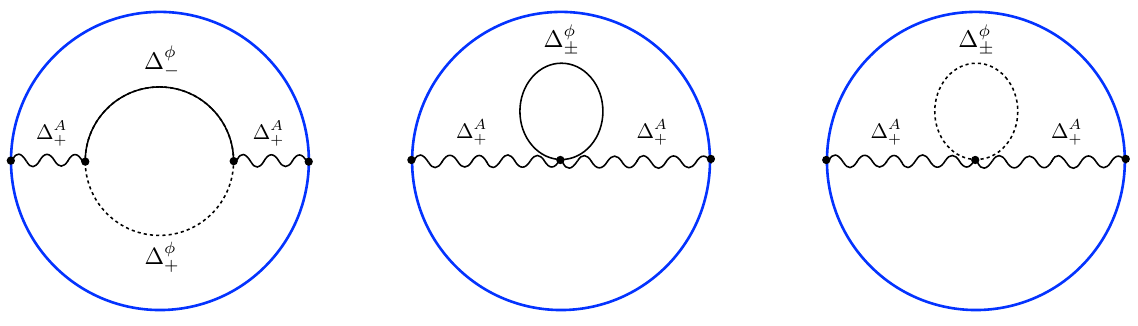}
    \caption{EAdS Witten diagrams contributing non-trivially to the $\Delta^A_+$ falloff of the one-loop dS gauge field two-point function. The blue circle denotes the EAdS boundary, wavy lines the external gauge field, and solid and dotted lines the two real scalar fields.}
    \label{fig::loops}
\end{figure}

\vskip 4pt
For each fixed $\alpha=\pm$, the cubic term inside the square brackets
in \eqref{eq::LoopAdaptedLagrangian} generates a bubble in which both
internal scalar propagators carry the boundary-condition label
$\alpha$. The corresponding contribution is
\begin{align}
   \left(I^{\mathrm{B},\,\alpha}_{\Delta^A_I \Delta^A_J}\right)_{ij}
    &=
    4\,
    \kappa_{\Delta^\varphi_\alpha}^{\,2}
    g_{I\alpha}g_{J\alpha}
    A_{\Delta^\varphi_\alpha,\Delta^\varphi_\alpha}^{(d)}
    \left(T_{\Delta^A_I \Delta^A_J}\right)_{ij}\\&=
    2\,
    \kappa_{\Delta^\varphi_\alpha}^{\,2}
    g_{I\alpha}g_{J\alpha}
    G_{\Delta^\varphi_\alpha}(x,x)
    \left(T_{\Delta^A_I \Delta^A_J}\right)_{ij},
\label{eq::DiagonalScalarBubble}
\end{align}
where in the second equality we used\footnote{As explained in section~\ref{sec::loopssQEDEAdS}, this relation applies to
both gauge field boundary conditions.}
\begin{equation}
    A_{\Delta^\varphi_\alpha,\Delta^\varphi_\alpha}^{(d)}
    =
    \frac{1}{2}
    G_{\Delta^\varphi_\alpha}^{(d)}(x,x).
\label{eq::DiagonalScalarLoopIdentity}
\end{equation}
The second term inside the square brackets in
\eqref{eq::LoopAdaptedLagrangian} generates the corresponding
tadpole contribution,
\begin{equation}
    \left( I^{\mathrm{T},\,\alpha}_{\Delta^A_I \Delta^A_J}\right)_{ij}
    =
    -2\,
    \kappa_{\Delta^\varphi_\alpha}^{\,2}
    g_{I\alpha}g_{J\alpha}
    G_{\Delta^\varphi_\alpha}(x,x)
    \left(T_{\Delta^A_I \Delta^A_J}\right)_{ij}.
\label{eq::DiagonalScalarTadpole}
\end{equation}
Comparing \eqref{eq::DiagonalScalarBubble} and
\eqref{eq::DiagonalScalarTadpole}, we obtain the same cancellation expected from flat-space analysis (see e.g. Appendix~\ref{subsec::sqedflatren})
\begin{equation}
    \left(I^{\mathrm{B},\,\alpha}_{\Delta^A_I \Delta^A_J} +
    I^{\mathrm{T},\,\alpha}_{\Delta^A_I \Delta^A_J}\right)_{ij}
    =
    0,
    \qquad
    \alpha=\pm,
    \qquad
    I,J=\pm.
\label{eq::DiagonalScalarCancellation}
\end{equation}
Thus, for each fixed scalar boundary-condition label $\alpha$, the
bubble generated by the cubic interaction cancels exactly against the
tadpole generated by its accompanying quartic interaction. The
cancellation occurs independently of
the boundary conditions imposed on the external gauge fields. Together with the absence of tadpoles generated by
$\mathcal{L}_{\mathrm{mixed}}^{(4)}$, this leaves only the terms
proportional to $g_{\mathrm{mix}}$ as the source of possible mass-like contributions at one loop. The corresponding diagrams are displayed in figure \ref{fig::loops}. 

\vskip 4pt
The cubic $g_{\mathrm{mix}}$ vertex contains one
$A_{\Delta^A_+}$ field, while the corresponding quartic vertex
contains two. Two insertions of the former, or one insertion of the
latter, can therefore contribute only to the Dirichlet-Dirichlet
two-point function. It follows that
\begin{equation}
    \delta M_{\Delta^A_-\Delta^A_-}^{2}
    =
    \delta M_{\Delta^A_+ \Delta^A_-}^{2}
    =
    \delta M_{\Delta^A_- \Delta^A_+}^{2}
    =
    0.
\end{equation}
Hence no one-loop mass insertion involving the Neumann gauge field is generated. This is consistent with gauge invariance.  As discussed in section \ref{subsec::NeumannProjection}, the Neumann branch is associated with the dynamical boundary gauge field $\tilde a_i$, whose scaling dimension $\Delta^A_-=1$ is protected by the boundary gauge symmetry and whose gauge-invariant content is carried by the field strength $f_{ij}$. A local mass-like deformation involving $A_{\Delta^A_-}$ would not be gauge invariant.

\paragraph{Dirichlet mass correction}

It remains to evaluate the terms proportional to
$g_{\mathrm{mix}}$ in \eqref{eq::LoopAdaptedLagrangian}. Both terms can contribute only to the Dirichlet-Dirichlet two-point
function: the quartic term contains two $A_{\Delta^A_+}$ fields,
whereas two insertions of the cubic term supply the two Dirichlet
external legs.

\vskip 4pt
The quartic term proportional to $g_{\mathrm{mix}}^2$ generates the
remaining tadpole contribution,
\begin{align}
    \left(I^{\mathrm{T,\,mix}}_{\Delta^A_+\Delta^A_+}\right)_{ij}
    =
    2g_{\mathrm{mix}}^2
    \bigg[
        \kappa_{\Delta^\varphi_+}^{\,2}
        G_{\Delta^\varphi_+}(x,x)+
        \kappa_{\Delta^\varphi_-}^{\,2}
        G_{\Delta^\varphi_-}(x,x)
    \bigg]
    \left(T_{\Delta^A_+\Delta^A_+}\right)_{ij}.
\label{eq::dStadpole}
\end{align}
The cubic interaction proportional to $g_{\mathrm{mix}}$ generates
the bubble in which the two internal scalar propagators carry
opposite boundary-condition labels:
\begin{equation}
    \left(I^{\mathrm{B,\,mix}}_{\Delta^A_+\Delta^A_+}\right)_{ij}
    =
    8g_{\mathrm{mix}}^2
    \kappa_{\Delta^\varphi_+}
    \kappa_{\Delta^\varphi_-}
    A_{\Delta^\varphi_+,\Delta^\varphi_-}^{(d)}
    \left(T_{\Delta^A_+\Delta^A_+}\right)_{ij}.
\label{eq::MixedScalarBubble}
\end{equation}
The mixed scalar loop coefficient, evaluated in
section~\ref{sec::scalarloops}, is
\begin{equation}
    A_{\Delta^\varphi_+,\Delta^\varphi_-}^{(d)}
    =
    \frac{1}{2d}
    \left[
        \Delta^\varphi_+
        G_{\Delta^\varphi_+}^{(d)}(x,x)
        +
        \Delta^\varphi_-
        G_{\Delta^\varphi_-}^{(d)}(x,x)
    \right].
\label{eq::Aplusminus}
\end{equation}
Consequently,
\begin{align}
    \left(I^{\mathrm{B,\,mix}}_{\Delta^A_+\Delta^A_+}\right)_{ij}
    =
    \frac{
        4g_{\mathrm{mix}}^2
        \kappa_{\Delta^\varphi_+}
        \kappa_{\Delta^\varphi_-}
    }{d}
    \bigg[
        &\Delta^\varphi_+
        G_{\Delta^\varphi_+}(x,x)+
        \Delta^\varphi_-
        G_{\Delta^\varphi_-}(x,x)
    \bigg]
   \left(T_{\Delta^A_+\Delta^A_+}\right)_{ij}.
\label{eq::dSbubble}
\end{align}
Adding \eqref{eq::dStadpole} and \eqref{eq::dSbubble} therefore gives
\begin{equation}
    \left(I^{\mathrm{1-loop}}_{\Delta^A_+\Delta^A_+}\right)_{ij}
    =
    \frac{
        2g_{\mathrm{mix}}^2
        \kappa_{\Delta^\varphi_+}
        \kappa_{\Delta^\varphi_-}
    }{d}
    \left(
        \Delta^\varphi_+
        -
        \Delta^\varphi_-
    \right)
    \left[
        G_{\Delta^\varphi_+}(x,x)
        -
        G_{\Delta^\varphi_-}(x,x)
    \right]
    \left(T_{\Delta^A_+\Delta^A_+}\right)_{ij}.
\label{eq::DirichletLoopInGmix}
\end{equation}
Comparing \eqref{eq::DirichletLoopInGmix} with
\eqref{eq::GeneralMassInsertionContribution}, and combining this
result with the cancellation
\eqref{eq::DiagonalScalarCancellation}, the one-loop mass matrix in
the basis $\{A_{\Delta^A_+},A_{\Delta^A_-}\}$ takes the form
\begin{equation}
    \delta M_{\Delta^A_I\Delta^A_J}^{2}
    =
    \begin{pmatrix}
        \delta M_{\Delta^A_+\Delta^A_+}^{2} & 0
        \\
        0 & 0
    \end{pmatrix}_{\!IJ},
\label{eq::oneLoopGaugeMassMatrix}
\end{equation}
where
\begin{equation}
    \delta M_{\Delta^A_+\Delta^A_+}^{2}
    =
    -
    \frac{
        2g_{\mathrm{mix}}^2
        \kappa_{\Delta^A_+}
        \kappa_{\Delta^\varphi_+}
        \kappa_{\Delta^\varphi_-}
    }{d}
    \left(
        \Delta^\varphi_+
        -
        \Delta^\varphi_-
    \right)
    \left[
        G_{\Delta^\varphi_+}(x,x)
        -
        G_{\Delta^\varphi_-}(x,x)
    \right].
\label{eq::inducedVectorMass}
\end{equation}
Thus only the Dirichlet gauge field acquires a mass correction at one loop, while the Neumann gauge field remains massless. As reviewed in the next section, this Dirichlet mass correction was previously computed in \cite{Sleight:2025dmt} by extracting the anomalous
dimension of the boundary current from the deformed Ward identities and the associated multiplet recombination.\footnote{For scalar QED with a conformally coupled scalar in AdS$_4$, the corresponding photon mass was obtained earlier in appendix~E of \cite{Sadekov:2023ivd}. For scalars of generic mass, an analytic expression valid for general $d$ was derived in \cite{Sleight:2025dmt} using a boundary multiplet recombination argument. Its $d=3$ specialisation was subsequently verified in \cite{Ankur:2026ylr} by numerically evaluating the finite part of the
dimensionally regulated bubble and tadpole contributions in AdS. Here we give a diagrammatic derivation for general $d$, based on a fully analytic evaluation of the bubble and tadpole contributions in (A)dS.}

\section{Interpretation}
\label{sec::interpretation}

In this section we discuss the physical implications of the one-loop mass matrix \eqref{eq::oneLoopGaugeMassMatrix}. It gives a mass
correction to the Dirichlet field $A_{\Delta^A_+}$, while the Neumann field $A_{\Delta^A_-}$ remains massless. In EAdS, the Dirichlet correction realises Schwinger mass generation, which is discussed in section \ref{subsec::schweads}. In dS, it modifies the late-time electric response without changing the leading magnetic falloff, which we discuss in section \ref{subsec::electscreen}. Since the corrected dimensions no longer form a shadow pair, the effect cannot arise from a local Proca deformation in dS. Its interpretation within the Schwinger-Keldysh effective action is discussed in section \ref{subsec::SKinterp}. Finally, in section \ref{sec::retarded} we show that the in-in branch sum produces a retarded kernel and relate its discontinuity to the two-particle cut
of the scalar bubble. This gives a dispersive interpretation of the anomalous dimension.

\subsection{Schwinger mass generation in EAdS}
\label{subsec::schweads}

From the EAdS perspective, the mass shift
\eqref{eq::inducedVectorMass} for the Dirichlet gauge field $A_{\Delta^A_+}$ is generated dynamically by the scalar vacuum polarisation. It therefore realises a Schwinger-type mechanism for mass generation \cite{Schwinger:1962tn,Schwinger:1962tp}, in which the longitudinal mode is supplied by a two-scalar state rather than by a classical Higgs condensate. Closely related one-loop mechanisms induced by AdS boundary conditions were studied in
\cite{Rattazzi:2009ux,Porrati:2009dy}.

\vskip 4pt
From the boundary perspective, the same mechanism is encoded in an anomalous dimension for the spin-one current $J_i$ and its multiplet recombination with the shadow-pair scalar composite identified in \cite{Sleight:2025dmt}. In the following we recover the anomalous dimension from the induced EAdS mass and then return to its interpretation in terms of multiplet recombination.

\vskip 4pt
In the free theory, the Dirichlet gauge field
$A_{\Delta^A_+}$ is massless, with
\begin{equation}\label{freedimspin1}
    \left.\Delta^A_+\right|_{e=0}=d-1.
\end{equation}
The induced mass \eqref{eq::inducedVectorMass} shifts the scaling dimension of the boundary
spin-one operator $J_i$ sourced by the Dirichlet field
$A_{\Delta^A_+}$ away from the conserved-current value \eqref{freedimspin1}:
\begin{equation}
    \Delta^A_+=d-1+\gamma_J,
    \qquad
    \gamma_J=O(e^2).
\label{eq::gammaJDefinition}
\end{equation}
The mass-dimension relation for a spin-one field in
EAdS$_{d+1}$ gives
\begin{equation}
    M_{\Delta^A_+\Delta^A_+}^2
    =
    \left(\Delta^A_+-1\right)
    \left(\Delta^A_+-d+1\right).
\label{eq::vectorMassDimension}
\end{equation}
Expanding around the massless value $\Delta^A_+=d-1$ gives, for
$d\neq2$,
\begin{equation}
    \delta M_{\Delta^A_+\Delta^A_+}^2
    =
    (d-2)\gamma_J
    +
    O(e^4).
\label{eq::massGammaRelation}
\end{equation}
Combining this relation with \eqref{eq::inducedVectorMass}, we find 
\begin{equation}
    \gamma_J
    =-
    \frac{
        2g_{\mathrm{mix}}^2
        \kappa_{\Delta^A_+}
\kappa_{\Delta^\varphi_+}
\kappa_{\Delta^\varphi_-}
    }{
        d(d-2)
    }
    \left(
        \Delta^\varphi_+
        -
        \Delta^\varphi_-
    \right)
    \left[
        G_{\Delta^\varphi_+}(x,x)
        -
        G_{\Delta^\varphi_-}(x,x)
    \right].
\label{eq::gammaJGeneral}
\end{equation}
Parametrising the free theory scaling dimensions for the scalar fields as
\begin{equation}
    \Delta^\varphi_\pm
    =
    \frac d2\pm i\mu,
\end{equation}
and substituting the coincident propagators \eqref{eq:tadpole}, we
obtain
\begin{equation}
\gamma_J
=
\frac{
e^2
\Gamma(1+i\mu)
\Gamma(1-i\mu)
\Gamma\left(\frac d2+i\mu\right)
\Gamma\left(\frac d2-i\mu\right)
}{
2\pi^{\frac d2+1}
\Gamma\left(2-\frac d2\right)
\Gamma(d+1)
}.
\label{eq::gammaJNu}
\end{equation}
This agrees with the anomalous dimension obtained independently
in \cite{Sleight:2025dmt} via a multiplet recombination argument, providing a non-trivial check of the explicit one-loop computation.\footnote{The massless limit $\mu\to i\frac d2$ is the exceptional point
where $\gamma_J$ diverges and higher loops of the light scalar must be
resummed. At fixed order the photon acquires a secularly growing mass squared
$\propto e^2\langle\varphi^2\rangle$ \cite{Prokopec:2002jn}. The
stochastic resummation of the leading infrared logarithms equilibrates it to a finite, coupling-independent value of order $H^2$
\cite{Prokopec:2007ak}, where $H$ is the Hubble scale.}

\vskip 4pt
The Schwinger mass-generation mechanism has a simple
representation-theoretic description: the shadow-pair scalar
composite supplies the longitudinal mode required to complete the
long spin-one multiplet. For non-zero $\gamma_J$, the current $J_i$
has dimension $d-1+\gamma_J$ and generates the long spin-one representation
$D(d-1+\gamma_J,1)$.\footnote{$D(\Delta,s)$ denotes the
$SO(d+1,1)$ representation, or equivalently the conformal multiplet
generated by a boundary primary of scaling dimension $\Delta$ and
spin $s$.} As $\gamma_J\to0$, this representation decomposes as
\begin{equation}
    \lim_{\gamma_J\to0}
    D(d-1+\gamma_J,1)
    =
    D(d-1,1)
    \oplus
    D(d,0).
\label{eq::vectorMultipletDecomposition}
\end{equation}
Here $D(d-1,1)$ is the short representation generated by the
conserved current, while $D(d,0)$ has the quantum numbers of its null
descendant $\partial^iJ_i$.

\vskip 4pt
In the present theory, the required $D(d,0)$ state is supplied by a composite of the shadow pair of boundary operators ${\cal O}_{\Delta^\varphi_+}$ and ${\cal O}_{\Delta^\varphi_-}$ associated with the two scalar
boundary conditions. Since their scaling dimensions obey the shadow relation $\Delta^\varphi_++\Delta^\varphi_-=d$, the following scalar composites have dimension $d$ at the free point:
\begin{equation}
    {\cal O}_{\Delta^\varphi_+}
    {\cal O}^{\dagger}_{\Delta^\varphi_-},
    \qquad
    {\cal O}^{\dagger}_{\Delta^\varphi_+}
    {\cal O}_{\Delta^\varphi_-}.
\end{equation}
For non-zero coupling $e$, the free-theory null descendant $\partial^iJ_i$ mixes with the dimension-$d$ shadow-pair
composites, and the conservation equation is deformed to
\begin{equation}
    \partial^iJ_i
    \sim
    e
    \left(
        {\cal O}^{\dagger}_{\Delta^\varphi_+}
        {\cal O}_{\Delta^\varphi_-}
        -
        {\cal O}_{\Delta^\varphi_+}
        {\cal O}^{\dagger}_{\Delta^\varphi_-}
    \right)
    +
    O(e^2).
\label{eq::currentRecombination}
\end{equation}
This mixing recombines the short current multiplet with the
dimension-$d$ multiplet of the shadow-pair composite, producing the long spin-one multiplet $D(d-1+\gamma_J,1)$.

\vskip 4pt
The one-loop calculation presented here in the EAdS reformulation gives a complementary perspective. It recovers the same anomalous dimension from the induced mass of the Dirichlet gauge field $A_{\Delta^A_+}$ in EAdS, and shows how the scalar state identified above appears in the vacuum polarisation. As shown in the previous section, for each $\alpha=\pm$, the bubble in which both scalars have the same boundary condition $\Delta^\varphi_\alpha$ cancels against the corresponding tadpole. By contrast, the surviving bubble contains one scalar of dimension $\Delta^\varphi_+$ and one of dimension
$\Delta^\varphi_-$ in the loop, and hence propagates the two-particle state of dimension $\Delta^\varphi_++\Delta^\varphi_-=d$. This is precisely the scalar state that supplies the longitudinal mode in the multiplet recombination. See figure \ref{fig::loops}. Together with the tadpole proportional to $g_{\mathrm{mix}}^2$, the
mixed bubble produces the one-loop mass shift \eqref{eq::inducedVectorMass}. Since the mixed cubic coupling involves only $A_{\Delta^A_+}$, this
mechanism does not generate a mass insertion involving
$A_{\Delta^A_-}$. \vskip 4pt

\vskip 4pt
\paragraph{Vanishing for even boundary dimensions.} Note that for even $d\geq4$ the anomalous dimension \eqref{eq::gammaJNu} vanishes, since
$1/\Gamma\left(2-\frac{d}{2}\right)$ has zeros at
$d=4,6,\ldots\,$. This zero has a direct Schwinger-Keldysh
interpretation. In the EAdS reformulation, the sine factor \eqref{general EAdS Lagrangian scalar loops} multiplying
each vertex arises from combining the contributions from the two
branches of the in-in contour. As discussed in \cite{Abhishek:2025oki}, at a
zero of the corresponding sine factor the non-local part of the late-time correlator vanishes at separated points, provided the
associated EAdS Witten diagram is infrared finite.

\vskip 4pt
Indeed, for the mixed-current cubic vertex in \eqref{eq::LoopAdaptedLagrangian} responsible for generating the anomalous dimension, we have
\begin{equation}
    g_{\mathrm{mix}}
    \equiv
    2e
    \sqrt{
        \left|
            c_{\Delta^A_+}^{\mathrm{dS-AdS}}
        \right|
    }
    \sqrt{
        \left|
            c_{\Delta^\varphi_+}^{\mathrm{dS-AdS}}
        \right|
        \left|
            c_{\Delta^\varphi_-}^{\mathrm{dS-AdS}}
        \right|
    }
    \sin\left(\frac{d\pi}{2}\right).
\end{equation}
For even $d$, the zero of the sine factor is
partially compensated by the Dirichlet gauge field normalisation, where\footnote{The singularity of this coefficient reflects the degeneracy of the two EAdS solutions used in the dS-EAdS decomposition when $\Delta^A_+-\Delta^A_-=d-2$ is an even integer \cite{Abhishek:2025oki}.}
\begin{equation}
    \left|c^{\mathrm{dS-AdS}}_{\Delta^A_\pm}\right|
    =
    \frac{1}
    {2\left|\sin\!\left(\frac{\pi d}{2}\right)\right|}.
\end{equation}
Consequently,
\begin{equation}
    g_{\mathrm{mix}}
    \propto
    \frac{\sin\!\left(\frac{\pi d}{2}\right)}
    {\sqrt{\left|\sin\!\left(\frac{\pi d}{2}\right)\right|}},
\label{eq::gmixHalfSine}
\end{equation}
and hence in the expression \eqref{eq::gammaJNu} for the anomalous dimension we have
\begin{equation}
    \kappa_{\Delta^A_+}g_{\mathrm{mix}}^2
    \propto
    \sin\!\left(\frac{\pi d}{2}\right).
\end{equation}

\vskip 4pt
The case $d=2$ is special, where the simple pole at $d=2$ in \eqref{eq::gammaJGeneral} compensates the sine zero. For even $d\geq4$, by contrast, the EAdS diagram is infrared finite and one uncancelled power of $\sin\!\left(\frac{\pi d}{2}\right)$ remains. The one-loop Schwinger mass therefore vanishes only for even $d\geq4$. These zeros are also the points at which $\gamma_J$ changes sign, separating regions of electric screening and anti-screening.

\subsection{Late-time electric screening in dS} 
\label{subsec::electscreen}

The anomalous dimension $\gamma_J$ changes the late-time falloff of the $\Delta^A_+$ branch of the dS gauge field. We now determine how this affects the electric and magnetic components of the field strength, and how the sign of $\gamma_J$ distinguishes electric screening from anti-screening.

\vskip 4pt
In the EAdS reformulation, the fields
$A_{\Delta^A_-}$ and $A_{\Delta^A_+}$ encode the two independent
late-time branches of the dS gauge field. At the free point,
\begin{equation}
    \Delta^A_-=1,
    \qquad
    \Delta^A_+=d-1.
\end{equation}
The operator-valued late-time expansion of the spatial components
therefore takes the schematic form
\begin{equation}
    A_i(\eta \to 0^-,\vec{x})
    \sim
    \tilde a_i(\vec{x})
    +
    (-\eta)^{d-2}J_i(\vec{x}),
\label{eq::MaxwellFalloffs}
\end{equation}
where $\tilde a_i$ is the boundary gauge field operator associated
with the $\Delta^A_-=1$ branch, while $J_i$ is the conserved current
associated with the $\Delta^A_+=d-1$ branch. 

\vskip 4pt
To identify which field-strength components are affected by a shift
of the $\Delta^A_+$ dimension, it is useful to separate the
contributions carried by the two independent boundary data
$\tilde a_i$ and $J_i$. At the free point,
\begin{align}
    F_{ij}
    &\sim
    \partial_i\tilde a_j-\partial_j\tilde a_i
    +
    (-\eta)^{d-2}
    \left(
        \partial_iJ_j-\partial_jJ_i
    \right),
\label{eq::magneticFalloffs}
\\
    \left.F_{\eta i}\right|_{\Delta^A_+}
    &\sim
    -(d-2)(-\eta)^{d-3}J_i.
\label{eq::electricFalloffFree}
\end{align}
Here $\left.F_{\eta i}\right|_{\Delta^A_+}$ denotes the contribution carrying
the independent boundary data $J_i$. The electric field also receives
contributions from descendants of $\tilde a_i$, whose coefficients are
fixed by the leading data $\tilde a_i$. For $d>2$, the protected
$\Delta^A_-=1$ branch controls the leading magnetic field, whereas the
$\Delta^A_+=d-1$ branch controls the independent electric contribution
and contributes only subleadingly to the magnetic field.\footnote{For nonperturbative bounds on the late-time falloffs of locally measured electric and magnetic correlators in unitary parity-preserving QFTs on dS$_4$, see \cite{Loparco:2025azm}.}

\begin{figure}[t]
    \centering
    \includegraphics[width=\textwidth]{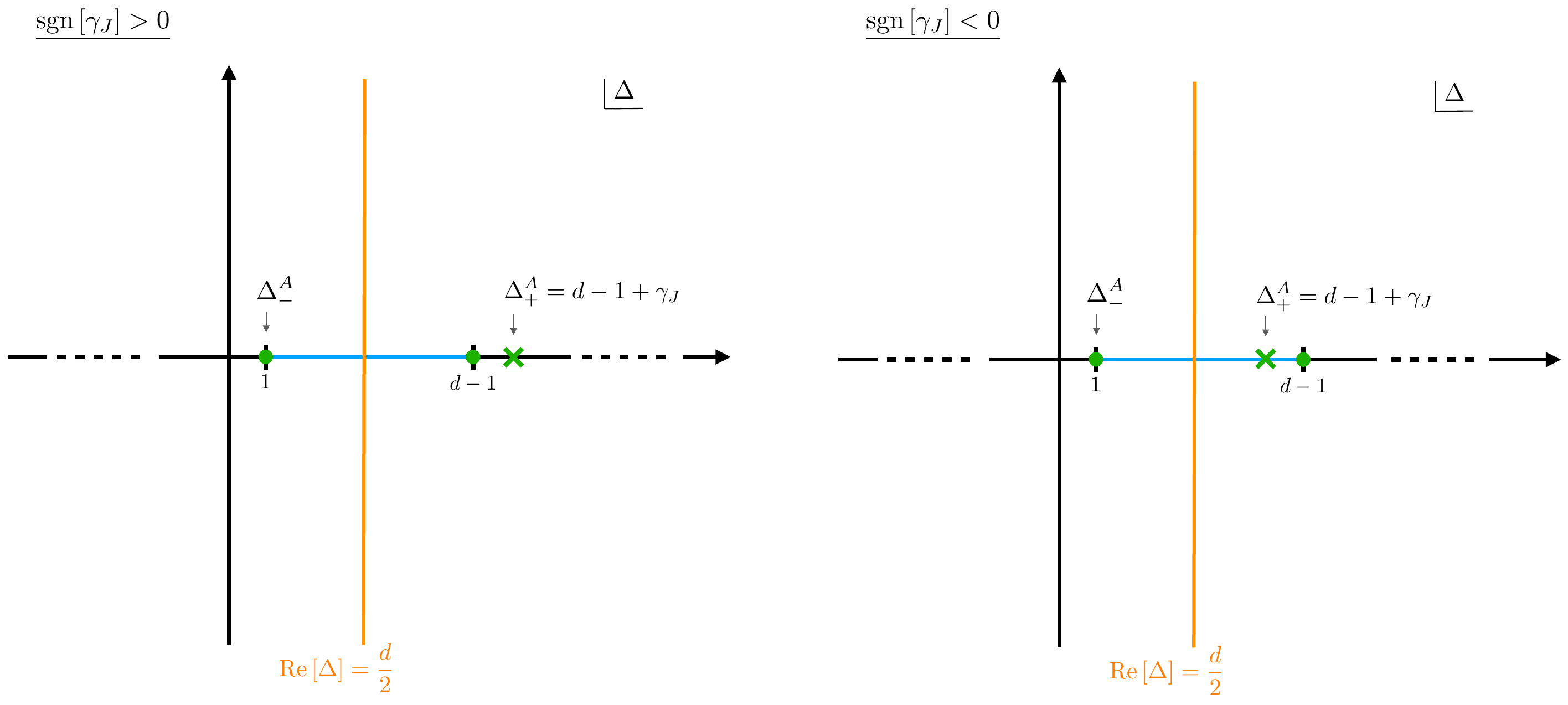}
    \caption{One-loop deformation $\gamma_{J}$ of the Dirichlet gauge field scaling dimension $\Delta^A_+$ away from the massless value. For $\gamma_J<0$ (RHS), the dimension moves into the spin-one complementary series, while for $\gamma_J>0$ (LHS) it moves outside the unitary range. The Neumann dimension $\Delta^A_-=1$, associated with the boundary gauge field, is protected by gauge symmetry. Cf. the unitary classification in figure \ref{fig::UIRs}.}
    \label{fig::anom}
\end{figure}

\vskip 4pt
The loop-induced mass shifts the scaling dimension of the operator $J_i$ to
\begin{equation}
    \Delta^A_+
    =
    d-1+\gamma_J.
\end{equation}
Accordingly, the contribution of the $\Delta^A_+$ branch to the
late-time gauge field becomes
\begin{equation}
    \left.A_i(\eta,\vec{x})\right|_{\Delta^A_+}
    \sim
    (-\eta)^{\Delta^A_+-1}J_i(\vec{x})
    =
    (-\eta)^{d-2+\gamma_J}J_i(\vec{x}).
\label{eq::anomalousGaugeFalloffdS}
\end{equation}
The same anomalous scaling appears directly in the equal-time two-point function. Conformal symmetry fixes the two-point function of the spin-one primary $J_i$ to be 
\begin{equation} \left\langle J_i(\vec y_1)J_j(\vec y_2)\right\rangle = C_J\, \frac{I_{ij}(\vec y_{12})} {\left(\vec y_{12}^{\,2}\right)^{\Delta^A_+}}, \qquad \vec y_{12}=\vec y_1-\vec y_2, \end{equation} 
where $I_{ij}$ is the inversion tensor \eqref{eq::InversionTensor}. It follows from \eqref{eq::anomalousGaugeFalloffdS} that the contribution of the $\Delta^A_+$ mode to the late-time gauge-field two-point function is 
\begin{equation}
    \left.
    \left\langle
        A_i(\eta,\vec y_1)A_j(\eta,\vec y_2)
    \right\rangle
    \right|_{\Delta^A_+}
    \sim
    (-\eta)^{2(d-2)}\,C_J\,
    \frac{I_{ij}(\vec y_{12})}
    {\left(\vec y_{12}^{\,2}\right)^{d-1}}
    \left(
        \frac{\eta^2}{\vec y_{12}^{\,2}}
    \right)^{\gamma_J}.
\end{equation}
The dependence controlled by the anomalous dimension is contained in the final factor. To one-loop order, 
\begin{equation} \left( \frac{\eta^2}{\vec y_{12}^{\,2}} \right)^{\gamma_J} = 1+ \gamma_J\log \frac{\eta^2}{\vec y_{12}^{\,2}} + O(e^4). 
\end{equation} 
In our calculation, the spatial logarithm arises upon renormalising the bulk integral
$(T_{\Delta_+^A\Delta_+^A})_{ij}$, defined in
\eqref{eq::GeneralMassInsertionTensor}, which has a logarithmic
divergence at the AdS boundary for $\Delta_+^A=d-1$ and $d>2$;
see also appendix~A.2 of \cite{Giombi:2017hpr} and eq. (15) of \cite{Banados:2022nhj} which includes the leading coincident-point contribution.
At fixed time, the logarithm in the two-point function gives
the first-order correction to the spatial scaling.
At fixed comoving separation, it becomes large as $\eta\to0^-$
and fixed-order perturbation theory eventually breaks down.
The shifted power resums the corresponding leading secular
logarithms.

\vskip 4pt
The contribution of the $\Delta^A_+$ branch to both field-strength
components acquires the additional late-time factor
$(-\eta)^{\gamma_J}$:
\begin{align}
    \left.F_{\eta i}\right|_{\Delta^A_+}
    &\sim
    -(d-2+\gamma_J)
    (-\eta)^{d-3+\gamma_J}J_i,
\label{eq::anomalousElectricFalloffdS}
\\
    \left.F_{ij}\right|_{\Delta^A_+}
    &\sim
    (-\eta)^{d-2+\gamma_J}
    \left(
        \partial_iJ_j-\partial_jJ_i
    \right).
\label{eq::anomalousMagneticSubleadingFalloffdS}
\end{align}
Thus, up to the finite change in the prefactor of $F_{\eta i}$, both
contributions carried by this branch are multiplied relative to their
free-theory behaviour by $(-\eta)^{\gamma_J}$. For the magnetic
components, this modifies only the subleading contribution
proportional to $J_i$: the leading term
$\partial_i\tilde a_j-\partial_j\tilde a_i$, carried by the protected
$\Delta^A_-$ branch, remains unchanged. By contrast, the contribution
to the electric field carrying the independent boundary data $J_i$ is
entirely supplied by the $\Delta^A_+$ branch. Whenever $\gamma_J>0$,
this contribution decays more rapidly than in the free theory. We
refer to this enhanced late-time decay as the electric screening
induced by the Schwinger mass. For $\gamma_J<0$, the same factor
enhances the late-time electric contribution, corresponding to
electric anti-screening.

\vskip 4pt
For principal-series scalars ($\mu\in\mathbb R$) the sign of $\gamma_J$ is controlled by
$\Gamma(2-d/2)$, which is directly related to
the sign of the kinetic term of the Dirichlet gauge field. Indeed, 
\begin{equation}
    \operatorname{sgn}\left[
        \frac{1}{\Gamma\left(2-\frac d2\right)}
    \right]
    =
    -
    \operatorname{sgn}\left[
        c^{\mathrm{dS-AdS}}_{\Delta^A_+}
    \right].
\end{equation}
Thus, for principal-series scalars, electric screening occurs precisely when the Dirichlet gauge field carries the negative-sign kinetic term in the EAdS reformulation. In particular, this is the case in
dS$_4$.

\vskip 4pt
For perturbatively small $\gamma_J<0$, the corrected dimension $\Delta^A_+ = d-1+\gamma_J$ moves into the spin-one complementary-series range and is therefore
compatible with a unitary dS representation. For $\gamma_J>0$, it instead moves beyond the endpoint $\Delta^A_+=d-1$ and no longer labels a unitary dS representation. See figure~\ref{fig::anom}. 

\vskip 4pt
As shown in section~\ref{sec::dSloop}, no one-loop mass insertion
involving the Neumann field $A_{\Delta^A_-}$ is generated. Consequently,
$\Delta^A_-=1$ remains unshifted at this order, and the leading
magnetic contribution
\begin{equation}
    F_{ij}
    \sim
    \partial_i\tilde{a}_j
    -
    \partial_j\tilde{a}_i
\end{equation}
retains its free-theory falloff. The scalar loop therefore modifies
the electric response encoded by the $A_{\Delta^A_+}$ branch, while
leaving the leading magnetic sector encoded by
$A_{\Delta^A_-}$ unchanged.

\vskip 4pt
As discussed in more detail in the next section, the breaking of the shadow symmetry and the fact that $\Delta^A_+$ may lie outside the unitary range are not in tension with unitarity of the underlying theory. The deformed dimension $\Delta^A_+=d-1+\gamma_J$ should be read as a feature of the mass-like projection of the one-loop effective action, not as the dimension of a stable single-particle state---in analogy with flat space resonances.

\subsection{Schwinger-Keldysh interpretation and resonances}
\label{subsec::SKinterp}

The one-loop result derived above is asymmetric: it generates a mass
insertion for the Dirichlet EAdS gauge field $A_{\Delta^A_+}$, but no
insertion involving the Neumann field $A_{\Delta^A_-}$. This has a
direct analogue in the flat-space description of a resonance.\footnote{Particle decay in dS scalar theories, including its kinematics and associated lifetimes, was analysed in \cite{Bros:2010rku}. In the resummed scalar two-point functions of \cite{Marolf:2010nz}, the two principal-series poles acquire a common real displacement and hence no longer obey the shadow relation; this displacement was interpreted as the opening of a particle-decay channel.} The
quadratic part of the one-loop effective action takes the schematic
form
\begin{equation}
\Gamma_{\mathrm{flat}}^{(2)}(p^2)
=
\frac{1}{2}
\left[
p^2+m^2+e^2\Pi(p^2)
\right].
\end{equation}
Near its dressed pole at $p^2=-s_\star$ it can be expanded as
\begin{equation}
\Gamma_{\mathrm{flat}}^{(2)}(p^2)
=
\frac{1}{2}
\left[
Z_\star^{-1}(p^2+s_\star)
+
O\!\left((p^2+s_\star)^2\right)
\right],
\end{equation}
where $s_\star$ is the complex pole position, its real part is the squared mass
$M_\star^{2}$ of the resonance, and its imaginary part is fixed by the width. Retaining only the leading term in this expansion gives a local non-Hermitian description of the resonant sector. It correctly captures the pole position and
width, while omitting the non-resonant and multiparticle contributions
required for the complete unitary theory.

\vskip 4pt
We shall see that the asymmetric EAdS mass correction produces the
corresponding phenomenon when recast in the dS Schwinger-Keldysh
effective action. 

\vskip 4pt
Recall that
\begin{equation}
    \delta M_{\Delta^A_+\Delta^A_+}^{2}
    =
    (d-2)\gamma_J
    +
    O(e^4),
    \qquad
    \gamma_J=O(e^2),
\label{eq::DirichletMassShiftGammaJ}
\end{equation}
so that the corresponding quadratic insertion in EAdS is
\begin{equation}
    \delta\mathcal{L}_{\mathrm{EAdS}}^{(2)}
    =
    \frac{1}{2}
    \kappa_{\Delta^A_+}
    \delta M_{\Delta^A_+\Delta^A_+}^{2}\,
    A_{\Delta^A_+}\cdot A_{\Delta^A_+}.
\label{eq::DirichletMassInsertionEAdS}
\end{equation}
No mass insertion involving the Neumann field
$A_{\Delta^A_-}$ is generated. Therefore, at this order,
\begin{equation}
    \Delta^A_+
    =
    d-1+\gamma_J,
    \qquad
    \Delta^A_-=1.
\label{eq::CorrectedGaugeFieldDimensions}
\end{equation}
The resulting dimensions no longer form a shadow pair. This is
already an indication that \eqref{eq::DirichletMassInsertionEAdS}
cannot arise from an ordinary local Proca mass deformation in dS,
which would deform both late-time behaviours of the same local
equation of motion while preserving their shadow symmetry.

\vskip 4pt
To make this correspondence precise, we introduce the
Schwinger-Keldysh effective action for the gauge field, obtained by
integrating out the charged scalar fields in the S-K path integral
\eqref{SK generating scalar QED} while keeping $A_+$ and $A_-$ fixed:\footnote{The Bunch-Davies state at the initial end of the contour is imposed
by the usual $i\epsilon$ prescription, while the identification of
$\phi_+$ and $\phi_-$ at $\eta=\eta_0$ implements the sum over the
common final scalar configuration.}
\begin{multline}
    \exp\left\{
        i\Gamma_{\mathrm{SK}}
        \left[
            A_+,A_-
        \right]
    \right\}
    \equiv{}
    \int_{\phi_+(\eta_0,\vec{x})=\phi_-(\eta_0,\vec{x})}
    \mathcal{D}\phi_+
    \mathcal{D}\phi_- \\
    \times  \exp\bigg\{
        iS_{\mathrm{dS}}
        \left[
            A_+,\phi_+
        \right]
        -
        iS_{\mathrm{dS}}
        \left[
            A_-,\phi_-
        \right]
    \bigg\}.
\label{eq::SKEffectiveActionDefinition}
\end{multline}
The integration over $\phi_\pm$ in \eqref{eq::SKEffectiveActionDefinition} is the field-theoretic Feynman-Vernon influence-functional construction
\cite{Feynman:1963fq}. Since the scalar action is quadratic, this functional integral is
Gaussian and yields the one-loop scalar determinant, exact to all
orders in the background fields $A_+$ and $A_-$. Its part quadratic
in the gauge fields is precisely the sum of the scalar bubble and
tadpole contributions computed in the previous sections using the EAdS reformulation. Expanding the effective action in powers of the gauge fields,
\begin{equation}
    \Gamma_{\mathrm{SK}}
    = \Gamma_{\mathrm{SK}}^{(2)}
    +
    O(A^3),
    \qquad
    \Gamma_{\mathrm{SK}}^{(n)}
    =
    O(A^n),
\label{eq::SKEffectiveActionExpansion}
\end{equation}
where the superscript $(n)$ counts powers of the gauge field. The quantity $\Gamma_{\mathrm{SK}}^{(2)}$ denotes the part quadratic in the gauge fields and in general is bilocal:
\begin{equation}
    \Gamma_{\mathrm{SK}}^{(2)}
    =
    \frac{1}{2}
    \sum_{\epsilon,\epsilon^\prime=\pm}
    \int {\rm d}^{d+1}x \sqrt{-g}
    \int {\rm d}^{d+1}x^\prime \sqrt{-g^\prime}\,
    (A_\epsilon)_\mu(x)
    \Pi_{\epsilon \epsilon^\prime}^{\mu\nu}(x,x';e)
    (A_{\epsilon^\prime})_\nu(x').
\label{eq::GeneralQuadraticSKEffectiveAction}
\end{equation}
Here $\Pi_{\epsilon \epsilon^\prime}^{\mu\nu}(x,x';e)$ denotes the full inverse
two-point kernel of the scalar-integrated effective action. Since the scalar action is
quadratic, this is one-loop exact:
\begin{equation}
    \Pi_{\epsilon \epsilon^\prime}^{\mu\nu}(x,x';e)
    =
    \Pi_{\epsilon \epsilon^\prime}^{[0]\,\mu\nu}(x,x')
    +
    e^2\Pi_{\epsilon \epsilon^\prime}^{[1]\,\mu\nu}(x,x'),
\label{eq::SKKernelExpansion}
\end{equation}
where the square brackets label the order in the coupling. The kernel
$\Pi_{\epsilon \epsilon^\prime}^{[0]\,\mu\nu}$ is the free inverse propagator,
whereas
$\Pi_{\epsilon \epsilon^\prime}^{[1]\,\mu\nu}$ is the generally bilocal
one-loop scalar vacuum-polarisation kernel. The calculation of the
preceding sections determines its matrix elements between the free
external modes $A_{\Delta^A_\pm}$ and, in particular, the coefficient
of the mass-insertion tensor structure. In the following we retain
only this projected information, rather than analytically continuing
the complete bilocal kernel back to dS. 

\vskip 4pt
At fixed order $e^2$, we project the scalar-induced bilocal kernel
onto the two free Maxwell modes,
\begin{equation}
    \Delta^A_-
    =
    1,
    \qquad
    \Delta^A_+
    =
    d-1.
\end{equation}
The coefficients of the mass-insertion tensor structure defined in \eqref{eq::GeneralMassInsertionTensor} form the matrix
$\delta M_{\Delta^A_I\Delta^A_J}^2$, with $I,J=\pm$. The corresponding projection of the quadratic
Schwinger-Keldysh effective action can be parametrised as\footnote{Equation~\eqref{eq::MassLikeSKAnsatz} is not the most general
quadratic Schwinger-Keldysh effective action, which contains the
bilocal kernels in \eqref{eq::GeneralQuadraticSKEffectiveAction}.
It is the minimal ansatz required to reproduce
\eqref{eq::DirichletMassInsertionEAdS}.}
\begin{multline}
    i\Gamma_{\mathrm{SK}}^{(2)}\left[A_+,A_-\right]
    \supset
    \int_{-\infty}^{\eta_0}
    \frac{{\rm d}\eta\,{\rm d}^d\vec{x}}{(-\eta)^{d+1}}
    \bigg[
        \,i f_3(e^2)
        \left(
            A_+\cdot A_+
            -
            A_-\cdot A_-
        \right)\\
        +f_1(e^2)
        \left(
            A_+\cdot A_+
            +
            A_-\cdot A_-
        \right)
        + f_2(e^2)
        A_+\cdot A_-
    \bigg].
\label{eq::MassLikeSKAnsatz}
\end{multline}
The term proportional to $f_3$ has the standard difference-of-actions form. The term proportional to $f_1$ contains the two same-branch quadratic terms with the same sign, whereas the
term proportional to $f_2$ couples the two branches. The last term denotes a bilinear coupling between the two branches of the
in-in contour at equal values of $|\eta|$.\footnote{More explicitly,
the cross-branch bilinear is defined by
\begin{equation}
    \int_{-\infty}^{\eta_0}
    \frac{{\rm d}\eta_+}{(-\eta_+)^{d+1}}
    \int_{-\infty}^{\eta_0}
    \frac{{\rm d}\eta_-}{(-\eta_-)^{d+1}}
    \int {\rm d}^d\vec{x}\,
    \left(
        |\eta_+||\eta_-|
    \right)^{\frac{d+1}{2}}
    \delta\left(
        |\eta_+|-|\eta_-|
    \right) A_+(\eta_+,\vec{x})
    \cdot
    A_-(\eta_-,\vec{x}).
\end{equation}
It is local in the spatial coordinates and identifies points on the
two branches with equal $|\eta|$.}
The coefficients $f_i(e^2)$ are real and, since the gauge field is
massless in the free theory, their mass-like contributions begin at
one loop:
\begin{equation}
    f_i(e^2)=O(e^2),
    \qquad
    i=1,2,3.
\label{eq::SKCoefficientOrder}
\end{equation}
In the EAdS reformulation these correspond to the following terms in the quadratic Lagrangian
\begin{multline}
    \delta\mathcal{L}_{\mathrm{EAdS}}^{(2)}
    =
    -
    \sum_{I=\pm}
    \bigg[
        \kappa_{\Delta^A_I}f_3
        +
        \left|
            c_{\Delta^A_I}^{\mathrm{dS-AdS}}
        \right|
        \left(
            f_2
            +
            2f_1
            \cos\left(\frac{\pi d}{2}\right)
        \right)
    \bigg]
    A_{\Delta^A_I}\cdot A_{\Delta^A_I}
    \\
    +
    2
    \sqrt{
        \left|
            c_{\Delta^A_+}^{\mathrm{dS-AdS}}
        \right|
        \left|
            c_{\Delta^A_-}^{\mathrm{dS-AdS}}
        \right|
    }
    \left[
        f_2
        \cos\left(\frac{\pi d}{2}\right)
        +
        2f_1
    \right]
    A_{\Delta^A_+}\cdot A_{\Delta^A_-}.
\label{eq::MassLikeSKAnsatzEAdS}
\end{multline}
Requiring the coefficients of
$A_{\Delta^A_-}\cdot A_{\Delta^A_-}$ and
$A_{\Delta^A_+}\cdot A_{\Delta^A_-}$ to vanish, while matching the
coefficient of $A_{\Delta^A_+}\cdot A_{\Delta^A_+}$ to
\eqref{eq::DirichletMassInsertionEAdS}, gives
\begin{equation}
    f_1
    =
    \frac{\delta M_{\Delta^A_+\Delta^A_+}^{2}}{4}
    \cot\left(\frac{\pi d}{2}\right),
    \qquad
    f_2
    =
    -\frac{\delta M_{\Delta^A_+\Delta^A_+}^{2}}{2}
    \csc\left(\frac{\pi d}{2}\right),
    \qquad
    f_3
    =
    -\frac{\delta M_{\Delta^A_+\Delta^A_+}^{2}}{4},
\label{eq::SKCoefficientSolution}
\end{equation}
up to corrections of order $e^4$. Substituting
\eqref{eq::SKCoefficientSolution} into
\eqref{eq::MassLikeSKAnsatz}, the local quadratic ansatz
reproducing the projected mass insertions becomes
\begin{multline}
   i \Gamma_{\mathrm{SK}}^{(2)}\left[A_+,A_-\right]
    \supset
    \frac{\delta M_{\Delta^A_+\Delta^A_+}^{2}}{4}
    \int_{-\infty}^{\eta_0}
    \frac{{\rm d}\eta\,{\rm d}^d\vec{x}}{(-\eta)^{d+1}}
    \bigg[
        i(A_-\cdot A_--A_+\cdot A_+)\\
        +\cot\left(\frac{\pi d}{2}\right)
        \left(
            A_+\cdot A_+
            +
            A_-\cdot A_-
        \right)\\
        -
        2\csc\left(\frac{\pi d}{2}\right)
        A_+\cdot A_-
    \bigg]
    +
    O(e^4).
\label{eq::MatchedMassLikeSKAction}
\end{multline}
It is useful to analyse this result in the Keldysh basis \cite{Keldysh:1964ud},
\begin{equation}
    A_r
    \equiv
    \frac{1}{2}
    \left(
        A_++A_-
    \right),
    \qquad
    A_a
    \equiv
    A_+-A_-.
\label{eq::KeldyshBasis}
\end{equation}
Equation~\eqref{eq::MatchedMassLikeSKAction} then becomes
\begin{multline}
    i\Gamma_{\mathrm{SK}}^{(2)}\left[A_r,A_a\right]
    \supset
    \frac{\delta M_{\Delta^A_+\Delta^A_+}^{2}}{2}
    \int_{-\infty}^{\eta_0}
    \frac{{\rm d}\eta\,{\rm d}^d\vec{x}}{(-\eta)^{d+1}}
    \bigg[
        -i
        A_r\cdot A_a
        -\tan\left(\frac{\pi d}{4}\right)
        A_r\cdot A_r\\
        +
        \frac{1}{4}
        \cot\left(\frac{\pi d}{4}\right)
        A_a\cdot A_a
    \bigg]
    +
    O(e^4).
\label{eq::MatchedMassLikeSKKeldysh}
\end{multline}
The terms proportional to $A_r\cdot A_a$ and $A_a\cdot A_a$ in
\eqref{eq::MatchedMassLikeSKKeldysh} have the usual form of response
and noise contributions, respectively. Since only the mass-like
projection has been retained, their coefficients represent only part
of the complete response and noise kernels. Both terms vanish when
$A_a=0$. The $A_r\cdot A_r$ term, on the other hand, remains non-zero
in this limit. Indeed, when the same gauge field configuration is placed on the two
branches, the forward and backward scalar evolutions cancel:
\begin{equation}
\exp\left\{
i\Gamma_{\mathrm{SK}}[A,A]
\right\}
=
\langle\mathrm{BD}\rvert
U_A^\dagger U_A
\lvert\mathrm{BD}\rangle
=
1,
\label{eq::SKForwardBackwardCancellation}
\end{equation}
where $U_A$ denotes the scalar time-evolution operator in the
background gauge field $A$. Since $A_a=0$ is equivalent to
$A_+=A_-=A_r$, this implies
\begin{equation}
\Gamma_{\mathrm{SK}}[A_r,A_a=0]
=
0,
\label{eq::SKNormalisationCondition}
\end{equation}
and therefore forbids an $A_r\cdot A_r$ term in the exact quadratic
kernel. The appearance of this term in
\eqref{eq::MatchedMassLikeSKKeldysh} is a direct consequence of the
asymmetric mass correction. Reproducing a non-zero Dirichlet
insertion while setting the Neumann and mixed insertions to zero
requires the structures proportional to $f_1$ and $f_2$. In the
Keldysh basis, these generate both the noise-like $A_a\cdot A_a$ term
and the $A_r\cdot A_r$ term.

\vskip 4pt
Its apparent conflict with
\eqref{eq::SKNormalisationCondition} reflects the restriction to the
mass-like projection of the complete bilocal one-loop kernel. The non-zero $A_r\cdot A_r$ term shows that the local matching
ansatz does not by itself satisfy the Schwinger-Keldysh
normalisation identity. Equation \eqref{eq::SKNormalisationCondition} requires the total $rr$ kernel to vanish in the complete influence functional.
The present calculation determines the projected mass
insertions without reconstructing this complete kernel. This mirrors the
flat-space pole approximation described at the beginning of the
section: the isolated resonant contribution may take a non-Hermitian form, while the complete theory remains unitary. This point will be confirmed by the analysis of the next section, where we
shall argue on dispersive grounds that the Dirichlet pole should indeed be
identified with a quasinormal mode/resonance.

\subsection{Retarded structure and absorptive origin of the anomalous dimension}
\label{sec::retarded}

The one-loop mass matrix \eqref{eq::oneLoopGaugeMassMatrix} has a simple interpretation in terms of the analytic structure of the scalar bubble. Summing over the two branches of the in-in contour converts the shadow-symmetric EAdS bubble coefficient into the retarded dS kernel of Bros-Moschella \cite{Bros:1995js}. The resulting sine factors cancel the shadow poles, while the discontinuity across the principal-series produces the double cut of the scalar loop internal lines. At the gauge-field point, the tadpoles cancel in the difference between the Dirichlet and Neumann kernels, and boundary gauge invariance sets the complete Neumann mass correction to zero.
The Dirichlet mass correction, and hence the anomalous dimension $\gamma_J$, is therefore determined entirely by the discontinuity.\footnote{This parallels type-A higher-spin theory in AdS$_4$, where the anomalous dimensions of the higher-spin currents vanish at the free $O(N)$ fixed point, while those at the critical fixed point are encoded in the difference between the two scalar boundary conditions and can be expressed in terms of bulk cuts \cite{Giombi:2017hpr}.}

\paragraph{The EAdS resolvent and bubble.}
To establish this relation, we first make the pole structure of the EAdS bubble manifest, before incorporating the sum over the in-in branches. Since the bulk-to-boundary propagators are eigenfunctions of the EAdS Laplacian, they diagonalise convolution with the EAdS Green function: 
\begin{align}
\int_{\mathrm{EAdS}}\mathrm{d}^{d+1}y\,\sqrt{g(y)}\,
G_{\Delta}(x,y)K_{\Delta_3}(y,\vec y_2)
&=
C^{(d)}_{\Delta;\Delta_3}
K_{\Delta_3}(x,\vec y_2),
\label{resolvent}
\\
C^{(d)}_{\Delta;\Delta_3}
&\equiv
\frac{1}{(\Delta-\Delta_3)(\Delta+\Delta_3-d)}\,.
\end{align}
More generally, by EAdS isometries, $K_{\Delta_3}$ is an eigenfunction of convolution with any invariant kernel $f(\xi)$ (see appendix~\ref{subsec::expectisomsc}), with eigenvalue given by \eqref{eq:xikformula} applied term by term to the series expansion of $f$. For the Green function this eigenvalue is \eqref{resolvent}. We refer to $C^{(d)}_{\Delta;\Delta_3}$ as the resolvent eigenvalue. Defining
\begin{equation}
{\cal C}(\Delta)\equiv\Delta(d-\Delta),
\end{equation}
it can be written as
\begin{equation}
C^{(d)}_{\Delta;\Delta_3}
=
\frac{1}{{\cal C}(\Delta_3)-{\cal C}(\Delta)}\,,
\end{equation}
which is the eigenvalue of the inverse kinetic operator on the mode $K_{\Delta_3}$. Since
${\cal C}(\Delta_3)={\cal C}(d-\Delta_3)$, the single pole in the Casimir variable corresponds to the shadow-related pair $\Delta_3=\Delta$ and $\Delta_3=d-\Delta$.

\vskip 4pt
Inserting the propagator-product decomposition \eqref{321} into
\eqref{eq::scalarBubbleReducedForm}, the bubble eigenvalue becomes a
sum of resolvent eigenvalues,\footnote{For $d<3$ the series converges.
For $d\geq3$ it determines the bubble up to a polynomial in
the Casimir ${\cal C}(\Delta_3)$, corresponding to the local ultraviolet
terms discussed in section~\ref{subsub:UVandSDexp}.}
\begin{align}
C^{(d)}_{\Delta_1,\Delta_2;\Delta_3}
&=
\sum_{k=0}^{\infty}
a^{(d)}_{\Delta_1,\Delta_2}(k)\,
C^{(d)}_{\Delta_{12}+2k;\Delta_3}
\nonumber\\
&=
\sum_{k=0}^{\infty}
\frac{a^{(d)}_{\Delta_1,\Delta_2}(k)}
{\left(\Delta_{12}+2k-\Delta_3\right)
 \left(\Delta_{12}+2k-d+\Delta_3\right)}\,,
\qquad
\Delta_{12}\equiv\Delta_1+\Delta_2.
\label{eq::Cresolvent}
\end{align}
This expression is invariant under $\Delta_3\to d-\Delta_3$ and has
two shadow-related families of poles,
\begin{equation}
\Delta_3=\Delta_{12}+2k,
\qquad
\Delta_3=d-\Delta_{12}-2k,
\qquad
k=0,1,2,\ldots\,,
\end{equation}
corresponding respectively to the double-trace dimensions and their
shadows.

\vskip 4pt
All dependence on the internal bubble kernel is contained in this eigenvalue. Attaching a second bulk-to-boundary propagator therefore gives the factorised form fixed by EAdS isometries (see appendix~\ref{subsec::expectisomsc}):
\begin{multline}
\int_{\mathrm{EAdS}}\mathrm{d}^{d+1}x\,\sqrt{g(x)}\,
K_{\Delta_3'}(x,\vec y_1)
\int_{\mathrm{EAdS}}\mathrm{d}^{d+1}y\,\sqrt{g(y)}\,
G_{\Delta_1}(x,y)G_{\Delta_2}(x,y)
K_{\Delta_3}(y,\vec y_2)
\\
=
C^{(d)}_{\Delta_1,\Delta_2;\Delta_3}\,
{\cal T}_{\Delta_3'\Delta_3}(\vec y_1,\vec y_2),
\end{multline}
where 
\begin{equation}
{\cal T}_{\Delta_3'\Delta_3}(\vec y_1,\vec y_2)
\equiv
\int_{\mathrm{EAdS}}\mathrm{d}^{d+1}x\,\sqrt{g(x)}\,
K_{\Delta_3'}(x,\vec y_1)
K_{\Delta_3}(x,\vec y_2).
\label{eq::twoPtContact}
\end{equation}
For $\Delta_3, \Delta_3'$ on the principal series, this integral evaluates to
\begin{equation}
{\cal T}_{\Delta_3'\Delta_3}(\vec y_1,\vec y_2)
=
2\pi\,\delta(\Delta_3+\Delta_3'-d)\,
\delta^{(d)}(\vec y_{12})
+
2\pi\,\delta(\Delta_3-\Delta_3')\,
\frac{c_{\Delta_3}}
     {(\vec y_{12}^{\,2})^{\Delta_3}},
\end{equation}
where $c_{\Delta_3}$ is defined in
\eqref{eq::sc2ptnorm}.

\vskip 4pt
\paragraph{The retarded dS resolvent.}
We now turn to dS and incorporate the sum over the two branches of the in-in contour. As we show below, summing over the branch of an integrated point selects the retarded dS two-point kernel. The rotation to EAdS makes its analytic structure explicit: it expresses the retarded resolvent in
terms of the shadow-symmetric EAdS resolvent eigenvalues derived above, multiplied by phases associated with the two in-in branches. The interference of these phases produces sine factors that cancel the shadow poles and leave the retarded pole prescription. We first demonstrate this
for a single propagator, before relating the result to the full in-in two-point function and applying the same construction term by term to the bubble decomposition \eqref{eq::Cresolvent}.

\vskip 4pt
Let $b=\pm$ denote the branch of the integrated point and $a=\pm$ that of
the other bulk point. Using the rotation to EAdS reviewed in
section \ref{subsec::rottoeads}, the dS counterpart of
\eqref{resolvent} is 
\begin{align}
\sum_{b=\pm}
(b\,i)
\int_{\mathrm{dS}_b}\mathrm{d}^{d+1}y&\,\sqrt{-g(y)}\,
G_{ab}(x,y)\,
K^{b}_{\Delta_3}(y,\vec y_2)
\nonumber\\
&=
\sum_{\sigma=\pm}
c^{\mathrm{dS\text{-}AdS}}_{\Delta_{\sigma}}\,
c^{\mathrm{dS\text{-}AdS}}_{\Delta_{3}}\,
e^{-ia\frac{\pi}{2}\Delta_{\sigma}}\,
\frac{
2\sin\left[
\frac{\pi}{2}
\left(\Delta_{\sigma}+\Delta_{3}-d\right)
\right]
}{
\left(\Delta_{\sigma}-\Delta_{3}\right)
\left(\Delta_{\sigma}+\Delta_{3}-d\right)
}
K_{\Delta_{3}}(x,\vec y_2)
\nonumber\\
&\equiv
C^{(R)}_{\Delta;\Delta_{3}}\,
K^{a}_{\Delta_{3}}(x,\vec y_2).
\label{eq::dSresolvent}
\end{align}
Here
\begin{equation}
\Delta_+\equiv\Delta,
\qquad
\Delta_-\equiv d-\Delta,
\end{equation}
are the two falloffs of the internal field, while
$\Delta_{3}$ is that of the external field. The final equality defines
\begin{equation}
C^{(R)}_{\Delta;\Delta_3}
=
\sum_{\sigma=\pm}
c^{\mathrm{dS\text{-}AdS}}_{\Delta_\sigma}\,
e^{a\frac{i\pi}{2}(\Delta_3-\Delta_\sigma)}\,
2\sin\left[
\frac{\pi}{2}
(\Delta_\sigma+\Delta_3-d)
\right]
C^{(d)}_{\Delta_\sigma;\Delta_3}.
\label{eq::CRsingle}
\end{equation}
Each EAdS resolvent in this sum has poles at
$\Delta_3=\Delta_\sigma$ and $\Delta_3=d-\Delta_\sigma$. The sine
factor vanishes at the latter and therefore cancels the shadow pole.

\vskip 4pt
The superscript $(R)$ reflects the causal meaning of the in-in branch sum. Indeed, the same weighted sum applied to a contour-ordered propagator gives
\begin{align}
\sum_{b=\pm}(b\,i)\,G_{ab}(x,y)
&=
i\left[G_{a+}(x,y)-G_{a-}(x,y)\right]
\nonumber\\
&=
i\,\theta\!\left(x^0-y^0\right)
\left\langle
\left[\phi(x),\phi(y)\right]
\right\rangle
\equiv
G^{(R)}(x,y).
\label{eq::rprop}
\end{align}
For $x^0<y^0$, the two contour-ordered propagators in the first line coincide and cancel. For $x^0>y^0$, their difference is the commutator. The result is therefore independent of the branch $a$ of the unintegrated point. This is the largest-time identity \cite{tHooft:1973wag}. Accordingly, $C^{(R)}_{\Delta;\Delta_3}$ is the spectral eigenvalue of the retarded propagator. The sinusoidal factor appearing after the rotation to EAdS is the EAdS image of this causal branch sum.

\vskip 4pt
The cancellation of the shadow poles and the resulting pole prescription are particularly transparent on the principal series. We set
\begin{equation}
\Delta_\sigma
=
\frac d2+i\sigma\mu+\epsilon,
\qquad
\Delta_3
=
\frac d2+i\nu,
\qquad
\mu,\nu>0,
\end{equation}
where the real shift $\epsilon>0$ places the EAdS integrals inside their
convergence region. One finds
\begin{align}
C^{(R)}_{\Delta;\Delta_3}
&=-\frac{1}{\sinh(\pi\mu)}
\left[e^{-a\frac{\pi}{2}(\nu-\mu)}
\frac{\sinh\left[\frac{\pi}{2}(\nu+\mu)\right]}{(\nu-\mu+i\epsilon)(\nu+\mu-i\epsilon)}
-e^{-a\frac{\pi}{2}(\nu+\mu)}
\frac{\sinh\left[\frac{\pi}{2}(\nu-\mu)\right]}{(\nu-\mu-i\epsilon)(\nu+\mu+i\epsilon)}
\right]
\nonumber\\
&=\frac{1}{\mu^2-(\nu+i\epsilon)^2}.
\label{eq::CRprincipal}
\end{align}
Each term in the first expression has one pole in each half-plane. At the pole in the upper half-plane, however, the corresponding hyperbolic sine vanishes, leaving an $O(\epsilon)$ residue. Hence $C^{(R)}_{\Delta;\Delta_3}$ is analytic in the upper half of the $\nu$-plane. Equivalently, $C^{(R)}_{\Delta;\Delta_3}$ is the spherical Fourier-Laplace transform, or convolution eigenvalue, of the retarded
propagator on dS introduced in \cite{Bros:1990cu,Bros:1992dr,Bros:1995js}. Here it arises directly from the EAdS representation of the in-in branch sum.

\vskip 4pt
So far, we have performed the in-in branch sum only at the integrated point $y$. We now attach the second external leg and sum over the branch of $x$, thereby reconstructing the full in-in two-point function. We then apply the same construction to the scalar bubble.

\paragraph{The in-in two-point function.}
The full in-in two-point function is obtained by attaching a second external leg to the half-diagram and summing over its branch. Letting $\Delta_3'$ and $\Delta_3$ denote the dimensions of the two external
legs, we have
\begin{multline}
\mathcal I^{\,\mathrm{in\text{-}in}}_{\Delta;\Delta_3'\Delta_3}
(\vec y_1,\vec y_2)
\equiv
\sum_{a,b=\pm}(a\,i)(b\,i)
\int_{\mathrm{dS}_a}\mathrm{d}^{d+1}x\,\sqrt{-g(x)}
\int_{\mathrm{dS}_b}\mathrm{d}^{d+1}y\,\sqrt{-g(y)}
\\
\times
K^{a}_{\Delta_3'}(x,\vec y_1)\,
G_{ab}(x,y)\,
K^{b}_{\Delta_3}(y,\vec y_2)\,.
\label{eq::CompleteInInTwoPoint}
\end{multline}
Using the half-diagram result \eqref{eq::dSresolvent}, this reduces
immediately to
\begin{equation}
\mathcal I^{\,\mathrm{in\text{-}in}}_{\Delta;\Delta_3'\Delta_3}
(\vec y_1,\vec y_2)
=
C^{(R)}_{\Delta;\Delta_{3}}
\sum_{a=\pm}(a\,i)
\int_{\mathrm{dS}_a}\mathrm{d}^{d+1}x\,\sqrt{-g(x)}\,
K^a_{\Delta_3'}(x,\vec y_1)\,
K^a_{\Delta_{3}}(x,\vec y_2)\,.
\label{eq::fullFromHalfDiagram}
\end{equation}
All dependence on the internal propagator is now contained in $C^{(R)}_{\Delta;\Delta_{3}}$. It remains to perform the integral over $x$ and the sum over its branch $a$. After rotating this integral to EAdS according to section \ref{subsec::rottoeads}, the sum over in-in branches gives
\begin{equation}
\sum_{a=\pm}(a\,i)\,e^{-ia\frac{\pi}{2}(\Delta'_{3}+\Delta_{3}-d)}
=
2\sin\!\left[\frac{\pi}{2}\left(\Delta'_{3}+\Delta_{3}-d\right)\right].
\end{equation}
It follows that
\begin{equation}
\mathcal I^{\,\mathrm{in\text{-}in}}_{\Delta;\Delta_3'\Delta_3}(\vec y_1,\vec y_2)
=
c^{\mathrm{dS\text{-}AdS}}_{\Delta'_{3}}\,c^{\mathrm{dS\text{-}AdS}}_{\Delta_{3}}\,
2\sin\!\left[\frac{\pi}{2}\left(\Delta'_{3}+\Delta_{3}-d\right)\right]
C^{(R)}_{\Delta;\Delta_{3}}\,
{\cal T}_{\Delta'_{3}\,\Delta_{3}}(\vec y_1,\vec y_2)\,.
\label{eq::CompleteInInTwoPointEAdS}
\end{equation}
The complete in-in two-point function is obtained by summing \eqref{eq::CompleteInInTwoPointEAdS} over the shadows of both external dimensions,
\begin{equation}
\widehat{\mathcal I}^{\,\mathrm{in\text{-}in}}_{\Delta;\Delta_3'\Delta_3}
\equiv
\sum_{\tau,\bar\tau=\pm}
\mathcal I^{\,\mathrm{in\text{-}in}}_{\Delta;\Delta'_{3,\bar\tau}\Delta_{3,\tau}}\,,
\label{eq::shadowSum}
\end{equation}
since the boundary limit of each dS propagator contains both falloffs \eqref{SKbubo}, where $\Delta_{3,+}\equiv\Delta_3$, $\Delta_{3,-}\equiv d-\Delta_3$ and similarly for $\Delta_3'$. We will perform this sum below and show that it is responsible for the cancellation of the contact terms.

\vskip 4pt
In the following we take the external dimensions on
the principal series,
\begin{equation}
\Delta_{3,\tau}=\frac d2+i\tau\nu,
\qquad
\Delta'_{3,\bar\tau}=\frac d2+i\bar\tau\bar\nu,
\qquad
\nu,\bar\nu>0.
\end{equation}
Equation \eqref{eq::twoPtContact} then gives
\begin{equation}
{\cal T}_{\Delta'_{3,\bar\tau}\,\Delta_{3,\tau}}
(\vec y_1,\vec y_2)
=
2\pi\,
\delta\!\left(\bar\tau\bar\nu+\tau\nu\right)
\delta^{(d)}(\vec y_{12})
+
2\pi\,
\delta\!\left(\bar\tau\bar\nu-\tau\nu\right)
\frac{c_{\Delta_{3,\tau}}}
     {\left(\vec y_{12}^{\,2}\right)^{\Delta_{3,\tau}}}\,.
\label{eq::principalPairing}
\end{equation}
For $\nu,\bar\nu>0$, the first term is supported at
$\bar\tau=-\tau$ and $\bar\nu=\nu$. On this support, the external
dimensions are shadow-related,
\begin{equation}
\Delta'_{3,\bar\tau}=d-\Delta_{3,\tau},
\end{equation}
and the interference factor in
\eqref{eq::CompleteInInTwoPointEAdS} vanishes:
\begin{equation}
2\sin\!\left[
\frac{\pi}{2}
\left(
\Delta'_{3,\bar\tau}
+\Delta_{3,\tau}
-d
\right)
\right]
=0.
\end{equation}
The local contribution therefore drops out of the complete in-in sum.

\vskip 4pt
The second term in \eqref{eq::principalPairing} is supported at
$\bar\tau=\tau$ and $\bar\nu=\nu$. The external dimensions then
coincide,
\begin{equation}
\Delta'_{3,\bar\tau}=\Delta_{3,\tau},
\end{equation}
and the interference factor becomes
\begin{equation}
2\sin\!\left[
\pi\left(\Delta_{3,\tau}-\frac d2\right)
\right]
=
-\frac{1}{
c^{\mathrm{dS\text{-}AdS}}_{\Delta_{3,\tau}}
}.
\end{equation}
Substituting the surviving non-local contribution into \eqref{eq::CompleteInInTwoPointEAdS}, we obtain
\begin{equation}
\widehat{\mathcal I}^{\,\mathrm{in\text{-}in}}_{\Delta;\Delta_3'\Delta_3}
(\vec y_1,\vec y_2)
=
-2\pi\,\delta(\Delta_3-\Delta_3')
\sum_{\tau=\pm}
c^{\mathrm{dS\text{-}AdS}}_{\Delta_{3,\tau}}\,
C^{(R)}_{\Delta;\Delta_{3,\tau}}\,
\frac{c_{\Delta_{3,\tau}}}
     {\left(\vec y_{12}^{\,2}\right)^{\Delta_{3,\tau}}}\,,
\label{eq::InInIsDisc}
\end{equation}
which is entirely expressed in terms of the retarded resolvent. For real $\nu$ the two terms are complex conjugates of each other, so that the result is real, as required for the equal-time correlator of a real field. We now apply this relation to the one-loop scalar bubble. 

\vskip 4pt
\paragraph{The scalar bubble.}
The preceding argument applies without change when the single
propagator is replaced by the scalar bubble. We define its retarded
eigenvalue by 
\begin{multline}
\sum_{b=\pm}(b\,i)
\int_{\mathrm{dS}_b}\mathrm{d}^{d+1}y\,\sqrt{-g(y)}\,
G_{\Delta_1,ab}(x,y)\,
G_{\Delta_2,ab}(x,y)\,
K^b_{\Delta_3}(y,\vec y_2)
\\
=
C^{(R)}_{\Delta_1,\Delta_2;\Delta_{3}}\,
K^a_{\Delta_{3}}(x,\vec y_2).
\label{eq::retardedBubbleDefinition}
\end{multline}
It follows immediately from \eqref{eq::InInIsDisc} that the
corresponding contribution to the full in-in two-point function is
\begin{equation}
\widehat{\mathcal I}^{\,\mathrm{in\text{-}in}}_{\Delta_1,\Delta_2;
\Delta_3'\Delta_3}
(\vec y_1,\vec y_2)
=
-2\pi\,\delta(\Delta_3-\Delta_3')
\sum_{\tau=\pm}
c^{\mathrm{dS\text{-}AdS}}_{\Delta_{3,\tau}}\,
C^{(R)}_{\Delta_1,\Delta_2;\Delta_{3,\tau}}\,
\frac{c_{\Delta_{3,\tau}}}
     {\left(\vec y_{12}^{\,2}\right)^{\Delta_{3,\tau}}}.
\label{eq::fullInInBubble}
\end{equation}
It therefore remains to determine the retarded bubble eigenvalue.

Choosing $a=+$ in \eqref{eq::retardedBubbleDefinition}, and using the
rotation to EAdS, gives
\begin{multline}
C^{(R)}_{\Delta_1,\Delta_2;\Delta_3}
=
\sum_{\sigma_1,\sigma_2=\pm}
c^{\mathrm{dS\text{-}AdS}}_{\Delta_{1,\sigma_1}}\,
c^{\mathrm{dS\text{-}AdS}}_{\Delta_{2,\sigma_2}}\,
e^{\frac{i\pi}{2}
\left(
\Delta_3-\Delta_{1,\sigma_1}-\Delta_{2,\sigma_2}
\right)}
\\
\times
2\sin\!\left[
\frac{\pi}{2}
\left(
\Delta_{1,\sigma_1}
+\Delta_{2,\sigma_2}
+\Delta_3-d
\right)
\right]
C^{(d)}_{\Delta_{1,\sigma_1},
\Delta_{2,\sigma_2};\Delta_3},
\label{eq::retardedBubbleEAdS}
\end{multline}
where
\begin{equation}
\Delta_{i,+}=\Delta_i,
\qquad
\Delta_{i,-}=d-\Delta_i.
\end{equation}
The result is independent of the choice of the external branch $a$,
as follows from the defining retarded combination in
\eqref{eq::retardedBubbleDefinition}.

\vskip 4pt
For each choice of the internal falloffs, the EAdS coefficient has the
resolvent expansion \eqref{eq::Cresolvent}. Its poles occur at
\begin{equation}
\Delta_3
=
\Delta_{1,\sigma_1}
+\Delta_{2,\sigma_2}
+2k
\end{equation}
and at their shadows,
\begin{equation}
\Delta_3
=
d-\Delta_{1,\sigma_1}
-\Delta_{2,\sigma_2}
-2k.
\end{equation}
At a shadow pole, the sine in
\eqref{eq::retardedBubbleEAdS} becomes
\begin{equation}
\sin(-\pi k)=0,
\end{equation}
and cancels the pole. The retarded bubble therefore retains only the
double-trace poles
\begin{equation}
\Delta_3
=
\Delta_{1,\sigma_1}
+\Delta_{2,\sigma_2}
+2k,
\qquad
k=0,1,2,\ldots\,.
\end{equation}

The same causal interpretation applies to the scalar bubble. Replacing
$G_{\Delta,ab}(x,y)$ in \eqref{eq::rprop} by
$G_{\Delta_1,ab}(x,y)G_{\Delta_2,ab}(x,y)$ gives the retarded
one-loop kernel, whose spectral eigenvalue is
$C^{(R)}_{\Delta_1,\Delta_2;\Delta_3}$. In terms of
\begin{equation}
\nu=-i\left(\Delta_3-\frac d2\right),
\end{equation}
as above, the cancellation of the shadow poles implies that all poles of
$C^{(R)}_{\Delta_1,\Delta_2;\Delta_3}$ lie in the lower half
$\nu$-plane. The retarded bubble is therefore analytic for
$\operatorname{Im}\nu>0$ and admits a dispersion relation.\footnote{For $d<3$, the dispersion relation is unsubtracted. For
$d\geq3$, it holds up to a subtraction polynomial in the Casimir
${\cal C}(\Delta_3)$, corresponding to the local ultraviolet terms
discussed following \eqref{eq::Cresolvent}.}${}^{,}$\footnote{Continuing the internal dimensions to the complementary series can move poles into the upper half-plane. In that case, the spectral representation must include the corresponding discrete complementary-series contributions. Equation \eqref{eq::CRdispersion} below is written for the regime in which these additional contributions are absent.}

\vskip 4pt
As in the single-propagator case, $C^{(R)}_{\Delta_1,\Delta_2;\Delta_3}$
is the spherical Laplace transform of the invariant retarded bubble
kernel. The EAdS bubble
coefficient is itself symmetric under
$\Delta_3\to d-\Delta_3$. The retarded prescription is supplied by the
sum over the in-in branches. In the EAdS representation, this causal
information is encoded by the sinusoidal vertex factors, which cancel
the shadow poles and produce the required half-plane analyticity.

\vskip 4pt
\paragraph{The discontinuity and the double cut.} We now take the discontinuity of the retarded bubble across the principal-series. With a normalization convenient for the spectral
representation, we define
\begin{equation}
\operatorname{disc}_{\Delta_3}
C^{(R)}_{\Delta_1,\Delta_2;\Delta_3}
\equiv
\frac{\Delta_3-\frac d2}{2\pi}
\left[
C^{(R)}_{\Delta_1,\Delta_2;\Delta_3}
-
C^{(R)}_{\Delta_1,\Delta_2;d-\Delta_3}
\right].
\label{eq::retardedDisc}
\end{equation}
For $\Delta_3=\frac d2+i\nu$, this combination is proportional to the imaginary part of the retarded kernel and therefore isolates its absorptive part. We now show that it is given by the double cut of the scalar loop. Any local term polynomial in the Casimir $\mathcal{C}(\Delta_3)$ cancels in \eqref{eq::retardedDisc}.

\begin{figure}[h]
    \centering

    \includegraphics[
        width=\textwidth,
        height=0.3\textheight,
        keepaspectratio
    ]{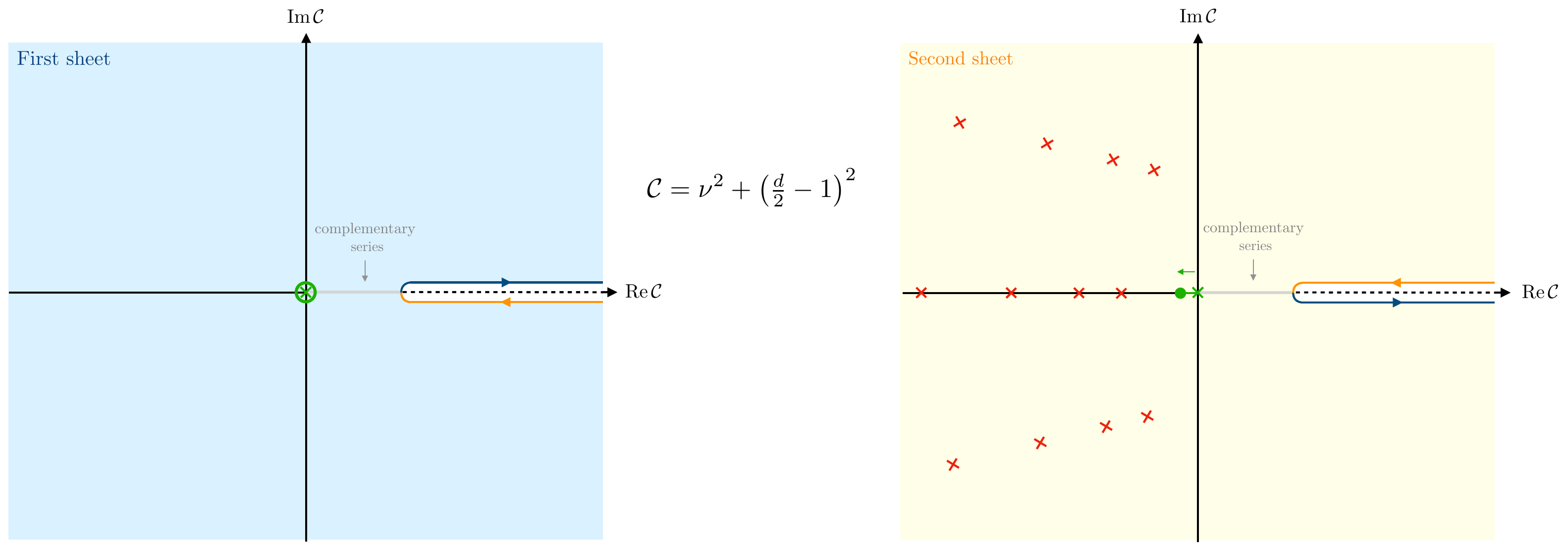} 
    \caption{Analytic structure in the complex spin-one Casimir plane, $\mathcal C=\nu^2+(d/2-1)^2$, for $d>2$ and principal-series internal scalars $\Delta^\varphi_\pm=d/2\pm i\mu$.
Photon poles are shown in green and two-particle poles of
the retarded gauge-field bubble in red.
The first and second sheets correspond to the upper and
lower half $\nu$-planes, respectively, with
$\Delta^A=d/2+i\nu$, and are joined along the principal-series
cut $\mathcal C\geq(d/2-1)^2$.
The free Neumann and Dirichlet poles both have
$\mathcal C=0$, but lie on the first and second sheets,
respectively.
The Neumann pole remains fixed, while the scalar loop shifts
the Dirichlet pole on the second sheet.
The arrow shows this displacement for $\gamma_J>0$.
The two-particle poles are shown at their leading-order
positions, with their placement in the Casimir plane schematic.}\label{fig:poles2}

    \vspace{1em}

    \includegraphics[
        width=\textwidth,
        height=0.3\textheight,
        keepaspectratio
    ]{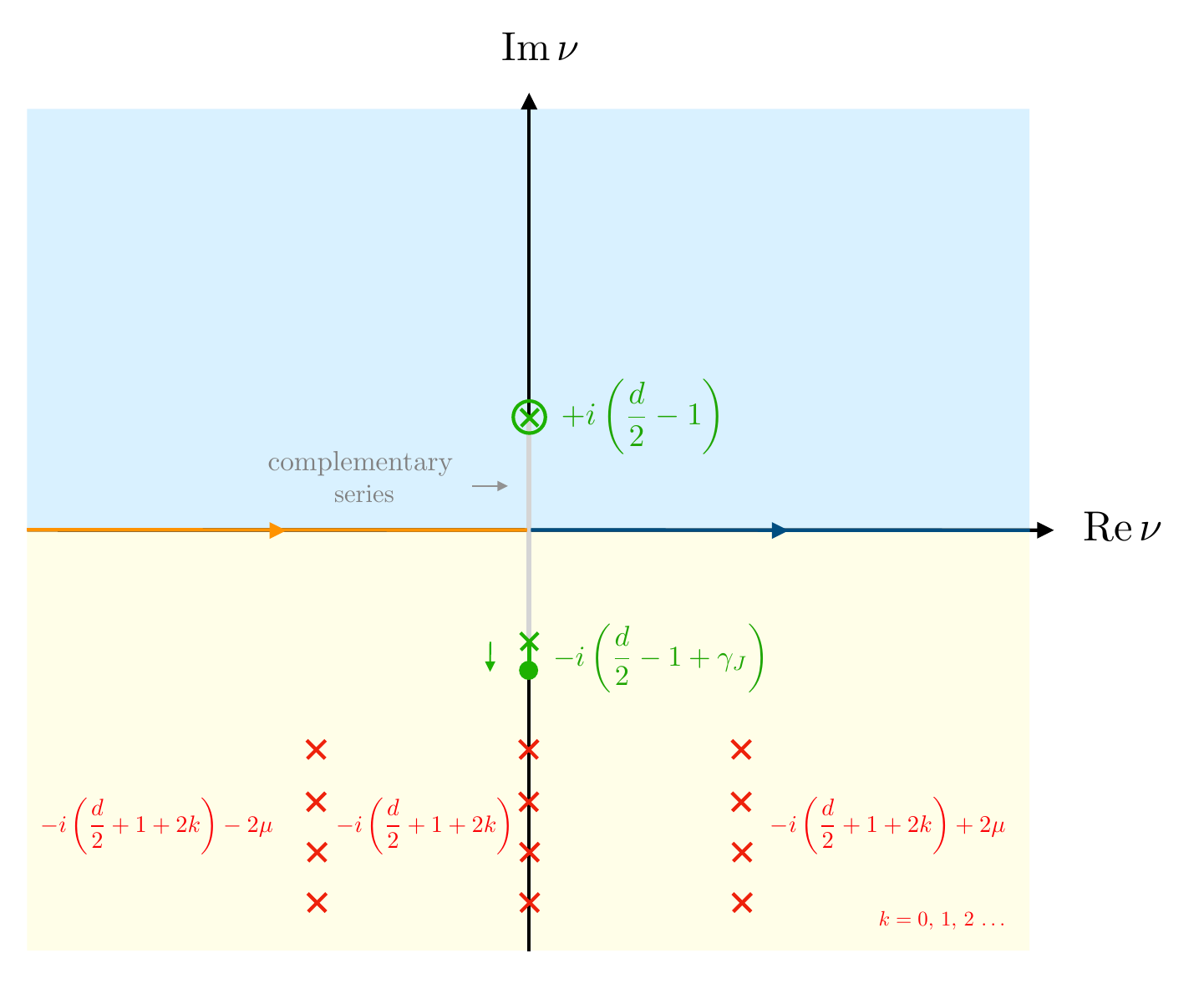}
    \caption{The pole structure in the complex $\nu$-plane, with $\Delta^A=d/2+i\nu$.
Green markers denote photon poles and red crosses denote
the leading-order two-particle poles of the retarded
gauge-field bubble.
The principal series lies on the real axis.
For principal-series internal scalars, the bubble poles lie
in the lower half-plane and the retarded bubble is analytic
for $\operatorname{Im}\nu>0$.
The protected Neumann pole lies at
$\nu=i(d/2-1)$.
The scalar loop shifts the Dirichlet pole from
$\nu=-i(d/2-1)$ to $\nu=-i(d/2-1+\gamma_J)$,
corresponding to $\Delta^A_+=d-1+\gamma_J$.
The arrow shows this displacement for $\gamma_J>0$.}\label{fig:poles1}
\end{figure}

\vskip 4pt
Term by term in the sum over the internal boundary conditions in EAdS, the phase and sine appearing in
\eqref{eq::retardedBubbleEAdS} satisfy
\begin{equation}
2e^{\frac{i\pi}{2}
(\Delta_3-\Delta_1-\Delta_2)}
\sin\!\left[
\frac{\pi}{2}
(\Delta_1+\Delta_2+\Delta_3-d)
\right]
=
\frac{1}{i}
\left[
e^{i\pi\left(\Delta_3-\frac d2\right)}
-
e^{-i\pi\left(\Delta_1+\Delta_2-\frac d2\right)}
\right].
\label{eq::phaseSineSplit}
\end{equation}
The second term on the right-hand side is independent of $\Delta_3$
and cancels in \eqref{eq::retardedDisc}. This gives
\begin{multline}
\operatorname{disc}_{\Delta_3}
C^{(R)}_{\Delta_1,\Delta_2;\Delta_3}
=
\frac{\Delta_3-\frac d2}{\pi}
\sin\!\left[
\pi\left(\Delta_3-\frac d2\right)
\right]
\\
\times
\sum_{\sigma_1,\sigma_2=\pm}
c^{\mathrm{dS\text{-}AdS}}_{\Delta_{1,\sigma_1}}\,
c^{\mathrm{dS\text{-}AdS}}_{\Delta_{2,\sigma_2}}\,
C^{(d)}_{\Delta_{1,\sigma_1},
\Delta_{2,\sigma_2};\Delta_3}.
\label{eq::shadowOddCut}
\end{multline}
We define the EAdS double cut by 
\begin{equation}
\operatorname{Cut}_{\Delta_1\Delta_2}
C^{(d)}_{\Delta_1,\Delta_2;\Delta_3}
\equiv
\sum_{\sigma_1,\sigma_2=\pm}
c^{\mathrm{dS\text{-}AdS}}_{\Delta_{1,\sigma_1}}\,
c^{\mathrm{dS\text{-}AdS}}_{\Delta_{2,\sigma_2}}\,
C^{(d)}_{\Delta_{1,\sigma_1},
\Delta_{2,\sigma_2};\Delta_3}.
\label{eq::cutDef}
\end{equation}
Hence, 
\begin{equation}
\operatorname{disc}_{\Delta_3}
C^{(R)}_{\Delta_1,\Delta_2;\Delta_3}
=
-\frac{\Delta_3-\frac d2}
{2\pi\,
c^{\mathrm{dS\text{-}AdS}}_{\Delta_3}}\,
\operatorname{Cut}_{\Delta_1\Delta_2}
C^{(d)}_{\Delta_1,\Delta_2;\Delta_3}.
\label{eq::discAsCut}
\end{equation}
To relate this definition to a cut of the internal lines, recall that
the bubble coefficient is defined by
\begin{equation}
\int_{\mathrm{EAdS}}
\mathrm{d}^{d+1}y\,\sqrt{g(y)}\,
G_{\Delta_1}(x,y)G_{\Delta_2}(x,y)
K_{\Delta_3}(y,\vec y_2)
=
C^{(d)}_{\Delta_1,\Delta_2;\Delta_3}
K_{\Delta_3}(x,\vec y_2).
\label{eq::bubbleCoefficientRecall}
\end{equation}
Multiplying this relation by the two dS-EAdS coefficients and
summing over the internal boundary conditions gives
\begin{multline}
\operatorname{Cut}_{\Delta_1\Delta_2}
C^{(d)}_{\Delta_1,\Delta_2;\Delta_3}\,
K_{\Delta_3}(x,\vec y_2)
\\
=
\int_{\mathrm{EAdS}}
\mathrm{d}^{d+1}y\,\sqrt{g(y)}\,
K_{\Delta_3}(y,\vec y_2)
\sum_{\sigma_1,\sigma_2=\pm1}
c^{\mathrm{dS\text{-}AdS}}_{\Delta_{1,\sigma_1}}\,
c^{\mathrm{dS\text{-}AdS}}_{\Delta_{2,\sigma_2}}\,
G_{\Delta_{1,\sigma_1}}(x,y)
G_{\Delta_{2,\sigma_2}}(x,y).
\label{eq::cutAtKernelLevel}
\end{multline}
Using $c^{\mathrm{dS\text{-}AdS}}_{\Delta_{i,\sigma_i}}
=
\sigma_i\,
c^{\mathrm{dS\text{-}AdS}}_{\Delta_i}$, the sum in the integrand factorises:
\begin{multline}
\sum_{\sigma_1,\sigma_2=\pm1}
c^{\mathrm{dS\text{-}AdS}}_{\Delta_{1,\sigma_1}}\,
c^{\mathrm{dS\text{-}AdS}}_{\Delta_{2,\sigma_2}}\,
G_{\Delta_{1,\sigma_1}}(x,y)
G_{\Delta_{2,\sigma_2}}(x,y)
\\
 =
c^{\mathrm{dS\text{-}AdS}}_{\Delta_1}\,
c^{\mathrm{dS\text{-}AdS}}_{\Delta_2}\,
\bigl[G_{\Delta_{1,+}}(x,y)-G_{\Delta_{1,-}}(x,y)\bigr]
\bigl[G_{\Delta_{2,+}}(x,y)-G_{\Delta_{2,-}}(x,y)\bigr].
\label{eq::propagatorDoubleCut}
\end{multline}
For each internal line, the difference between the two boundary conditions cancels the bulk delta-function source and leaves the corresponding delta-function-normalisable harmonic function,\footnote{Note that the overall odd factor in $\nu$ is the Jacobian of the transformation that brings us from the Casimir plane to the $\nu$ plane in figure \ref{fig:poles1}.}
\begin{equation}
\Omega^{(d)}_{\nu}(x,y)
\equiv
\frac{i\nu}{2\pi}
\left[
G^{(d)}_{\frac d2+i\nu}(x,y)
-
G^{(d)}_{\frac d2-i\nu}(x,y)
\right]\equiv\operatorname{disc}_{\frac{d}{2}+i\nu}\,G_{\frac{d}{2}+i\nu}^{(d)}.
\label{eq::scalarHarmonic}
\end{equation}
Each internal Green function is replaced by an
on-shell harmonic function. Since this happens for both internal lines, it gives the double cut of the scalar bubble, in direct analogy with the Cutkosky double cut in flat space.

\paragraph{Gauge-field kinematics.} We now specialise to the gauge-field
two-point function. Under the dimension shift
\eqref{eq::vectorBubbleScalarBubbleRelation}, the spin-one bubble in
boundary dimension $d$ is expressed as a scalar bubble in boundary
dimension
\begin{equation}
D=d+2,
\end{equation}
with internal dimensions
$\Delta^\varphi_++1$ and $\Delta^\varphi_-+1$. The Dirichlet and Neumann gauge field kernels correspond respectively to the external
scalar dimensions
\begin{equation}
\Delta_3=d,
\qquad
\Delta_3=2.
\end{equation}
These are shadow-related in $D$ dimensions. It follows that
\begin{equation}
\left.
\operatorname{disc}_{\Delta_3}
C^{(R),(d+2)}_{\Delta^\varphi_++1,\,
\Delta^\varphi_-+1;\Delta_3}
\right|_{\Delta_3=d}
=
\frac{d-2}{4\pi}
\left[
C^{(R),(d+2)}_{\Delta^\varphi_++1,\,
\Delta^\varphi_-+1;d}
-
C^{(R),(d+2)}_{\Delta^\varphi_++1,\,
\Delta^\varphi_-+1;2}
\right].
\label{eq::gaugeDiscontinuity}
\end{equation}

The tadpole contributions are independent of $\Delta_3$ and therefore
cancel in the difference between the two gauge-field kernels. Moreover,
the complete Neumann mass correction vanishes by boundary gauge
invariance. Consequently, the Dirichlet mass correction is equal to
the Dirichlet-Neumann difference. Using the dimension-shift relation,
\begin{align}
(d-2)\gamma_J
&=
\delta M_{\mathrm D}^2-\delta M_{\mathrm N}^2
\nonumber\\
&=
8\pi e^2
\left[
C^{(R),(d+2)}_{\Delta^\varphi_++1,\,
\Delta^\varphi_-+1;d}
-
C^{(R),(d+2)}_{\Delta^\varphi_++1,\,
\Delta^\varphi_-+1;2}
\right]
\nonumber\\
&=
\frac{32\pi^2e^2}{d-2}
\left.
\operatorname{disc}_{\Delta_3}
C^{(R),(d+2)}_{\Delta^\varphi_++1,\,
\Delta^\varphi_-+1;\Delta_3}
\right|_{\Delta_3=d}.
\end{align}
We therefore obtain
\begin{equation}
\gamma_J
=
\frac{32\pi^2e^2}{(d-2)^2}
\left.
\operatorname{disc}_{\Delta_3}
C^{(R),(d+2)}_{\Delta^\varphi_++1,\,
\Delta^\varphi_-+1;\Delta_3}
\right|_{\Delta_3=d}.
\label{eq::gammaJIsCut}
\end{equation}
By \eqref{eq::discAsCut}, the anomalous dimension is therefore fixed
entirely by the double cut of the scalar loop.

\paragraph{Evaluation of the double cut.} It remains to evaluate the double cut. Once both internal propagators have been replaced by harmonic
functions, the cut can be evaluated using their plane-wave/split
representation \cite{Bros:1995js,Penedones:2010ue}
\begin{equation} 
\Omega_{\nu}(x_1,x_2)
=
\frac{1}{4\pi}
\int_{\partial\mathrm{EAdS}}\mathrm{d}^{d}\vec z\,
K_{\frac d2+i\nu}(x_1,\vec z)\,
K_{\frac d2-i\nu}(x_2,\vec z).
\label{eq::harmonicSplit}
\end{equation}
Introducing the scalar three-point Witten diagram
\begin{equation}
\mathcal{M}^{\mathrm{3pt}}_{\Delta_1,\Delta_2,\Delta_3}
(\vec z_1,\vec z_2,\vec z_3)
\equiv
\int_{\mathrm{EAdS}}\mathrm{d}^{d+1}x\,\sqrt{g}\,
K_{\Delta_1}(x,\vec z_1)\,
K_{\Delta_2}(x,\vec z_2)\,
K_{\Delta_3}(x,\vec z_3),
\label{eq::scalarThreePointWitten}
\end{equation}
the double-cut contribution satisfies 
\begin{multline}
\operatorname{Cut}_{\Delta_1\Delta_2}
C^{(d)}_{\Delta_1,\Delta_2;\Delta_3}\,
\mathcal{T}_{\Delta_3\Delta_3}(\vec y_1,\vec y_2)
\\
=
\frac{
c^{\mathrm{dS\text{-}AdS}}_{\Delta_1}\,
c^{\mathrm{dS\text{-}AdS}}_{\Delta_2}
}{
(2\Delta_1-d)(2\Delta_2-d)
}
\int_{\partial\mathrm{EAdS}}
\mathrm{d}^d\vec z_1\,\mathrm{d}^d\vec z_2\,
\mathcal{M}^{3\mathrm{pt}}_{\Delta_3,\Delta_1,\Delta_2}
(\vec y_1,\vec z_1,\vec z_2)
\\
\times
\mathcal{M}^{3\mathrm{pt}}_{\Delta_3,d-\Delta_1,d-\Delta_2}
(\vec y_2,\vec z_1,\vec z_2).\label{eq::cutSewnThreePoint}
\end{multline}
The integrations over $\vec z_1$ and $\vec z_2$ sew the two three-point diagrams along their shadow-related legs. Factoring out the external two-point structure and using \cite[eq.~(2.14)]{Giombi:2017hpr}, one obtains
\begin{multline}
\operatorname{Cut}_{\Delta_1\Delta_2}
C^{(d)}_{\Delta_1,\Delta_2;\Delta_3}
=
\frac{1}{
32\,\pi^{2+\frac d2}\,
\Gamma\left(\tfrac d2\right)
\Gamma(\Delta_3)\Gamma(d-\Delta_3)}
\\
\times
\Gamma\left(\tfrac{\Delta_1+\Delta_2+\Delta_3-d}{2}\right)
\Gamma\left(\tfrac{\Delta_1+\Delta_2-\Delta_3}{2}\right)
\Gamma\left(\tfrac{\Delta_1+\Delta_3-\Delta_2}{2}\right)
\Gamma\left(\tfrac{\Delta_2+\Delta_3-\Delta_1}{2}\right)
\\
\times
\Gamma\left(\tfrac{d+\Delta_1-\Delta_2-\Delta_3}{2}\right)
\Gamma\left(\tfrac{d+\Delta_2-\Delta_1-\Delta_3}{2}\right)
\Gamma\left(\tfrac{d+\Delta_3-\Delta_1-\Delta_2}{2}\right)
\Gamma\left(\tfrac{2d-\Delta_1-\Delta_2-\Delta_3}{2}\right).
\label{eq::cutClosedForm}
\end{multline}
Thus, although the individual EAdS bubble coefficients are expressed
in terms of a ${}_7F_6$, their weighted alternating sum, and hence the
discontinuity of the retarded bubble, reduces to the closed
Gamma-function expression \eqref{eq::cutClosedForm}.

\vskip 4pt
We now restrict the cut to the gauge field kinematics. Writing
\begin{equation}
\Delta^\varphi_\pm=\frac d2\pm i\mu,
\end{equation}
the dimensions entering the scalar bubble in boundary dimension
$D=d+2$ are
\begin{equation}
\Delta_1=\frac D2+i\mu,
\qquad
\Delta_2=\frac D2-i\mu,
\qquad
\Delta_3=D-2=d.
\end{equation}
Substituting these values into \eqref{eq::cutClosedForm} gives
\begin{equation}
\operatorname{Cut}_{\Delta^\varphi_++1,\,
\Delta^\varphi_-+1}
C^{(d+2)}_{\Delta^\varphi_++1,\,
\Delta^\varphi_-+1;\,d}
=
\frac{
\mu\,
\Gamma\left(\tfrac d2\right)
\Gamma\left(\tfrac d2+i\mu\right)
\Gamma\left(\tfrac d2-i\mu\right)}
{16d\,\pi^{2+\frac d2}\,
\Gamma(d)\sinh(\pi\mu)}.
\label{eq::cutGaugeGamma}
\end{equation}
On the other hand, evaluating the split representation
\eqref{eq::harmonicSplit} at coincident points gives
\cite[eq.~(2.52) and app.~B]{Giombi:2017hpr}
\begin{equation}
\Omega^{(d)}_\mu(x,x)
=
\frac{
\mu\sinh(\pi\mu)\,
\Gamma\left(\tfrac d2\right)
\Gamma\left(\tfrac d2+i\mu\right)
\Gamma\left(\tfrac d2-i\mu\right)}
{4\,\pi^{2+\frac d2}\Gamma(d)}.
\label{eq::harmonicCoincident}
\end{equation}
Therefore,
\begin{equation}
\operatorname{Cut}_{\Delta^\varphi_++1,\,
\Delta^\varphi_-+1}
C^{(d+2)}_{\Delta^\varphi_++1,\,
\Delta^\varphi_-+1;\,d}
=
\frac{\Omega^{(d)}_\mu(x,x)}
{4d\,\sinh^2(\pi\mu)},
\label{eq::cutTadpole}
\end{equation}
and we recover
\begin{equation}
\gamma_J
=
-\,\frac{4\pi e^2\sin\left(\frac{\pi d}{2}\right)}
{d\,(d-2)\,\sinh^2(\pi\mu)}\,
\Omega^{(d)}_{\mu}(x,x).
\label{eq::gammaJHarmonic}
\end{equation}
The explicit factor $\sin(\pi d/2)$ makes the vanishing at even $d \geq 4$ manifest. This agrees with the boundary
multiplet recombination calculation of \cite{Sleight:2025dmt}. The
double cut isolates the two bulk scalar constituents of the recombination
operator
${\cal O}_{\Delta^\varphi_+}{\cal O}_{\Delta^\varphi_-}$, showing
directly how the bulk loop calculation reproduces the boundary result.

\vskip 4pt
\paragraph{Dispersion relation.}
The closed form \eqref{eq::cutClosedForm} also determines the
non-local part of the retarded bubble at general external dimension.
We define the K\"all\'en-Lehmann density by\footnote{For $\Delta_1=\Delta_2$ and scalar external legs, the density \eqref{eq::rhoKL}, evaluated with the closed form \eqref{eq::cutClosedForm}, coincides with the two-particle K\"all\'en-Lehmann density of \cite{Bros:2010rku}, eq. (6.25) there, with $\rho_{\rm KL}(\nu)=\nu\rho_{\rm BEM}(\nu)$ in their conventions. Through the dimension shift \eqref{eq::vectorBubbleScalarBubbleRelation}, the same expression at $D=d+2$ gives the corresponding density for spin-one external legs, which is the case relevant here.}
\begin{equation}
\rho_{\mathrm{KL}}(\nu)
\equiv
-
\left.
\operatorname{disc}_{\Delta_3}
C^{(R)}_{\Delta_1,\Delta_2;\Delta_3}
\right|_{\Delta_3=\frac d2+i\nu}
=
\frac{\nu}{\pi}\,
\operatorname{Im}
C^{(R)}_{\Delta_1,\Delta_2;\frac d2+i\nu}.
\label{eq::rhoKL}
\end{equation}
Analyticity of the retarded kernel in the upper half $\nu$-plane then
gives the dispersion relation
\begin{equation}
C^{(R)}_{\Delta_1,\Delta_2;\Delta_3}
=
2\int_0^\infty\mathrm{d}\nu\,
\frac{\rho_{\mathrm{KL}}(\nu)}
{{\cal C}\!\left(\frac d2+i\nu\right)
 -{\cal C}(\Delta_3)}
+
P_{\mathrm{UV}}\!\left({\cal C}(\Delta_3)\right),
\label{eq::CRdispersion}
\end{equation}
Here $P_{\mathrm{UV}}$ is a polynomial in the quadratic Casimir
representing the local ultraviolet terms. It is absent for $d<3$,
where the dispersion relation is unsubtracted. For $d\geq3$, it
contains the subtraction ambiguities discussed below
\eqref{eq::Cresolvent}.

\vskip 4pt
Equation \eqref{eq::CRdispersion} is the K\"all\'en-Lehmann
representation of the retarded kernel \cite{Bros:1995js}. It expresses
the kernel as a continuous superposition of the single-particle
resolvents encountered in \eqref{eq::dSresolvent}, weighted by the
two-particle density \eqref{eq::rhoKL}. The factor of $\nu$ in
\eqref{eq::rhoKL} is the Jacobian associated with the change of
variables
\begin{equation}
{\cal C}\left(\frac d2+i\nu\right)
=
\frac{d^2}{4}+\nu^2.
\end{equation}
As a check, for a free field of dimension
$\Delta=\frac{d}{2}+i\mu$ we have
\begin{equation}
\rho_{\mathrm{KL}}(\nu)
=
\frac12\,\delta(\nu-\mu),
\end{equation}
and \eqref{eq::CRdispersion} reduces to
\begin{equation}
C^{(R)}_{\Delta;\Delta_3}
=
\frac{1}
{{\cal C}(\Delta)-{\cal C}(\Delta_3)}.
\end{equation}

\vskip 4pt
The dispersion integral in \eqref{eq::CRdispersion} gives the
first-sheet value of the kernel, obtained by continuation from
$\operatorname{Re}\Delta_3<d/2$. In the Casimir plane (see figure \ref{fig:poles2}), the principal
series maps to a branch cut. Its two sides correspond to the
shadow-related values
\begin{equation}
\Delta_3=\frac d2+i\nu,
\qquad
d-\Delta_3=\frac d2-i\nu,
\end{equation}
which have the same quadratic Casimir. By the definition
\eqref{eq::retardedDisc}, the difference between the two boundary
values is
\begin{equation}
C^{(R)}_{\Delta_1,\Delta_2;\Delta_3}
-
C^{(R)}_{\Delta_1,\Delta_2;d-\Delta_3}
=
\frac{2\pi}{\Delta_3-\frac d2}\,
\operatorname{disc}_{\Delta_3}
C^{(R)}_{\Delta_1,\Delta_2;\Delta_3}.
\end{equation}
Analytic continuation through the cut therefore takes the kernel from
the first sheet to the second.

\vskip 4pt
This distinction is essential for the gauge-field application. The
dimension-shifted scalar bubble has boundary dimension $D=d+2$, for
which
\begin{equation}
{\cal C}_D(d)={\cal C}_D(2)=2d,
\qquad
{\cal C}_D(\Delta)=\Delta(D-\Delta).
\end{equation}
Thus the Dirichlet and Neumann evaluation points have the same
quadratic Casimir. For $d>2$, since $2<D/2<d$, the Neumann point
$\Delta_3=2$ lies on the first sheet, whereas the Dirichlet point
$\Delta_3=d$ lies on the second. The Dirichlet value is therefore
obtained by continuing the Neumann value through the principal-series
cut, and their difference is fixed by the analytic continuation of the
discontinuity entering \eqref{eq::gammaJIsCut}.

\vskip 4pt
The same EAdS coefficients $C^{(d)}_{\Delta_1,\Delta_2;\Delta_3}$ thus admit two different interpretations. In AdS, the double-trace poles \eqref{eq::Cresolvent} are the spectrum, they lie at real scaling dimensions, i.e. on the imaginary $\nu$ axis, and there is a single sheet. In dS, the in-in branch sum selects the retarded combination by cancelling the shadow poles. For principal-series internal fields, the surviving double-trace poles lie in the lower half $\nu$-plane and describe quasinormal modes. In the Casimir plane,
the principal series maps to a branch cut, and these poles lie
on the second sheet.

\vskip 4pt
More generally, poles of the retarded kernel in the lower half
$\nu$-plane have the interpretation of dS quasinormal modes: they
govern contributions that decay exponentially in cosmic time and do
not correspond to states in the first-sheet K\"all\'en-Lehmann
spectrum. In the free theory, the poles at
$\Delta_3=\Delta_{12}+2k$ reproduce the two-particle dS quasinormal
tower \cite{Lopez-Ortega:2006aal,Jafferis:2013qia}. Once interactions are included,
also their positions shift to
$\Delta_{12}+2k+\gamma_k$, where $\gamma_k$ denotes the corresponding
anomalous dimension. In the one-loop gauge-field two-point function,
boundary gauge invariance leaves the Neumann pole unshifted, while the
scalar loop shifts the Dirichlet pole to
$\Delta^A_+=d-1+\gamma_J$, which remains on the second sheet for either
sign of a perturbative $\gamma_J$. The shifted Dirichlet pole is
therefore a dS resonance and governs the anomalous late-time falloff of
the $\Delta^A_+$ mode \cite{Sleight:2025dmt}. This analytic structure
is illustrated in figure \ref{fig:poles1}.

\section{Conclusion}

In this work, we have provided an explicit, fully analytic one-loop bulk derivation of the anomalous dimension of the dS boundary current, first identified in \cite{Sleight:2025dmt} through the deformed boundary Ward
identities and the associated multiplet recombination.

\vskip 4pt
The computation is the dS analogue of the one-loop vacuum polarisation of
scalar QED. In flat space gauge invariance forces the vacuum polarisation
to be transverse, so that it renormalises the charge but generates no
photon mass. In dS the same diagram, through the bubble in which the two
charged scalars obey opposite boundary conditions, generates a mass-like
term for the Dirichlet mode, and, as for a self-energy with an
absorptive part in flat space, the dressed pole shifts on the second sheet, where it describes a resonance rather than a stable particle.

\vskip 4pt
In AdS a conserved current saturates the
unitarity bound and its dimension is in general protected. In dS the shadow structure
of the late-time spectrum supplies the composite operator needed for
multiplet recombination, and the current acquires an anomalous dimension.
In the EAdS reformulation this is a Schwinger-type mass for the Dirichlet
gauge field, generated by the bubble in which the two charged scalars obey
opposite boundary conditions. The Neumann mode, associated with the boundary gauge
field, remains protected by boundary gauge invariance. The same mechanism
was argued in \cite{Sleight:2025dmt} to apply to higher-spin currents and
to the stress tensor, whose Dirichlet-type modes should likewise be read as
dS resonances. The present analysis provides the explicit spin-one
realisation of this picture.

\vskip 4pt
Along the way we have clarified several aspects of the in-in formalism in
dS. The sine interference factors of the EAdS dictionary
\cite{Sleight:2019mgd,Sleight:2019hfp,Sleight:2020obc,Sleight:2021plv},
once summed over the two branches of the contour, reconstruct the dS retarded
Green function, in agreement with the spectral framework of Bros-Moschella \cite{Bros:1990cu,Bros:1995js}, of which the EAdS reformulation
offers a systematic diagrammatic realisation. This clarifies a sharp
difference between the analytic structure of the two spacetimes. In AdS
the spectrum is discrete and there is a single physical sheet, and all
singularities are poles on the imaginary $\nu$ axis, i.e.\ at real scaling
dimensions. In dS, the principal series corresponds to a branch cut in the
complex Casimir plane. The first and second sheets are represented
by the upper and lower half-planes of $\nu$, respectively. The
retarded kernel is analytic in the upper half-plane, apart from
possible complementary-series poles, while its quasinormal poles
lie in the lower half-plane. It is this structure that
manifests the dispersive and absorptive nature of dS. The retarded kernel
obeys a dispersion relation, its discontinuity across the principal series
is a two-particle cut, which we have shown to collapse to a closed product
of Gamma functions, and the anomalous dimension of the current is entirely
determined by the analytic continuation of this cut.

\vskip 4pt
The same physics is visible in the Schwinger-Keldysh effective action.
The mass-like projection that reproduces the asymmetric Dirichlet
correction is necessarily non-Hermitian. In the Keldysh basis it contains
a noise-like $A_a\cdot A_a$ term and an $A_r\cdot A_r$ term that violates
the normalisation condition $\Gamma_{\rm SK}[A,A]=0$, i.e.\ trace
preservation. Both are artefacts of isolating the resonant sector from the
complete bilocal kernel, which restores unitarity, exactly as the pole
approximation of a flat-space resonance is non-Hermitian while the full
theory is not.

\vskip 4pt
This makes the notion of quasinormal mode sharp in the expanding patch,
directly at the level of late-time in-in correlators and without reference
to a horizon. Quasinormal modes are the poles of the retarded kernel on the
second sheet. In the free theory the two-particle poles reproduce the known
dS quasinormal spectrum, and interactions shift them by their one-loop
anomalous dimensions. From this perspective, operators that in AdS saturate
the unitarity bound, like conserved currents and, by the argument of
\cite{Sleight:2025dmt}, higher-spin currents, become in dS quasinormal
modes located on the second sheet, while the first sheet retains only singularities on the real Casimir axis, such as the protected Neumann mode of
scalar QED or, by the same argument, the boundary graviton \cite{Sleight:2025dmt}. The explicit one-loop analysis for an external graviton will be reported elsewhere.

\vskip 4pt
Finally, we have explicitly shown how the discontinuity of the retarded kernel across the principal series is the cut of the diagram in the sense of Cutkosky. Taking the discontinuity with respect to the external spectral variable, each internal propagator is replaced by an on-shell harmonic function, and the result reduces to factorised three-point Witten diagrams. Cutting rules for cosmological
observables have been formulated in momentum space, first for the
wavefunction of the universe \cite{Goodhew:2020ece,Melville:2021lst} and then also for in-in correlators \cite{Cespedes:2020xqq,Goodhew:2021oqg}. The cut derived here is however of a different nature. It is a statement about the spectral variable of the
retarded kernel and it follows directly from the dispersion relation while it
carries the same physical picture as the flat-space cut of a self-energy.
Since it relies only on the retarded structure of the in-in branch sum,
we expect it to extend to more complicated diagrams.

\section*{Acknowledgments}

EB thanks Kostas Skenderis, and CS and MT thank Ugo Moschella,
for valuable discussions. This research was supported by the European Union (ERC grant ``HoloBoot'', project number 101125112),\footnote{Views and opinions expressed are however those of the author(s) only and do not necessarily reflect those of the European Union or the European Research Council. Neither the European Union nor the granting authority can be held responsible for them.} and by the INFN initiative STEFI.

\newpage

\begin{appendix}

\section{Hypergeometric relations}\label{app:hyper}

In this appendix we collect the hypergeometric identities used to evaluate
the loop integrals in the main text. These include transformation
and summation formulas, reduction identities, and a contiguous
relation for extracting ultraviolet poles.

\subsection{Transformations}

The expansion of a hypergeometric $_2F_1(a,b;c;z)$ around $z=1$ is a linear combination of two series, one starting at $(1-z)^0$ and the other at $(1-z)^{c-a-b}$. For arbitrary values of $a$, $b$ and $c$, it is given by the transformation
\begin{align}\label{eq:hyptr1}
    _2F_1\left(\begin{matrix}
        a,\ b\\
        c
    \end{matrix}\,;\,z\right) = &\,\frac{\Gamma(c)\Gamma(c-a-b)}{\Gamma(c-a)\Gamma(c-b)}{}_2F_1\left(\begin{matrix}
        a,\ b\\
        a+b-c+1
    \end{matrix}\,;\,1-z\right)\\
    &+ \frac{\Gamma(c)\Gamma(a+b-c)}{\Gamma(a)\Gamma(b)}(1-z)^{c-a-b}{}_2F_1\left(\begin{matrix}
        c-a,\ c-b\\
        c-a-b+1
    \end{matrix}\,;\,1-z\right)\,.\nonumber
\end{align}
When $a+b-c=n$ with $n$ a positive integer, the transformation becomes ill-defined and is modified to
\begin{align}\label{eq:hyptr2}
    _2F_1\left(\begin{matrix}
        a,\ b\\
        c
    \end{matrix}\,;\,z\right) = &\,\frac{\Gamma(c)\Gamma(n)}{\Gamma(a)\Gamma(b)}(1-z)^{-n}\sum_{i=0}^{n-1}\frac{(c-a)_i(c-b)_i}{(1-n)_i\,i!}(1-z)^i\\
    &-\frac{\Gamma(c)(-1)^n}{\Gamma(c-a)\Gamma(c-b)}\sum_{i=0}^\infty\frac{(a)_i(b)_i}{(n+i)!\,i!}(1-z)^i\big[\ln(1-z)+\psi(a+i)+\psi(b+i)\nonumber\\
    &\hspace{6.9cm}-\psi(1+i)-\psi(n+1+i)\big]\,.\nonumber
\end{align}

\subsection{Summation formulas at $z=1$}
\label{app::z=1}

We collect here the hypergeometric summation formulas at unit argument used in
the main text. The identities are stated in their domains of convergence and
extended elsewhere by analytic continuation.

\paragraph{Gauss summation formula.}
\begin{equation}\label{eq:2f1unit}
    {}_2F_1\left(
    \begin{matrix}
        a,\ b\\
        c
    \end{matrix}
    ;1
    \right)
    =
    \frac{\Gamma(c)\Gamma(c-a-b)}
    {\Gamma(c-a)\Gamma(c-b)} .
\end{equation}

\paragraph{Rogers-Dougall summation for a very-well-poised ${}_5F_4(1)$.} 
\begin{align}
&{}_5F_4\left(
\begin{matrix}
a,\ 1+\frac a2,\ b,\ c,\ e\\
\frac a2,\ 1+a-b,\ 1+a-c,\ 1+a-e
\end{matrix}
;1
\right)
\nonumber\\
&\qquad =
\frac{
\Gamma(1+a-b)
\Gamma(1+a-c)
\Gamma(1+a-e)
\Gamma(1+a-b-c-e)
}
{
\Gamma(1+a)
\Gamma(1+a-c-e)
\Gamma(1+a-b-e)
\Gamma(1+a-b-c)
}.
\label{eq:RogersDougall5F4}
\end{align}
See \cite[Sec. 4.4, Eq. (1)]{BaileyGHS}.

\paragraph{Two-parameter very-well-poised ${}_7F_6(1)$
summation.}

The special function required for the opposite-boundary-condition
bubble is
\begin{multline}
{}_7F_6\left(
\begin{matrix}
A,\ 1+\frac A2,\ A+B,\ A-B,\ \frac12,\ A-1,\ 1\\
\frac A2,\ 1-B,\ 1+B,\ A+\frac12,\ 2,\ A
\end{matrix}
;1
\right)
\\
=
\frac{
\Gamma\left(A+\frac12\right)
\Gamma\left(\frac32-A\right)
}
{\pi A}
\Gamma(1+B)\Gamma(1-B)
\\
\times
\left[
\frac{1}
{\Gamma(A-B)\Gamma(B-A+2)}
+
\frac{1}
{\Gamma(A+B)\Gamma(2-A-B)}
\right].
\label{eq:VeryWellPoised7F6}
\end{multline}
The common upper and lower parameter $A$ could be cancelled,
reducing the series to a ${}_6F_5(1)$. We retain the
very-well-poised ${}_7F_6(1)$ form because it is the form obtained
directly from the scalar bubble coefficient and to which Bailey's
transformation applies.

\vskip 4pt
\textit{Proof.}
Bailey's non-terminating Whipple transformation
\cite[Sec.~4.4, Eq.~(4)]{BaileyGHS} gives 
\begin{multline}
{}_7F_6\left(
\begin{matrix}
A,\ 1+\frac A2,\ A+B,\ A-B,\ \frac12,\ A-1,\ 1\\
\frac A2,\ 1-B,\ 1+B,\ A+\frac12,\ 2,\ A
\end{matrix}
;1
\right)
 =
\frac{2A-1}{A}\,
{}_3F_2\left(
\begin{matrix}
1-A,\ A-1,\ 1\\
1-B,\ 1+B
\end{matrix}
;1
\right)\\
-
\frac{
2\Gamma\left(A+\frac12\right)
\Gamma(1-B)
\Gamma(1+B)
\Gamma\left(\frac32-A\right)
}
{
A\,
\Gamma(1-A)
\Gamma(A-1)
\Gamma\left(\frac32-B\right)
\Gamma\left(\frac32+B\right)
}
{}_3F_2\left(
\begin{matrix}
\frac32-A,\ A-\frac12,\ 1\\
\frac32-B,\ \frac32+B
\end{matrix}
;1
\right).
\label{eq:BaileyReductionSpecial}
\end{multline}
The two remaining ${}_3F_2(1)$ series are summed by \cite{LavoieGrondinRathie1996}
\begin{align}
{}_3F_2\left(
\begin{matrix}
1-A,\ A-1,\ 1\\
1-B,\ 1+B
\end{matrix}
;1
\right)=
-\frac{
B\left((A-1)\sin(\pi A)+B\sin(\pi B)\right)
}
{
\left((A-1)^2-B^2\right)\sin(\pi B)
},
\label{eq:ContiguousWhipple3F2}
\end{align}
and by Whipple's summation \cite[Sec.~3.4]{BaileyGHS}
\begin{align}
{}_3F_2\left(
\begin{matrix}
\frac32-A,\ A-\frac12,\ 1\\
\frac32-B,\ \frac32+B
\end{matrix}
;1
\right) =
\frac{
\frac{\pi}{2}
\Gamma\left(\frac32-B\right)
\Gamma\left(\frac32+B\right)
}
{
\Gamma\left(\frac{A-B+1}{2}\right)
\Gamma\left(\frac{A+B+1}{2}\right)
\Gamma\left(\frac{3-A-B}{2}\right)
\Gamma\left(\frac{3-A+B}{2}\right)
}.
\label{eq:Whipple3F2}
\end{align}
Substituting \eqref{eq:ContiguousWhipple3F2} and
\eqref{eq:Whipple3F2} into \eqref{eq:BaileyReductionSpecial} proves
\eqref{eq:VeryWellPoised7F6}.

\paragraph{Shifted ${}_3F_2(1)$ summations.} For a non-negative integer $m$ \cite{Karlsson:1971hypg},
\begin{equation}
    {}_3F_2\left(
    \begin{matrix}
        a,\ b,\ f+m\\
        c,\ f
    \end{matrix}
    ;1
    \right)
    =
    \frac{\Gamma(c)\Gamma(c-a-b)}
    {\Gamma(c-a)\Gamma(c-b)}
    {}_3F_2\left(
    \begin{matrix}
        -m,\ a,\ b\\
        a+b-c+1,\ f
    \end{matrix}
    ;1
    \right).
\label{eq:3f2PositiveShift}
\end{equation}
In particular, for $m=1$,
\begin{equation}
    {}_3F_2\left(
    \begin{matrix}
        a,\ b,\ f+1\\
        c,\ f
    \end{matrix}
    ;1
    \right)
    =
    \frac{\Gamma(c)\Gamma(c-a-b)}
    {\Gamma(c-a)\Gamma(c-b)}
    \left[
        1
        -
        \frac{ab}{(a+b-c+1)f}
    \right].
\label{eq:3f2PositiveShiftOne}
\end{equation}

\subsection{Reduction formulas}

Many relations for hypergeometric functions follow from relations between the Pochhammer symbols $(a)_k = \Gamma(a+k)/\Gamma(a)$. Consider one such relation
\begin{equation}\label{eq:abrel}
    (a)_k(b)_k = \frac{a}{a-b}(a+1)_k(b)_k + \frac{b}{b-a}(a)_k(b+1)_k\,,
\end{equation}
which can be easily proven by direct check. This is used to relate a hypergeometric of parameters $_pF_q(a,b,\dotsc)$ to a sum of its contiguous $_pF_q(a+1,b,\dotsc)$ and $_pF_q(a,b+1,\dotsc)$. It may be rearranged in two useful ways
\begin{align}
    \frac{(a)_k(b)_k}{(a+1)_k(b+1)_k} &= \frac{b}{b-a}\frac{(a)_k}{(a+1)_k}+\frac{a}{a-b}\frac{(b)_k}{(b+1)_k}\,,\label{eq:ab-bb1}\\
    \frac{(a)_k(b+1)_k}{(a+1)_k(b)_k} &= \frac{b-a}{b}\frac{(a)_k}{(a+1)_k}+\frac{a}{b}\,.\label{eq:ab-bb2}
\end{align}
When the combination of parameters on the LHS are present in a hypergeometric, the first formula reduces its order by 1, while the second formula reduces it by 1 and 2. Given their recursive structure, these two formulas can be used as the building blocks of a family of reduction formulas for Pochhammer symbols, and hence for hypergeometric functions, relevant when upper and lower parameters differ by an integer. This family of relations is constructed from the reiterative action of $(c)_k/(c+1)_k$ or $(c+1)_k/(c)_k$ in \eqref{eq:ab-bb1} and \eqref{eq:ab-bb2}, and by the application of these same formulas. For instance, after one iteration one has
\begin{align}
    \frac{(a)_k(b)_k(c)_k}{(a+1)_k(b+1)_k(c+1)_k} &= \frac{b}{b-a}\frac{c}{c-a}\frac{(a)_k}{(a+1)_k}+\frac{a}{a-b}\frac{c}{c-b}\frac{(b)_k}{(b+1)_k}+\frac{a}{a-c}\frac{b}{b-c}\frac{(c)_k}{(c+1)_k},\label{eq:abcrel}\\
    \frac{(a)_k(b)_k(c+1)_k}{(a+1)_k(b+1)_k(c)_k} &= \frac{b}{b-a}\frac{c-a}{c}\frac{(a)_k}{(a+1)_k}+\frac{a}{a-b}\frac{c-b}{c}\frac{(b)_k}{(b+1)_k}\,,\label{eq:abc+1rel}\\
    \frac{(a)_k(b+1)_k(c+1)_k}{(a+1)_k(b)_k(c)_k} &= \frac{b-a}{b}\frac{c-a}{c}\frac{(a)_k}{(a+1)_k}+\frac{a}{b}\frac{(c+1)_k}{(c)_k}+\frac{a}{c}\frac{b-a}{b}\,.
\end{align}
One can continue this process to find more relations involving an arbitrary number of parameters. Moreover, by properly choosing their values, relations with larger integer shifts between upper and lower parameters can be obtained. For instance, setting $c=b+1$ in \eqref{eq:abcrel} leads to a reduction formula for $(a)_k(b)_k/[(a+1)_k(b+2)_k]$.

\subsection{Extracting divergences}

To extract the ultraviolet poles of the ${}_3F_2(1)$ functions appearing in the scalar bubble, we use a contiguous relation that increases their parametric excess $b_1+b_2-a_1-a_2-a_3$ by one. Repeated application expresses
the result in terms of series that converge at unit argument, with the poles isolated in explicit rational prefactors. The relevant identity is
\begin{multline}\label{eq:3f2div}
    _3F_2\left(\begin{matrix}
        a_1,\ a_2,\ a_3\\
        b_1,\ b_2
    \end{matrix}\,;\,1\right) = \,\frac{(b_1-a_1)(b_1-a_2)(b_1-a_3)}{b_1(b_1-b_2)(b_1+b_2-a_1-a_2-a_3)}{}_3F_2\left(\begin{matrix}
        a_1,\ a_2,\ a_3\\
        b_1+1,\ b_2
    \end{matrix}\,;\,1\right)\\
    +\frac{(b_2-a_1)(b_2-a_2)(b_2-a_3)}{b_2(b_2-b_1)(b_1+b_2-a_1-a_2-a_3)}{}_3F_2\left(\begin{matrix}
        a_1,\ a_2,\ a_3\\
        b_1,\ b_2+1
    \end{matrix}\,;\,1\right)\,.
\end{multline}
To verify this relation, define
\begin{equation}
    s=b_1+b_2-a_1-a_2-a_3,
    \qquad
    t_k=
    \frac{(a_1)_k(a_2)_k(a_3)_k}
         {(b_1)_k(b_2)_k\,k!}.
\end{equation}
The difference between the $k$th summands on the right- and
left-hand sides of \eqref{eq:3f2div} is
\begin{equation}
    \frac{(k+1)t_{k+1}-kt_k}{s}.
\end{equation}
Summing over $k$ gives zero by telescoping for
$\operatorname{Re}s>0$, since $kt_k\to0$.
The identity extends elsewhere by meromorphic continuation.
Each shifted series has parametric excess $s+1$, so iteration
of the relation isolates the poles at non-positive integer
values of $s$.

\section{Isometry constraints on EAdS bubble diagrams}
\label{app::isometry}

In this appendix we use EAdS isometries to obtain the spacetime and tensor structures of the bubble integrals employed in sections~\ref{sec::scalarloops} and \ref{sec::loopssQEDEAdS}. To derive these, one needs the transformation properties of the different propagators entering the integrals.

\vskip 4pt
The scalar bulk-to-bulk propagator is invariant under simultaneous AdS transformations of its bulk points
\begin{equation}
    G_\Delta(x_1',x_2') = G_\Delta(x_1,x_2),
\end{equation}
while for the scalar and vector bulk-to-boundary propagators, following the language of \cite{Banados:2024kza}, the former transforms as a boundary scalar primary, while the latter transforms as a bulk vector in the bulk index and as a boundary vector primary in the boundary index
\begin{align}
    K_\Delta(x',\vec{y}^{\,\prime}) &= \Omega^{-\Delta}(\vec{y})K_\Delta(x,\vec{y}),\\
    K_{\Delta,\mu i}(x',\vec{y}^{\,\prime}) &= \frac{\partial x^\nu}{\partial x'^\mu}\Omega^{-\Delta}(\vec{y})R_i{}^j(\vec{y})K_{\Delta,\nu j}(x,\vec{y}).
\end{align}
For the explicit AdS transformations $x'^\mu=x'^\mu(x)$, together with the corresponding factors of $\Omega(\vec{y})$ and $R^i{}_j(\vec{y})$, we refer the reader to \cite{Banados:2024kza}.

\subsection{Scalar bubble integral}
\label{subsec::expectisomsc}

Consider the scalar bubble integral 
\begin{equation}
    \tilde{I}^\text{B}_{\Delta_1,\Delta_2;\Delta_3}(x_1,\vec y_2)
    \equiv
    \int_{x_2}
    G_{\Delta_1}(x_1,x_2)
    G_{\Delta_2}(x_1,x_2)
    K_{\Delta_3}(x_2,\vec y_2),
\label{eq::ScalarBubbleAppendixDefinition}
\end{equation}
where $\int_{x_2}\equiv\int\dd^{d+1}x_2\sqrt{g_2}$. EAdS covariance fixes its dependence on the bulk point $x_1$ and
the boundary point $\vec y_2$ to be
\begin{equation}
    \tilde{I}^\text{B}_{\Delta_1,\Delta_2;\Delta_3}(x_1,\vec y_2)
    =
    C_{\Delta_1,\Delta_2;\Delta_3}^{(d)}
    K_{\Delta_3}(x_1,\vec y_2).
\label{eq::ScalarBubbleIsometryForm}
\end{equation}

\vskip 4pt
We now verify \eqref{eq::ScalarBubbleIsometryForm} explicitly. We first perform the boundary translation that sends $\vec y_2$ to the origin and define
\begin{equation}
    x_1'
    \equiv
    \left(
        z_1,\vec x_1-\vec y_2
    \right).
\end{equation}
We then invert the bulk coordinates,
\begin{equation}
    x_1''
    \equiv
    \frac{x_1'}{(x_1')^2},
\end{equation}
translate the spatial part of $x_1''$ to the origin, and finally
perform a dilation that sends its radial coordinate $z_1''$ to
unity. Relabelling the integration variable $x_2$ after each
transformation gives
\begin{equation}
\begin{aligned}
    \tilde{I}^\text{B}_{\Delta_1,\Delta_2;\Delta_3}
    &=
    \int_{x_2}G_{\Delta_1}(x_1',x_2)
    G_{\Delta_2}(x_1',x_2)
    K_{\Delta_3}(x_2,\vec 0)
    \\
    &=
    c_{\Delta_3}
    \int_{x_2}G_{\Delta_1}(x_1'',x_2)
    G_{\Delta_2}(x_1'',x_2)
    z_2^{\Delta_3}
    \\
    &=
    c_{\Delta_3}
    \int_{x_2}G_{\Delta_1}
    \bigl(
        (z_1'',\vec 0),x_2
    \bigr)
    G_{\Delta_2}
    \bigl(
        (z_1'',\vec 0),x_2
    \bigr)
    z_2^{\Delta_3}
    \\
    &=
    c_{\Delta_3}
    \left(
        z_1''
    \right)^{\Delta_3}
    \int_{x_2}G_{\Delta_1}
    \bigl(
        (1,\vec 0),x_2
    \bigr)
    G_{\Delta_2}
    \bigl(
        (1,\vec 0),x_2
    \bigr)
    z_2^{\Delta_3}.
\end{aligned}
\label{eq::ScalarBubbleIsometrySequence}
\end{equation}
The final integral is independent of the external data and, by
definition, equals
$C_{\Delta_1,\Delta_2;\Delta_3}^{(d)}$. Moreover,
\begin{equation}
    z_1''
    =
    \frac{
        z_1
    }{
        z_1^2+
        \left|
            \vec x_1-\vec y_2
        \right|^2
    },
\end{equation}
and hence
\begin{equation}
    c_{\Delta_3}
    \left(
        z_1''
    \right)^{\Delta_3}
    =
    K_{\Delta_3}(x_1,\vec y_2).
\end{equation}
Consequently,
\begin{equation}
    \tilde{I}^\text{B}_{\Delta_1,\Delta_2;\Delta_3}(x_1,\vec y_2)
    =
    C_{\Delta_1,\Delta_2;\Delta_3}^{(d)}
    K_{\Delta_3}(x_1,\vec y_2),
\end{equation}
as stated in \eqref{eq::ScalarBubbleIsometryForm}.

\subsection{Scalar-vector bubble integral}
\label{subsec::expectisomGB}

Consider now the scalar-vector bubble integral
\begin{equation}
    \left(I^\text{B}_{\Delta_1,\Delta_2;\Delta_3}\right)_{ij}(\vec{y}_1,\vec{y}_2) =  \int_{x_1}\int_{x_2}K_{\Delta_3,\mu i}(x_1,\vec{y}_1)G_{\Delta_1}(x_1,x_2)\overset{\leftrightarrow}{\nabla}{}_1^\mu\overset{\leftrightarrow}{\nabla}{}_2^\nu G_{\Delta_2}(x_1,x_2)K_{\Delta_3,\nu j}(x_2,\vec{y}_2)\,.
\end{equation}
We use EAdS isometries to derive
\begin{equation}
   \left(I^\text{B}_{\Delta_1,\Delta_2;\Delta_3}\right)_{ij}(\vec{y}_1,\vec{y}_2)
    =
    4A_{\Delta_1,\Delta_2;\Delta_3}^{(d)}
    \left(T_{\Delta_3\Delta_3}\right)_{ij}(\vec{y}_1,\vec{y}_2),
    \qquad
    A_{\Delta_1,\Delta_2;\Delta_3}^{(d)}
    =
    A_{\Delta_2,\Delta_1;\Delta_3}^{(d)},
\label{eq::GaugeBubbleTarget}
\end{equation}
which shows that the bubble has the tensor structure of a local Proca mass insertion.

\vskip 4pt
To this end, it is useful to decompose the bubble integral as
\begin{align}
    \left(I^\text{B}_{\Delta_1,\Delta_2;\Delta_3}\right)_{ij} 
    \equiv I^{(1)}_{ij}+I^{(2)}_{ij}-I^{(3)}_{ij}-I^{(4)}_{ij}\,,
\end{align}
where 
\begin{align} \nonumber
    I^{(1)}_{ij} &=  \int_{x_1}\int_{x_2}K_{\Delta_3,\mu i}(x_1,\vec{y}_1)\nabla_1^\mu\nabla_2^\nu G_{\Delta_1}(x_1,x_2)G_{\Delta_2}(x_1,x_2)K_{\Delta_3,\nu j}(x_2,\vec{y}_2)\,,\\ \nonumber
    I^{(2)}_{ij} &= \int_{x_1}\int_{x_2}K_{\Delta_3,\mu i}(x_1,\vec{y}_1)G_{\Delta_1}(x_1,x_2)\nabla_1^\mu\nabla_2^\nu G_{\Delta_2}(x_1,x_2)K_{\Delta_3,\nu j}(x_2,\vec{y}_2)\,,\\ \nonumber
    I^{(3)}_{ij} &= \int_{x_1}\int_{x_2}K_{\Delta_3,\mu i}(x_1,\vec{y}_1)\nabla_1^\mu G_{\Delta_1}(x_1,x_2)\nabla_2^\nu G_{\Delta_2}(x_1,x_2)K_{\Delta_3,\nu j}(x_2,\vec{y}_2)\,,\\ \nonumber
    I^{(4)}_{ij} &= \int_{x_1}\int_{x_2}K_{\Delta_3,\mu i}(x_1,\vec{y}_1)\nabla_2^\nu G_{\Delta_1}(x_1,x_2)\nabla_1^\mu G_{\Delta_2}(x_1,x_2)K_{\Delta_3,\nu j}(x_2,\vec{y}_2)\,.
\end{align}

The form \eqref{eq::GaugeBubbleTarget} is a consequence of the following two claims.

\paragraph{Claim 1:} The vector integral
\begin{equation}
    \widetilde I_j(x_1,\vec y_2)
    \equiv
   \int_{x_2}\nabla^\nu_2 G_{\Delta_1}(x_1,x_2)
    G_{\Delta_2}(x_1,x_2)
    K_{\Delta_3,\nu j}(x_2,\vec y_2)
\end{equation}
vanishes:
\begin{equation}
    \widetilde I_j(x_1,\vec y_2)=0 .
\label{eq:claim1}
\end{equation}
The same statement holds with $\Delta_1\leftrightarrow\Delta_2$. It follows from \eqref{eq:claim1}, after acting with $\nabla_1^\mu$, that
\begin{equation}
    I^{(1)}_{ij}=-I^{(4)}_{ij},
    \qquad
    I^{(2)}_{ij}=-I^{(3)}_{ij}.
\end{equation}
Therefore 
\begin{equation}
    \left(I^\text{B}_{\Delta_1,\Delta_2;\Delta_3}\right)_{ij}
    =
    -2I^{(3)}_{ij}-2I^{(4)}_{ij}.
\label{eq:QEDloopAfterClaim1}
\end{equation}

\noindent {\bf Proof:} After a translation $x_2\to x_2+\vec y_2$ and an inversion
$x_2\to x_2/x_2^2$, one finds
\begin{align}
    \widetilde{I}_j &= \int_{x_2}z_2^2\partial_{2,\nu}G_{\Delta_1}(x_1',x_2)G_{\Delta_2}(x_1',x_2)K_{\Delta_3,\nu j}(x_2,\vec{0})\,,\\
    &= C_{\Delta_3,1}\int_{x_2}z_2^{\Delta_3+1}\partial_{2,j}G_{\Delta_1}(x_1'',x_2)G_{\Delta_2}(x_1'',x_2)\,,
\end{align}
where we called $x_1'' = x_1'/x_1'^2$, $x_1'=x_1-\vec{y}_2$, and used that under the inversion of $x_2$
\begin{equation}
    K_{\Delta_3,\nu j}(x_2,\vec{0})\partial_{2,\nu}\rightarrow C_{\Delta_3,1} x_2^2 z_2^{\Delta_3-1}I_{\nu j}(x_2) \times x_2^2 I_{\nu\rho}(x_2)\partial_{2,\rho} = C_{\Delta_3,1} x_2^4 z_2^{\Delta_3-1}\partial_{2,j}\,.
\end{equation}
Using the chain rule
\begin{equation}
    \partial_{2,j}G_{\Delta_1}(x_1'',x_2) = \partial_{2,j}\xi(x_1'',x_2)\partial_\xi G_{\Delta_1}(x_1'',x_2) = \frac{\xi^2}{z_1''z_2}(\vec{x}_1^{\,\prime\prime}-\vec{x}_2)_j\partial_\xi G_{\Delta_1}(x_1'',x_2)\,.
\end{equation}
Then after a translation $x_2\rightarrow x_2+\vec{x}_1^{\,\prime\prime}$
\begin{equation}
    \widetilde{I}_j = -\frac{C_{\Delta_3,1}}{z_1''}\int_{x_2}z_2^{\Delta_3}x_{2,j}\xi^2\partial_\xi G_{\Delta_1}((z_1'',\vec{0}),x_2)G_{\Delta_2}((z_1'',\vec{0}),x_2)\,.
\end{equation}
The remaining integral is odd in $x_{2,j}$, trivially vanishing as claimed
\begin{equation}
    \widetilde{I}_j = 0\,.
\end{equation}
\paragraph{Claim 2:} 
The tensor integral
\begin{equation}
    \widetilde I^\mu{}_j(x_1,\vec y_2)
    \equiv
    \int_{x_2}\nabla_1^\mu G_{\Delta_1}(x_1,x_2)
    \nabla_2^\nu G_{\Delta_2}(x_1,x_2)
    K_{\Delta_3,\nu j}(x_2,\vec y_2)
\end{equation}
is proportional to the vector bulk-to-boundary propagator:
\begin{equation}
    \widetilde I^\mu{}_j(x_1,\vec y_2)
    =
    -g^{\mu\nu}_1
    K_{\Delta_3,\nu j}(x_1,\vec y_2)
    A^{(d)}_{\Delta_1,\Delta_2;\Delta_3},
\label{eq:claim2}
\end{equation}
where $A^{(d)}_{\Delta_1,\Delta_2;\Delta_3}=A^{(d)}_{\Delta_2,\Delta_1;\Delta_3}$. This implies 
\begin{equation}
    I_{ij}^{(3)} = I_{ij}^{(4)}\ \implies\ \left(I^\text{B}_{\Delta_1,\Delta_2;\Delta_3}\right)_{ij} = -4I_{ij}^{(3)} = 4A^{(d)}_{\Delta_1,\Delta_2;\Delta_3}\left(T_{\Delta_3\Delta_3}\right)_{ij}\,.
\end{equation}
{\bf Proof:} following the same steps as before
\begin{align}
    \widetilde I^\mu{}_j &= z_1^2\int_{x_2}z_2^2\partial_{1,\mu}G_{\Delta_1}(x_1',x_2)\partial_{2,\nu}G_{\Delta_2}(x_1',x_2)K_{\Delta_3,\nu j}(x_2,\vec{0})\,,\\
    &= C_{\Delta_3,1}z_1^2\int_{x_2}z_2^{\Delta_3+1}\partial_{1,\mu}G_{\Delta_1}(x_1'',x_2)\partial_{2,j}G_{\Delta_2}(x_1'',x_2)\,.
\end{align}
Using the chain rule
\begin{align}
    \partial_{1,\mu}G_{\Delta_1}(x_1'',x_2) &= \frac{\partial x_1'^\rho}{\partial x_1^\mu}\frac{\partial x_1''^\sigma}{\partial x_1'^\rho}\frac{\partial\xi(x_1'',x_2)}{\partial x_1''^\sigma}\partial_\xi G_{\Delta_1}(x_1'',x_2)\,,\\
    &= \frac{1}{x_1'^2}I_{\mu\rho}(x_1')\left[\frac{\xi(1-\xi)}{z_1''}\delta_{\rho 0}-\frac{\xi^2}{z_1''z_2}(x_1''-x_2)_\rho\right]\partial_\xi G_{\Delta_1}(x_1'',x_2)\,,\\
    \partial_{2,j}G_{\Delta_2}(x_1'',x_2) &= \frac{\partial\xi(x_1'',x_2)}{\partial x_2^j}\partial_\xi G_{\Delta_2}(x_1'',x_2)\,,\\
    &= \frac{\xi^2}{z_1''z_2}(\vec{x}_1^{\,\prime\prime}-\vec{x}_2)_j\partial_\xi G_{\Delta_2}(x_1'',x_2)\,,
\end{align}
where we used
\begin{align}
    \partial_{1,\mu}\xi(x_1,x_2) &= \frac{\xi(1-\xi)}{z_1}\delta_{\mu0}-\frac{\xi^2}{z_1z_2}(x_1-x_2)_\mu\,,\\
    \partial_{2,\nu}\xi(x_1,x_2) &= \frac{\xi(1-\xi)}{z_2}\delta_{\nu0}+\frac{\xi^2}{z_1z_2}(x_1-x_2)_\nu\,,
\end{align}
leads to
\begin{multline}
    \widetilde I^\mu{}_j = \frac{C_{\Delta_3,1}z_1^2}{z_1''^2x_1'^2}I_{\mu\rho}(x_1')\int_{x_2}z_2^{\Delta_3}(\vec{x}_1^{\,\prime\prime}-\vec{x}_2)_j\xi^2\left[\xi(1-\xi)\delta_{\rho 0}-\frac{\xi^2}{z_2}(x_1''-x_2)_\rho\right]\\ \times \partial_\xi G_{\Delta_1}(x_1'',x_2)\partial_\xi G_{\Delta_2}(x_1'',x_2)\,.
\end{multline}
Translating $x_2\rightarrow x_2+\vec{x}_1^{\,\prime\prime}$ and dropping the terms odd in $x_{2,i}$
\begin{equation}
    \widetilde I^\mu{}_j = -\frac{C_{\Delta_3,1}z_1^2}{z_1''^2x_1'^2}I_{\mu i}(x_1')\int_{x_2}z_2^{\Delta_3-1}x_{2,i}x_{2,j}\xi^4\partial_\xi G_{\Delta_1}((z_1'',\vec{0}),x_2)\partial_\xi G_{\Delta_2}((z_1'',\vec{0}),x_2)\,.
\end{equation}
Rescaling $x_2\rightarrow z_1''x_2$ and writing everything in terms of $x_1$
\begin{equation}
    \widetilde I^\mu{}_j = -g_1^{\mu\nu}K_{\Delta_3,\nu i}(x_1,\vec{y}_2)\int_{x_2}z_2^{\Delta_3-1}x_{2,i}x_{2,j}\xi^4\partial_\xi G_{\Delta_1}((1,\vec{0}),x_2)\partial_\xi G_{\Delta_2}((1,\vec{0}),x_2)\,.
\end{equation}
The remaining integral is odd in $x_{2,i}$ and $x_{2,j}$ unless $i=j$
\begin{equation}\label{eq:Aconstant}
   \int_{x_2}z_2^{\Delta_3-1}x_{2,i}x_{2,j}\xi^4\partial_\xi G_{\Delta_1}((1,\vec{0}),x_2)\partial_\xi G_{\Delta_2}((1,\vec{0}),x_2) \equiv \delta_{ij}A^{(d)}_{\Delta_1,\Delta_2;\Delta_3}\,,
\end{equation}
for some constant $A^{(d)}_{\Delta_1,\Delta_2;\Delta_3}$ which can be explicitly seen to be symmetric in $\Delta_1$ and $\Delta_2$. Then
\begin{equation}
    \widetilde I^\mu{}_j(x_1,\vec{y}_2) = -g_1^{\mu\nu}K_{\Delta_3,\nu j}(x_1,\vec{y}_2)A^{(d)}_{\Delta_1,\Delta_2;\Delta_3}\,,
\end{equation}
as claimed.

\section{Comparison with flat space renormalisation}
\label{app::MaxwellRenormalisation}

The bubble coefficient $A^{(d)}_{\Delta_1,\Delta_2;\Delta_3}$ of
section~\ref{subsec::bubblediagramEAdSsQED} was computed for generic external
spin-one dimension $\Delta_3$. Away from the free Maxwell dimensions it is the
transverse vacuum polarisation form factor, whose divergent part encodes the
renormalisation of the Maxwell kinetic term. In this appendix we use it to
extract the corresponding counterterm and compare with the standard
flat-space renormalisation of scalar QED.
Let $\delta$ denote the scaling dimension of a transverse spin-one
harmonic in $\mathrm{EAdS}_{d+1}$, and define
\begin{equation}
    \lambda(\delta)
    =
    (\delta-1)(\delta-d+1).
\label{eq::SpinOneKineticEigenvalue}
\end{equation}
The two free Maxwell dimensions,
\begin{equation}
    \delta=1,
    \qquad
    \delta=d-1,
\end{equation}
are the two roots of $\lambda(\delta)=0$.
It is convenient to introduce the shifted scalar dimension
\begin{equation}
    \widehat{\Delta}=\delta+1.
\label{eq::ShiftedScalarDimension}
\end{equation}
The scalar mass associated with $\widehat{\Delta}$ in $d+2$ boundary
dimensions is then related to the spin-one kinetic eigenvalue
as
\begin{equation}
    \widehat{\Delta}
    \left(\widehat{\Delta}-d-2
    \right)
    =
    \lambda(\delta)-2d.
\label{eq::ShiftedScalarMassSpinOneEigenvalue}
\end{equation}
\vskip 4pt
Setting $\Delta_3=\delta$ in \eqref{eq::vectorBubbleScalarBubbleRelation},
\begin{equation}
    A^{(d)}_{\Delta_1,\Delta_2;\delta}
    =
    2\pi\,
    C^{(d+2)}_{\Delta_1+1,\Delta_2+1;\delta+1},
\label{eq::OffShellVectorScalarBubbleRelation}
\end{equation}
determines the complete transverse form factor from the scalar bubble
coefficient. Since the latter is invariant under $\Delta_3\to D-\Delta_3$ with
$D=d+2$, $A^{(d)}_{\Delta_1,\Delta_2;\delta}$ is invariant under
$\delta\to d-\delta$ and depends on $\delta$ only through the kinetic
eigenvalue $\lambda(\delta)$. We shall therefore write
$A^{(d)}_{\Delta_1,\Delta_2}(\lambda)\equiv A^{(d)}_{\Delta_1,\Delta_2;\delta}$
in the following; in particular the two roots of $\lambda(\delta)=0$ give the
same value, $A^{(d)}_{\Delta_1,\Delta_2}(0)=A^{(d)}_{\Delta_1,\Delta_2}$ in the
notation of \eqref{eq::Ashorthand}.

\paragraph{The divergent part of the transverse form factor.}
Without loss of generality for the divergent part, which is invariant under
$\Delta_i\to d-\Delta_i$, cf.\ \eqref{eq:Cpole5}, we focus on scalars with
equal boundary conditions propagating in the loop. Furthermore, the tadpole is a contact term, whose coefficient is independent of the external spin-one dimension and hence of the kinetic
eigenvalue $\lambda$. Gauge invariance forbids a mass for the gauge field,
exactly as in flat space, so the tadpole must be equal and opposite to the
on-shell value of the bubble at the Maxwell point $\lambda(\delta)=0$, as
verified explicitly in \eqref{eq::OrdinaryEAdSMassCancellation}.
The complete form factor, including the tadpole, therefore takes the form
\begin{equation}
    \Sigma_{\Delta}(\lambda)
    =4e^2\left[A^{(d)}_{\Delta,\Delta}(\lambda)-A^{(d)}_{\Delta,\Delta}(0)
    \right].
\label{eq::GaugeInvariantDiagonalFormFactor}
\end{equation}
The subtraction at $\lambda=0$ is the off-shell extension of the
bubble-tadpole cancellation found in \eqref{eq::DiagonalScalarCancellation}. One can then expand in $\lambda$ finding
\begin{equation}
    \Sigma_{\Delta}(\lambda)
    =
    \delta Z_A\,\lambda + c_2\lambda^2 + O(\lambda^3),
\label{eq::VectorFormFactorEFTExpansion}
\end{equation}
The absence of a constant term reflects the absence of mass renormalisation
implied by gauge invariance. The coefficient $\delta Z_A$ renormalises the
Maxwell kinetic term, while $c_2$ and the higher coefficients multiply possible higher-derivative
gauge-invariant operators.

\paragraph{Counterterm.}

The logarithmic divergence of the Maxwell kinetic term in four bulk
dimensions can be extracted directly from the scalar bubble pole at
boundary dimension $D=5$.\footnote{Since $D=d+2$ is a fixed function of $d$, a single regulator continues both sides of \eqref{eq::OffShellVectorScalarBubbleRelation}. In fact $d=3-2\epsilon$ becomes $D=5-2\epsilon$, and $D-5=d-3=-2\epsilon$.
The pole at $D=5$ is therefore the image of the four dimensional Maxwell divergence.
Consistently, the vector bubble in $\mathrm{EAdS}_{d+1}$ and the scalar bubble
in $\mathrm{EAdS}_{D+1}$ have the same superficial degree of divergence, both
converging for $d<1$, so that their pole families
$d=1,3,5,\ldots$ and $D=3,5,7,\ldots$ are in one-to-one correspondence. See
appendix~\ref{app::FlatDimensionShift}.}
Setting
\begin{equation}
    D=d+2,
    \qquad
    \Delta_3=\widehat{\Delta}=\delta+1,
\end{equation}
the external scalar mass entering \eqref{eq:Cpole5}, with $d$ replaced by $D$
throughout, is
\begin{equation}
    m_3^2
    =\widehat{\Delta}\left(\widehat{\Delta}-D\right)=\lambda(\delta)-2d.
\label{eq::ExternalScalarMassVectorEigenvalue}
\end{equation}
The part of the pole \eqref{eq:Cpole5} proportional to the spin-one
kinetic eigenvalue is therefore
\begin{equation}
    \left.
    C^{(D)}_{\Delta_1+1,\Delta_2+1;\widehat{\Delta}}
    \right|_{\mathrm{div}}
    \supset
    -
    \frac{
        \lambda
    }{
        192\pi^3(D-5)
    }.
\label{eq::ScalarBubbleKineticPole}
\end{equation}
Using
\eqref{eq::OffShellVectorScalarBubbleRelation}, one obtains
\begin{equation}
    \left.
    A^{(d)}_{\Delta_1,\Delta_2}(\lambda)
    \right|_{\mathrm{div}}
    \supset
    -
    \frac{
        \lambda
    }{
        96\pi^2(d-3)
    }.
\label{eq::VectorBubbleKineticPole}
\end{equation}
Inserting \eqref{eq::VectorBubbleKineticPole} into \eqref{eq::GaugeInvariantDiagonalFormFactor}, and noting that the subtraction at $\lambda=0$ does not affect the term linear in $\lambda$, one obtains for one complex scalar
\begin{equation}
    \Sigma_{\Delta}^{\mathrm{div}}(\lambda)
    =
    -
    \frac{
        e^2
    }{
        24\pi^2(d-3)
    }
    \lambda.
\label{eq::ScalarQEDVacuumPolarisationPoleD}
\end{equation}
\vskip 4pt
Writing
\begin{equation}
    d=3-2\epsilon,
\end{equation}
this becomes
\begin{equation}
    \Sigma_{\Delta}^{\mathrm{div}}(\lambda)
    =
    \frac{e^2}{48\pi^2\epsilon}
    \lambda.
\label{eq::ScalarQEDVacuumPolarisationPole}
\end{equation}
The corresponding minimal subtraction counterterm is then
\begin{equation}
    \delta Z_A
    =
    -\frac{e^2}{48\pi^2\epsilon}.
\label{eq::PhotonWaveFunctionCounterterm}
\end{equation}

\paragraph{Charge renormalisation and the beta function.}
The Abelian Ward identity relates the charge and gauge field
renormalisation factors:
\begin{equation}
    Z_e Z_A^{1/2}=1.
\label{eq::AbelianWardRenormalisation}
\end{equation}
It then follows that
\begin{equation}
    Z_e=1+\frac{e^2}{96\pi^2\epsilon} +  O(e^4),
\end{equation}
and the corresponding four-dimensional scalar QED beta function for one
complex scalar, obtained in the standard way from the $\epsilon$-pole of
$Z_e$, is
\begin{equation}
    \beta(e)=\frac{e^3}{48\pi^2}+O(e^5).
\label{eq::ScalarQEDBetaFunction}
\end{equation}
This provides a consistency check of the dimension-shifting
relation between the vector and scalar bubble coefficients. Ultraviolet
divergences are local and insensitive to the curvature, and must therefore
reproduce the flat-space result, as they do.

\section{Dimension-shifting relations in flat space}
\label{app::FlatDimensionShift}
The relation
\begin{equation}
    A^{(d)}_{\Delta_1,\Delta_2;\Delta_3}
    =
    2\pi\,
    C^{(d+2)}_{\Delta_1+1,\Delta_2+1;\,\Delta_3+1}
\label{eq::AppendixDimensionShift}
\end{equation}
expresses a two derivative vector bubble in $\mathrm{EAdS}_{d+1}$ through a non-derivative scalar bubble in $\mathrm{EAdS}_{d+3}$. Since the UV divergence relevant for the Maxwell kinetic term in four bulk dimensions is then read off from a pole located at boundary dimension $D=d+2=5$, that is in $d+3=6$ bulk dimensions, it is worth emphasising that this is not a peculiarity of the curved space computation. Indeed the same shift is present in the standard flat space evaluation of the scalar QED vacuum polarisation. In this appendix we recall the flat-space counterpart of \eqref{eq::AppendixDimensionShift}. Throughout, flat integrals are written in $d+1$ dimensions, the flat limit of $\mathrm{EAdS}_{d+1}$, so that the shifted integrals live in $d+3=D+1$ dimensions and a single $\epsilon$ regulates both descriptions: $d=3-2\epsilon$ is $d+1=4-2\epsilon$ and $d+3=6-2\epsilon$.

\subsection{The shift identity}
Let $f$ be any function of $\ell^2$ for which the integrals below converge. Writing $\ell^\mu\ell^\nu\to\delta^{\mu\nu}\ell^2/(d+1)$ under the angular average, and using the unit-sphere volumes $\Omega_{n}=2\pi^{n/2}/\Gamma\!\left(\frac n2\right)$ obeying
\begin{equation}
    \frac{\Omega_{d+1}}{\Omega_{d+3}}
    =
    \frac{d+1}{2\pi},
\label{eq::SphereRatio}
\end{equation}
one finds
\begin{equation}
    \int
    \frac{{\rm d}^{d+1}\ell}{(2\pi)^{d+1}}\,
    \ell^\mu\ell^\nu
    f(\ell^2)
    =
    \frac{\delta^{\mu\nu}}{d+1}
    \frac{\Omega_{d+1}}{(2\pi)^{d+1}}
    \int_0^\infty
    {\rm d}\ell\,
    \ell^{d+2}
    f(\ell^2)
    =
    2\pi\,
    \delta^{\mu\nu}
    \int
    \frac{{\rm d}^{d+3}\ell}{(2\pi)^{d+3}}\,
    f(\ell^2).
\label{eq::FlatTensorShift}
\end{equation}
The trace part of a rank-two tensor integral is therefore exactly $2\pi$ times the corresponding scalar integral in two dimensions higher. Identities of this type are the familiar dimension-shift relations of \cite{Bern:1993kr,Tarasov:1996br}. Note that the constant $2\pi$ in \eqref{eq::FlatTensorShift} is the same as in the EAdS propagator identity \eqref{eq:derivativeGshift}, and hence in \eqref{eq::AppendixDimensionShift}. Both originate from the ratio \eqref{eq::SphereRatio} of angular volumes in two adjacent dimensions.

\subsection{Scalar QED vacuum polarisation renormalisation revisited}
\label{subsec::sqedflatren}

Consider flat Euclidean scalar QED in $d+1$ dimensions with one complex scalar of mass $m$. The one-loop vacuum polarisation is the sum of a current bubble and a tadpole,
\begin{equation}
    \Pi^{\mu\nu}(p)
    =
    e^2
    \int
    \frac{{\rm d}^{d+1}\ell}{(2\pi)^{d+1}}
    \frac{
        (2\ell+p)^\mu(2\ell+p)^\nu
    }{
        \left(\ell^2+m^2\right)
        \left((\ell+p)^2+m^2\right)
    }
    -
    2e^2
    \delta^{\mu\nu}
    \int
    \frac{{\rm d}^{d+1}\ell}{(2\pi)^{d+1}}
    \frac{1}{\ell^2+m^2}.
\label{eq::FlatVacuumPolarisation}
\end{equation}
Changing variables as $\ell\to\ell-p/2$ the numerator of the bubble becomes $4\ell^\mu\ell^\nu$ which is the flat counterpart of the two-derivative structure of \eqref{eq:Aconstant}, and Feynman parametrisation gives
\begin{equation}
    \Pi^{\mu\nu}_{\rm bubble}
    =
    4e^2
    \int_0^1{\rm d}x
    \int
    \frac{{\rm d}^{d+1}\ell}{(2\pi)^{d+1}}
    \frac{
        \ell^\mu\ell^\nu
        +
        \left(x-\tfrac12\right)^2p^\mu p^\nu
    }{
        \left(\ell^2+M_x^2\right)^2
    },
    \qquad
    M_x^2=m^2+x(1-x)p^2 .
\end{equation}
One can now apply \eqref{eq::FlatTensorShift} obtaining
\begin{equation}
    \left.
        \Pi^{\mu\nu}_{\rm bubble}
    \right|_{\delta^{\mu\nu}}
    =
    8\pi e^2
    \delta^{\mu\nu}
    \int_0^1{\rm d}x\,
    B^{(d+3)}(M_x^2),
    \qquad
    B^{(d+3)}(M_x^2)
    \equiv
    \int
    \frac{{\rm d}^{d+3}\ell}{(2\pi)^{d+3}}
    \frac{1}{\left(\ell^2+M_x^2\right)^2},
\label{eq::FlatTracePart}
\end{equation}
which is the flat-space counterpart of \eqref{eq::AppendixDimensionShift}. The coefficient $A^{(d)}_{\Delta_1,\Delta_2;\Delta_3}$ was defined in \eqref{eq:Aconstant} precisely as the $\delta_{ij}$ component of the two-derivative bubble, and \eqref{eq::FlatTracePart} shows that in flat space this same component is $2\pi$ times a non-derivative bubble in two dimensions higher.
\vskip 4pt
For $d=3-2\epsilon$ the shifted integral lives in $d+3=6-2\epsilon$ dimensions, that is at boundary dimension $D=d+2=5-2\epsilon$, and
\begin{equation}
    B^{(6-2\epsilon)}(M_x^2)
    =
    \frac{\Gamma(-1+\epsilon)}
    {(4\pi)^{3-\epsilon}}
    \left(M_x^2\right)^{1-\epsilon}
    \;
    \underset{\epsilon\to0}{\longrightarrow}
    \;
    -\frac{M_x^2}{(4\pi)^3\epsilon},
\label{eq::SixDimensionalBubbleDiv}
\end{equation}
while the Feynman-parameter integral produces
\begin{equation}
    \int_0^1{\rm d}x\,M_x^2
    =
    m^2+\frac{p^2}{6}.
\label{eq::FeynmanParameterIntegral}
\end{equation}
The ubiquitous factor $p^2/6$ of the four-dimensional vacuum polarisation is thus nothing but the residue of the bubble in six dimensions and corresponds to the flat-space counterpart of the $-\tfrac13 m_3^2$ term in \eqref{eq:Cpole5}, under the identification $p^2=-m_3^2$.
\vskip 4pt
Collecting \eqref{eq::FlatTracePart}--\eqref{eq::FeynmanParameterIntegral} together with the tadpole in \eqref{eq::FlatVacuumPolarisation}, whose divergence is $-2e^2\delta^{\mu\nu}\times\left(-m^2/16\pi^2\epsilon\right)$, then gives
\begin{equation}
    \left.
        \Pi^{\mu\nu}
    \right|_{\delta^{\mu\nu},\,\mathrm{div}}
    =
    \underbrace{
        -\frac{e^2\left(m^2+\frac{p^2}{6}\right)}
        {8\pi^2\epsilon}\delta^{\mu\nu}
    }_{\text{bubble}}
    \;
    \underbrace{
        +\frac{e^2m^2}{8\pi^2\epsilon}\delta^{\mu\nu}
    }_{\text{tadpole}}
    =
    -\frac{e^2p^2}{48\pi^2\epsilon}\delta^{\mu\nu}.
\label{eq::FlatDivergenceAssembled}
\end{equation}
Two features of \eqref{eq::FlatDivergenceAssembled} have direct counterparts in the EAdS computation. First, the mass-dependent divergence cancels identically between the bubble and the tadpole. This is the flat-space version of the bubble-tadpole cancellation \eqref{eq::DiagonalScalarCancellation}, and it is what the subtraction at $\lambda=0$ in \eqref{eq::GaugeInvariantDiagonalFormFactor} implements off shell. Second, what survives is linear in $p^2$, the flat-space limit of the spin-one kinetic eigenvalue $\lambda$, and its coefficient is fixed entirely by the six-dimensional residue \eqref{eq::SixDimensionalBubbleDiv}. From \eqref{eq::FlatDivergenceAssembled} one reads off $\Pi_{\rm div}=-e^2/48\pi^2\epsilon$ and hence $\beta(e)=e^3/48\pi^2$, in agreement with \eqref{eq::ScalarQEDBetaFunction}.


\end{appendix}

\bibliographystyle{JHEP}
\bibliography{refs}

\end{document}